\documentclass[%
 reprint,
superscriptaddress,
nofootinbib,
 amsmath,amssymb,
 aps,
]{revtex4-2}

\usepackage{graphicx}
\usepackage{dcolumn}
\usepackage{bm}
\usepackage{amssymb}
\usepackage{xcolor}
\usepackage{placeins}
\usepackage[normalem]{ulem}
\definecolor{myNavy}{rgb}{0,0,.5} 
\usepackage[breaklinks=true]{hyperref}
\hypersetup{
    colorlinks=true,
    linkcolor=myNavy,
    filecolor=myNavy,
    urlcolor=myNavy,
    citecolor=myNavy
}
\usepackage{float}
\newcommand{\order}[2]{\text{$\mathcal{O}( #1^{ #2})$}}
\newcommand{\sci}[1]{\text{$ \times 10^{ #1}$}}

\newcommand{\eqbreak}{\nonumber\\&~}
\newcommand{\neweq}{\nonumber\\}
\newcommand{\gD}{g_\mathrm{D}}

\begin{document}

\preprint{APS/123-QED}
\title{Constraining dark-sector effects using gravitational waves \\from compact binary inspirals}

\author{Caroline B. Owen}%
\email{caroline.owen@unimib.it}
  \affiliation{Illinois Center for Advanced Studies of the Universe \& Department of Physics\\
   University of Illinois at Urbana-Champaign, Urbana, Illinois 61801, USA}%
\affiliation{
Dipartimento di Fisica “G. Occhialini”, Universitá degli Studi di Milano-Bicocca, \\ Piazza della Scienza 3, 20126 Milano, Italy}%
\affiliation{INFN, Sezione di Milano-Bicocca, 
Piazza della Scienza 3, 20126 Milano, Italy}
\author{Alexandria Tucker}%
  \affiliation{Illinois Center for Advanced Studies of the Universe \& Department of Physics\\
   University of Illinois at Urbana-Champaign, Urbana, Illinois 61801, USA}%
\author{Yonatan Kahn}%
  \affiliation{Illinois Center for Advanced Studies of the Universe \& Department of Physics\\
   University of Illinois at Urbana-Champaign, Urbana, Illinois 61801, USA}%
     \affiliation{Department of Physics, University of Toronto, Toronto, ON M5S 1A7, Canada}%
\author{Nicol\'as Yunes}%
  \affiliation{Illinois Center for Advanced Studies of the Universe \& Department of Physics\\
   University of Illinois at Urbana-Champaign, Urbana, Illinois 61801, USA}%

\date{\today}

\begin{abstract}
Gravitational wave observations are a powerful tool to constrain fundamental physics. This work considers dark matter that carries charge under a dark Abelian massive vector field. If such dark matter is bound inside coalescing neutron stars, the presence of the new force will modify the total energy of the binary, and the emission of dark radiation modes will impact the rate of inspiral. At leading order, the dark matter corrections introduce a Yukawa term to the potential energy and a dipole radiation mode. We calculate modifications to the binary inspiral waveform to first post-Newtonian order, incorporate these corrections into a state-of-the-art waveform model for neutron star binaries, and use this model in a Bayesian parameter estimation analysis on real data collected by current, second-generation ground-based detectors. Through this study, we place the first robust constraints on the Yukawa interaction strength and dipole emission parameter. We find that the strongest constraints are obtained from analysis of the GW170817 event, for which the Yukawa interaction strength is constrained to be less than $\mathcal{O}(10^{-2})$ and the dipole emission parameter is less than $\mathcal{O}(10^{-4})$ at 90\% credibility. 
\end{abstract}

\maketitle

\section{Introduction}\label{SEC:Intro}

The nature of dark matter (DM) is one of the most important questions in modern physics.
Despite overwhelming evidence for the existence of DM in our Universe, we have yet to detect it directly. 
Observations of the gravitational waves (GWs) produced by binaries containing neutron stars (NSs) provide a novel way to deepen our understanding of DM through its self-interactions \cite{Bertone:2019irm}.

As NSs move through galactic environments, they will collect DM through scattering, followed by gravitational capture~\cite{Goldman:1989nd,Bertone:2007ae,Kouvaris:2007ay,McDermott:2011jp,Guver:2012ba, Bertoni:2013bsa,Zurek:2013wia,Baryakhtar:2017dbj,Raj:2017wrv,Krall:2017xij,Cermeno:2017ejm,Croon:2017zcu,Kopp:2018jom,Alexander:2018qzg,Bell:2018pkk,Chen:2018ohx,Choi:2018axi,Camargo:2019wou,Bell:2019pyc,Joglekar:2019vzy,Garani:2019fpa,Hamaguchi:2019oev,Keung:2020teb,Garani:2020wge,Bell:2020lmm,Anzuini:2021lnv,Joglekar:2020liw,DeRocco:2022rze}. There are many observable effects arising from DM capture, including constraints from cooling, heating, or mini-black-hole formation inside the NS; see Ref.~\cite{Baryakhtar:2022hbu} for a review of such signatures.  Here, we consider the effects of DM self-interaction on the inspiral compact binaries \cite{Croon:2017zcu,Kopp:2018jom,Alexander:2018qzg,Bhattacharyya:2023kbh,Diedrichs:2023foj}. In particular, we focus on the case where DM interacts with itself through a dark Abelian massive vector field with mass $m_A$. A concrete model that realizes this scenario is ``atomic dark matter'' (aDM)~\cite{Kaplan:2009de,Kaplan:2011yj}, discussed in Sec.~\ref{SEC:Fields} below.

If the capture process leads to a net buildup of dark charge inside both members of a binary system, the presence of the new aDM force will modify the total energy and Kepler's third law for the orbital frequency of the binary. Furthermore, the addition of vector radiation modes will impact the rate at which that energy is carried away as the binary inspirals. 
If only one body in the binary carries dark charge, the total energy and Kepler's law will not be modified, but vector radiation modes will still occur. 
Therefore, while the no-hair theorems~\cite{Israel:1967wq,Israel:1967za,Carter:1971zc} dictate that a black hole (BH) cannot carry charge under the massive vector field, the presence of aDM bound inside the NS component(s) of an NS/NS or an NS/BH binary will imprint on the GW signal produced as the binary inspirals. The modifications to the waveform are parametrized by three dimensionless parameters:
\begin{equation}
    \label{eq:alpha-gamma-mu-def}
       \alpha = ~\frac{{\hbar c}}{G}\frac{q_1}{m_1}\frac{q_2}{m_2}, \quad 
       \gamma = \frac{\hbar c}{G}\left(\frac{q_1}{m_1}-\frac{q_2}{m_2}\right)^2,
       \quad 
       \mu = \frac{G m m_A}{\hbar c},
\end{equation}
where $q_1$ and $q_2$ are the net dark charges of the two compact objects in the binary, $m_1$ and $m_2$ are the compact object masses, $m = m_1 + m_2$ is the total mass, $G$ is Newton's constant, $\hbar$ is the reduced Planck's constant, and $c$ is the speed of light. 
Since the net dark charge on a BH must be zero, $\alpha=0$ for NS/BH binaries.
Such effects could, in principle, be detected or constrained in GW events observed by the LIGO-Virgo-KAGRA (LVK) Collaboration.

Previous work~\cite{Croon:2017zcu,Kopp:2018jom,Alexander:2018qzg} considered the signatures of dark vector forces in NS/NS binaries from the velocity-independent Yukawa force only, analogous to a dark Coulomb interaction. In particular, Ref.~\cite{Alexander:2018qzg} calculated an inspiral-only {\tt{TaylorF2}} waveform model, which incorporated modifications in the total energy and radiated power at leading post-Newtonian (PN) order.\footnote{The PN framework is an approximation technique to describe the inspiral of a compact binary when the gravitational fields are weak and the velocities are small relative to the speed of light. In the PN formalism, quantities are expanded in powers of the orbital velocity relative to the speed of light $(v/c)^{2}$ and $G m/(r c^2)$, where $r$ is the orbital separation. An $n$PN term is proportional to $c^{-2n}$ relative to the leading-order term.
Henceforth, we follow the standard conventions of PN literature, such as \cite{Poisson2014-cj,Blanchet:2013haa}.} The authors then used the stationary phase approximation (SPA) and a Fisher analysis to predict the degree to which aDM effects could be constrained or measured with future, next-generation GW observations~\cite{Alexander:2018qzg}. This study revealed that these future, high-sensitivity detectors should be able to constrain $\alpha < \mathcal{O}(10^{-5})$ and $\gamma < \mathcal{O}(10^{-7})$ with 1$\sigma$ credibility.

In this paper, we extend the work of~\cite{Alexander:2018qzg} by (i) calculating aDM modifications to higher PN order, (ii) incorporating them in a state-of-the-art waveform model for NS binaries, and (iii) using this model in a Bayesian parameter estimation analysis on real data collected by current, second-generation ground-based detectors to search for aDM effects in compact binaries.  We begin by calculating aDM modifications to the total energy (conservative sector) and radiated power (dissipative sector) of a compact binary to 1PN order. 

In the conservative sector, we first solve the field equation for the massive vector field in a PN expansion; we then infer corrections to the binary's equations of motion and, ultimately, compute the total energy in the quasicircular limit. We include not only the effects of the velocity-independent dark electric field but also the velocity-dependent magnetic field.  To make this problem tractable, we work in the long-range limit (i.e.~assuming the reduced Compton wavelength of the massive vector field $\lambda = \hbar/(m_A c)$ is much larger than the separation of the binary). 

In the dissipative sector, we compute the aDM modifications to the energy flux. While the dominant mode of gravitational radiation is quadrupolar, the presence of the aDM vector field activates an electric dipole radiation mode that dominates over the gravitational mode at low velocities, as it first enters at $-1$PN order. In addition to this electric dipole mode, we compute the electric quadrupole vector mode that first enters at Newtonian order.

With the aDM modifications to the orbital energy and the energy flux computed, we then apply the SPA to compute the frequency-domain GW strain (both the Fourier amplitude and phase) of the binary. 
We then incorporate the aDM corrections into the state-of-the-art NS binary waveform model {\tt {IMRPhenomD\_NRTidadlv2}} \cite{Dietrich:2019kaq}, which can model the inspiral and late inspiral of the binary. 

We then proceed to search for aDM modifications in GW data. We carry out a detailed Bayesian parameter estimation analysis on gravitational-wave data using the Bayesian inference library \texttt{Bilby} \cite{Ashton:2018jfp, Romero-Shaw:2020owr} with the nested sampling using \texttt{Dynesty} \cite{Speagle_2020}. We analyze the NS/NS binary events observed by the LVK Collaboration during its second (O2) and third (O3) observing runs, GW170817 \cite{LIGOScientific:2017vwq,LIGOScientific:2018hze} and GW190425 \cite{LIGOScientific:2020aai}. We also consider two NS/BH binary events, GW200105 and GW200115, observed during O3 \cite{LIGOScientific:2021qlt}. In all cases, we find no aDM effects, and therefore, we place the first robust constraints on this model. In particular, for GW170817, we find that $\alpha < \mathcal{O}(10^{-2})$ and $\gamma < \mathcal{O}(10^{-4})$, and for GW190425, we find $\alpha < \mathcal{O}(10^{-1})$ and $\gamma<\mathcal{O}(10^{-3})$ with 90\% credibility. For both NS/BH events, where $\alpha = 0$, we find that $\gamma < \mathcal{O}(10^{-1})$ with 90\% credibility. Finally, we perform an injection and recovery campaign using synthetic data to simulate observations during the fifth observing run (O5), scheduled for 2027 \cite{KAGRA:2013rdx}. We find that synthetic data of a GW170817-like signal observed with a simulated O5 detector network would place constraints on $\alpha$ and $\gamma$ that are a factor of three and an order of magnitude stronger, respectively. This demonstrates the potential for future detector networks to search even deeper for aDM effects and, if none are found, to place even more stringent constraints on its existence in NSs. 

This paper is organized as follows: In Sec. \ref{SEC:Fields}, we give the field equations that govern both the gravitational and dark-sector fields and define conventions that we will use during the calculation. We also provide further details on the astrophysical context of the aDM model. In Sec. \ref{SEC:EoM}, we compute the modified binary equation of motion. In Sec. \ref{SEC:Inspiral}, we compute corrections to the binding energy and radiated power and, ultimately, the modified frequency domain waveform. In Sec. \ref{SEC:DM_params}, we explore the aDM parameter space. In Sec. \ref{SEC:Bayes}, we present our Bayesian analysis and constraints on the aDM effects. Finally, we discuss our results and possible future work in Sec. \ref{SEC:discussion}.  

Throughout this paper, Greek indices $(\alpha,\beta,...)$ represent spacetime components with values 0, 1, 2, 3, and Latin indices $(i, j,...)$ represent spatial components with the values 1, 2, 3.
We also represent three-vectors with \textbf{bold} text and their magnitudes with \textit{italics}.
For example, $\mathbf{r}$ is a spatial vector with components $r^i$ and magnitude $r$.
We use the metric signature $(-,+,+,+)$, and Gaussian units with all factors of $G$, $c$, and $\hbar$ left explicit.  
\section{Dark Matter and Gravitational Fields}\label{SEC:Fields}
In this section, we give an overview of the aDM model, specify the dark sector and gravitational fields relevant to our analysis, and define the relevant quantities that influence the motion of the binary system at hand. We also introduce conventions that are used throughout the calculation. 

The aDM model consists of a Dirac fermion $\chi$ with mass $m_\chi$, charged under a U(1)$_D$ symmetry with gauge coupling $\gD$ and unit charge. We allow the gauge boson $A^\mu$ (which we will refer to as a dark photon) to have a nonzero mass $m_A$ but remain agnostic as to the source of the mass (St\"{u}ckelberg or spontaneous breaking of the gauge symmetry). If $\chi$ and its antiparticle comprised the galactic DM density, the DM captured in the NS would tend to annihilate rather than build up over time. Therefore, we assume that the cosmological history of this model includes a relic asymmetry that allows all the antiparticles to annihilate away~\cite{Petraki:2013wwa,Petraki:2014uza}. To ensure that the Universe does not have a net dark charge today, we imagine that there is a second Dirac fermion $\chi'$ with charge $-1$ under U(1)$_D$. The mass of $\chi'$ may be considerably different from that of $\chi$, leading to different capture rates and, therefore, to a net buildup of dark charge in the NS. In this work, we do not focus on the dynamics of the capture process, and therefore, $\chi'$ will play no role in our analysis. Because each NS will have a net charge of the same sign, the dark monopole force between NSs is purely repulsive.   See Refs.~\cite{Bansal:2022qbi,Roy:2023zar,Gemmell:2023trd} for recent studies on the cosmological and astrophysical consequences of the aDM model, including additional observables in NSs.\footnote{In particular, the name ``atomic dark matter'' for this model refers to the fact that $\chi$ and $\chi'$ may form bound states mediated by the vector force, analogous to the electromagnetic binding of the proton and electron in hydrogen. The effects of these bound states on the capture process are fertile ground for future work.}

The aDM Lagrangian density relevant to our analysis is thus
\begin{align}\label{EQ:lagrangian}
    \mathcal{L}_\mathrm{DS} =&~ -\frac{1}{4}F_{\mu\nu}F^{\mu\nu}-\frac{1}{2}\left(\frac{c}{\hbar} m_A \right)^2A^{\mu}A_{\mu}\eqbreak+\bar\chi\left(i\gamma^\mu D_\mu-\frac{c}{\hbar} m_\chi\right)\chi \,,
\end{align}
where $F_{\mu\nu}= \nabla_\mu A_\nu - \nabla_\nu A_\mu$ is the field strength tensor of the vector field and $D_\mu = \nabla_\mu +  ig_\mathrm{D} A_\mu $ is the gauge-covariant derivative. The aDM Lagrangian density is incorporated into the total Lagrangian density for our binary system 
\begin{align}
    \mathcal{L} = \mathcal{L}_\mathrm{EH}+\mathcal{L}_\mathrm{SM}+\mathcal{L}_\mathrm{DS}\,,
\end{align}
where $\mathcal{L}_\mathrm{EH}=  g^{\mu\nu}R_{\mu\nu}/(16\pi G)$ is the Einstein-Hilbert Lagrangian density of general relativity and $\mathcal{L}_\mathrm{SM}$ describes the standard model matter distribution of our system. In $\mathcal{L}_\mathrm{EH}$, $R_{\mu\nu}$ is the Ricci tensor associated with the metric tensor $g_{\mu\nu}$. 

The metric tensor does not distinguish between the DM fermion $\chi$ and Standard Model matter, so for our purposes, it is useful to regroup the Lagrangian density in the following way: 
\begin{align}
    \mathcal{L} = \mathcal{L}_\mathrm{EH}+\mathcal{L}_\mathrm{A}+\mathcal{L}_\mathrm{matter}
\end{align}
where 
\begin{align}
    \mathcal{L}_\mathrm{A} =&~ -\frac{1}{4}F_{\mu\nu}F^{\mu\nu}-\frac{1}{2}\frac{1}{\lambda^2} A^{\mu}A_{\mu}-\frac{4\pi}{c}J^\mu A_\mu\,,
\end{align}
is the Lagrangian density of the massive vector field $A^\mu$ and
\begin{align}
        \mathcal{L}_\mathrm{matter} =&~ \bar\chi\left(i\gamma^\mu \nabla_\mu-\frac{c}{\hbar} m_\chi\right)\chi + \mathcal{L}_\mathrm{SM}
\end{align}
describes both the dark and Standard Model matter that make up the binary components. 
We have written $\mathcal{L}_\mathrm{A}$ in terms of the reduced Compton wavelength of the dark vector field $\lambda = \hbar/(c m_A)$ and the dark four-current density $J^\mu=(c/4\pi) \gD \, \bar\chi\gamma^\mu\chi$.

The binary motion will be impacted by both the massive vector field $A^\alpha$ and the gravitational field $g_{\alpha\beta}$. After varying the action 
\begin{align}
    S = \int d^4x \, \sqrt{-g} \, \mathcal{L}
\end{align}
with respect to these fields, we obtain the Proca equation 
\begin{align}
        \nabla_\beta (\nabla^\alpha A^\beta - \nabla^\beta A^\alpha) + \frac{1}{\lambda^2}A^\alpha =&~
    \frac{4\pi}{c}J^\alpha\,,
\end{align}
and the Einstein equations 
\begin{align}
    \left(R_{\mu\nu} -\frac{1}{2}g^{\alpha\beta}R_{\alpha\beta} g_{\mu\nu}\right) = 8 \pi G T_{\mu\nu}\,,
\end{align}
where $T_{\mu \nu}$ is the total stress-energy tensor. In Sec.~\ref{SEC:Proca} below, we will solve the Proca equation for a binary system, so for now, let us focus on the Einstein equations.

The right-hand side of the Einstein equations contains the total stress-energy tensor, which we break up into two parts
\begin{align}
T_{\mu\nu} = T^\mathrm{A}_{\mu\nu}+T^\mathrm{matter}_{\mu\nu}\,,
\end{align}
such that 
\begin{align}
T^\mathrm{matter}_{\mu\nu} =&~ -2 \left[\frac{\delta}{g^{\mu\nu}}\mathcal{L}_\mathrm{matter}-\frac{1}{2}g_{\mu\nu}\mathcal{L}_\mathrm{matter}\right]\,,\neweq
T^\mathrm{A}_{\mu\nu} =&~ -2  \left[\frac{\delta}{g^{\mu\nu}}\mathcal{L}_\mathrm{A}-\frac{1}{2}g_{\mu\nu}\mathcal{L}_\mathrm{A}\right]\,. 
\end{align}

The contribution to the stress-energy tensor due to the massive vector field is 
\begin{align}
    T^\mathrm{A}_{\mu\nu} =&~F^\alpha_\mu F_{\alpha\nu}-\frac{1}{4}F_{\mu\nu}F^{\mu\nu}\eqbreak-\frac{1}{\lambda^2}\left(A_\mu A_\nu-\frac{1}{2}A_\alpha A^\alpha g_{\mu\nu}\right)\eqbreak
    +\frac{4\pi}{c}\left( A_\mu J_\nu+J_\mu A_\nu-J_\alpha A^\alpha g_{\mu\nu}\right).
\end{align}
The Einstein equations, without modifications due to additional fields, do an excellent job of describing all observed GW signals to date. Therefore, any deviation must be small (if at all present),  implying both $A^\alpha$ and $J^\alpha$ must be small. 
The term $T^\mathrm{A}_{\mu\nu}$ is quadratic in these fields and can, therefore, be disregarded. This choice allows us to consider the massive vector field as a test field on the standard PN binary background. 

After discarding $T^\mathrm{A}_{\mu\nu}$, the Einstein equations become 
\begin{align}
    \left(R_{\mu\nu} -\frac{1}{2}R g_{\mu\nu}\right) = 8 \pi G T^\mathrm{matter}_{\mu\nu}\,.
\end{align}
For a matter distribution that can be described as a perfect fluid  with mass density $\rho$, energy density $\epsilon$, pressure $p$, and normalized four-velocity field $u^{\mu}~\propto[c,v^i]$, we have
\begin{align}
        T^\mathrm{matter}_{\mu\nu} =&~ (\rho +\epsilon/c^2+p/c^2) u_{\mu}u_\nu+p g_{\mu\nu}\,.
\end{align}
The solution to the Einstein equations is then the standard PN expression
\begin{align}
    g_{00}=&~-1 +\frac{2}{c^2 }U + \frac{2}{c^4 }
    \left(\Psi-U^2\right) + \order{c}{-6}\,,\neweq
    g_{i0}=&~-\frac{4}{c^3}U_i+\order{c}{-5}\,,\neweq
    g_{ij}=&~\left(1 +\frac{2}{c^2 }U\right)\delta_{ij}+\order{c}{-4}\,,
\end{align}
where the potentials $U$, $U^i$ and $\Psi$ satisfy the following Poisson equations:
\begin{align}\label{EQ:Ueq}
    \nabla^2U =&~ - 4\pi G\rho\,,\neweq
    \nabla^2U^i =&~ - 4\pi G\rho v^i\,,\neweq
    \nabla^2\Psi=&~- 4\pi G\rho(\frac{3}{2}v^2-U+\epsilon/\rho +3 p/\rho)+\partial_t^2 U\,.
\end{align}

\begin{figure}
    \centering
    \includegraphics[width=.45\textwidth]{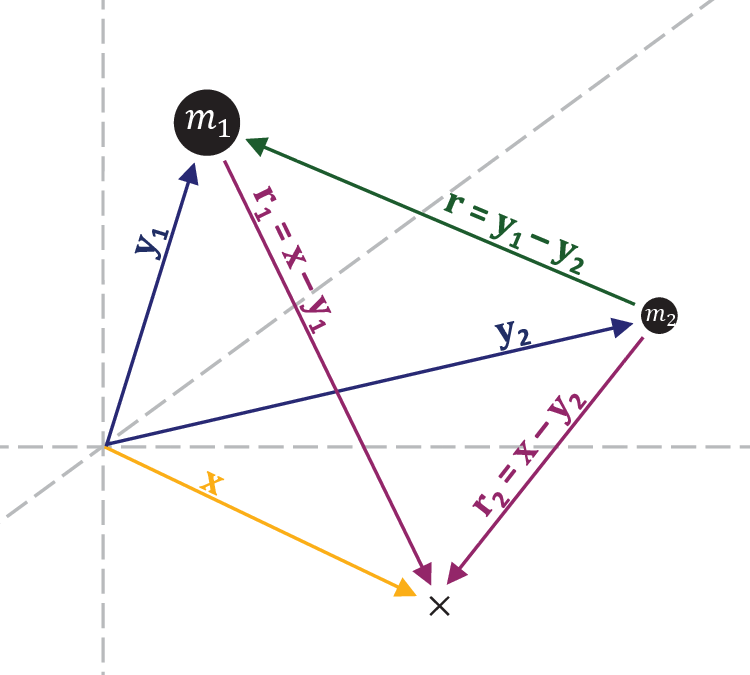}
    \caption{\label{FIG:vectors} Schematic illustration of the conventions of the spatial vectors used in this paper. The position of body $I$ relative to the origin of the coordinate system is given by $\mathbf{y}_I$. The coordinate point at which the fields are evaluated is $\mathbf{x}$. The position of the coordinate point relative to body $I$ is $\mathbf{r}_I  = \mathbf{x}-\mathbf{y}_I$. Finally, the position of body 2 relative to body 1 is $\mathbf{r}= \mathbf{y}_1-\mathbf{y}_2$. 
    }
   
    \label{fig:enter-label}
\end{figure}

Before we can solve the equations that define the PN potentials, we must introduce further notation that characterizes our binary system and that will be used throughout the remainder of the paper. 
We assume that compact body $I$ ($I = 1, 2$) has total mass $m_I$ and dark charge $q_I = \gD N_I$, where $N_I$ is the number of fermions $\chi$ that are bound inside the star.
If one of the two objects is a BH, then the charge for that object is zero by the no-hair theorems, and $q_I \to 0$ for that object.
In our choice of units, the DM charge $q_I$ is dimensionless.
Body $I$ is located at $\mathbf{y}_I $, travels with velocity $\mathbf{v}_I = d\mathbf{y}_I/dt$, and has acceleration $\mathbf{a}_I = d\mathbf{v}_I/dt$. 
The position of body $I$ relative to the field point $\mathbf{x}$ is $\mathbf{r}_I = \mathbf{x}-\mathbf{y}_I$.
The position of body 1 relative to body 2 is given by $\mathbf{r} = \mathbf{y}_1- \mathbf{y}_2 = \mathbf{r}_2- \mathbf{r}_1$. 
Figure \ref{FIG:vectors} illustrates these conventions.
From this, it follows that the relative velocity is $\mathbf{v} = d\mathbf{r}/dt=\mathbf{v}_1- \mathbf{v}_2$ and the relative acceleration is $\mathbf{a} = d\mathbf{v}/dt=\mathbf{a}_1- \mathbf{a}_2$. 
It is also useful to define unit vectors $\mathbf{n}_I =\mathbf{r}_I/r_I$ and $\mathbf{n} =\mathbf{r}/r$. 

We approximate the bodies in the binary as point particles, so $\epsilon=p=0$ and $\rho = m_1\delta^3(\mathbf{r}_1)+m_2\delta^3(\mathbf{r}_2)$, such that
\begin{align}
    U =&~ \frac{G m_1}{r_1}+ (1\leftrightarrow2)\,,\neweq
    U^i =&~ \frac{ Gm_1}{r_1}v_1^i+ (1\leftrightarrow2)\,,\neweq
    \Psi= &~\frac{G m_1}{r_1} \Bigg[2 v_1^2-\frac{1}{2}(\mathbf{n}_1\cdot\mathbf{v}_1)^2\eqbreak~~~~~~~~~~-\frac{G m_2}{r}\left(1-\frac{\mathbf{n}\cdot\mathbf{r}_1}{2r}\right)\Bigg]+ (1\leftrightarrow2)\,,
\end{align}
where $(1\leftrightarrow2)$ indicates the same quantity but with all particle labels exchanged.  Note that $\mathbf{n}$ changes sign under that exchange. 
We work with the metric in Cartesian spacetime coordinates $x^\mu = [c t, \mathbf{x}]=[c t, x,y,z]$.
In Appendix \ref{SEC:app_finitesize}, we discuss the applicability of working in the point-particle limit when dealing with a massive vector field. 

Another result of our choice to disregard the impact of the vector field on the gravitational field is that the presence of DM has a negligible impact on the binary's barycenter, which is derived from the conservation of the stress-energy tensor. When  the barycenter is taken to be the origin of the coordinate system, we have
\begin{align}\label{EQ:positions}
    \mathbf{y}_1 =& \frac{m_2}{m}\mathbf{r} + \frac{\eta \Delta}{2 c^2}\left(v^2 - \frac{G m}{r}\right)\mathbf{r}+\order{c}{-4}\,,\neweq
    \mathbf{y}_2 =& -\frac{m_1}{m}\mathbf{r} + \frac{\eta \Delta}{2 c^2}\left(v^2 - \frac{G m}{r}\right)\mathbf{r}+\order{c}{-4}\,,
\end{align}
where $m = m_1+m_2$  is the total binary mass, $\eta = m_1 m_2/m^2$ is the symmetric mass ratio, and $\Delta = \sqrt{1-4\eta}$ \cite{Poisson2014-cj,Blanchet:2013haa}.

\section{Binary Equation of Motion}\label{SEC:EoM}

Recall that we consider a binary in which each body is charged with DM of the variety described in the previous section. The first step in computing the modified binary inspiral waveform for this system is computing the modified binary equation of motion, i.e., $\mathbf{a}$ as a function of $\mathbf{r}$ and $\mathbf{v}$, as defined in the previous section. 
In this section, we do just that, keeping terms to 1PN order, or $\order{c}{-2}$ relative to the leading-order term.
To do so, we compute the Proca field produced by the dark charge on both bodies and then the acceleration of body 1 due to that Proca field and the curvature of the spacetime. 
Once we have done this, we use the problem's symmetries to infer the acceleration of the second body and, ultimately, the binary equation of motion. 
\subsection{PN solution to the field equation for the massive vector field}\label{SEC:Proca}
Recall that the massive vector field coupled to the DM charge is governed by the Proca equation
\begin{align}
    \nabla_\beta (\nabla^\alpha A^\beta - \nabla^\beta A^\alpha) + \frac{1}{\lambda^2}A^\alpha =&~
    \frac{4\pi}{c}J^\alpha,
\end{align}
where $A^\alpha = [\phi, A^i]$ is the Proca field four-potential, $J^\alpha$ is the source four-current density, $\lambda = \hbar/(c m_A)$ is the reduced Compton wavelength of the dark photon, and $\nabla^\alpha$ is the covariant derivative on the curved binary spacetime.
The Proca equation can be massaged into a more familiar form by commuting the covariant derivatives in the first term and picking up a Ricci tensor term, yielding
\begin{align}
\label{eq:proca-eq-massaged}
 \nabla^\alpha\nabla_\beta A^\beta+R^\alpha_{~\beta} A^\beta      -\nabla_\beta\nabla^\beta A^\alpha+\frac{1}{\lambda^2}A^\alpha= \frac{4\pi}{c}J^\alpha\,.
\end{align}
In the massless limit, where the Proca equation reduces to Maxwell's equation, the first term above is often set to zero via the Lorenz gauge condition $\nabla_\beta A^\beta=0$. 
However, this gauge condition cannot be applied to a massive field. 
In flat space, $\nabla_\beta A^\beta=0$ is recovered as a constraint due to DM charge conservation. 
For the \textit{nonflat} background of a binary system, we find that the Lorenz gauge can still be applied when computing corrections to the binary equation of motion to 1PN, as we show in detail in Appendix \ref{SEC:app_lorenz}. 
Applying $\nabla_\beta A^\beta=0$ leaves the following equation to solve
\begin{align}\label{EQ:proca}
    \nabla_\beta\nabla^\beta A^\alpha - R^\alpha_{~\beta} A^\beta - \frac{1}{\lambda^2}A^\alpha=- \frac{4\pi}{c}J^\alpha\,.
\end{align}

We set out to compute the Proca field produced by both bodies. The field is sourced by the current density 
\begin{align}
    J^\alpha =&~ q_1 u_1^\alpha \delta^3(\mathbf{r}_1) + q_2 u_2^\alpha \delta^3(\mathbf{r}_2) \,.  
\end{align}
For body $I$, the four-velocity is defined as $ u_I^\alpha=\gamma(v_I)v_I^\alpha$
where $v_I^\alpha = (c, \mathbf{v}_I)$ and 
\begin{align}
 \gamma(v_I^\alpha) =\left[-v_I^\alpha v_I^\beta g_{\alpha\beta}/c^2\right]^{-1/2}\,.   
\end{align}
We define a rescaled current density $\bar{J}^\alpha = J^\alpha/c$, absorbing the factor of $c^{-1}$ on the right-hand side of the Proca equation. 
The rescaled current density is 
\begin{align}\label{EQ:Jbar}
    \bar{J}^0 =&~  q_1  \left[ 1+\frac{1}{2c^2}\left(v_1^2+2U\right)+\order{c}{-4}\right]\delta^3(\mathbf{r}_1)+(1 \leftrightarrow 2) \,,\nonumber\\
    \bar{J}^i =&~  q_1\left[\frac{1}{c}v_1^i +\order{c}{-3}\right]\delta^3(\mathbf{r}_1)+(1 \leftrightarrow 2)\,.
\end{align}
To obtain the binary equation of motion through \order{c}{-2}, we must first compute the scalar potential $\phi$ through \order{c}{-2} and the vector potential $A^i$ through \order{c}{-1}. We therefore assume the following post-Newtonian Ansatz for the four-potential
\begin{align}
    A^\alpha = 
    A_{(0)}^\alpha+
    \frac{1}{c}A_{(1)}^\alpha+
    \frac{1}{c^2}A_{(2)}^\alpha\,,
\end{align}
where $A^\alpha_{(n)} = [\phi_{(n)}, A^i_{(n)}]$ is the coefficient of the \order{c}{-n} term of $A^\alpha$. Similarly, we can write
\begin{align}
    \bar J^\alpha = 
    \bar J_{(0)}^\alpha+
    \frac{1}{c}\bar J_{(1)}^\alpha+
    \frac{1}{c^2}\bar J_{(2)}^\alpha\,.
\end{align}
where the coefficients $\bar J^\alpha_{(n)}$ can be read off from Eq. \eqref{EQ:Jbar}.

We wish to solve Eq. (\ref{EQ:proca}) 
perturbatively in a PN expansion. After applying well-known results for the Christoffel symbols and Ricci tensor of this metric \cite{Poisson2014-cj}, we obtain the following three equations
\begin{align}
    &\left[\nabla^2 - \frac{1}{\lambda^2}\right]\phi_{(0)}=-4\pi \bar J_{(0)}^0\, ,
    \label{EQ:phi_0}
    \\
    &\left[\nabla^2 - \frac{1}{\lambda^2}\right]A^i_{(1)}= -4\pi \bar J_{(1)}^i\, ,
    \label{EQ:A} 
    \\
    &\left[\nabla^2 - \frac{1}{\lambda^2}\right]\phi_{(2)} = S\, ,\label{EQ:phi_2}
\end{align}
where we have defined the nonlinear source term %
\begin{align}
\label{EQsource}
   S=&~ \nabla^2(\phi_{(0)}U)+U\nabla^2\phi_{(0)}+\phi_{(0)}\nabla^2U \nonumber\\&~+ \partial_t^2\phi_{(0)} -4\pi \bar J^0_{(2)}\,,
\end{align}
and $\nabla^2$ is the flat Laplacian operator. 
In deriving these expressions, we have used the standard PN fact that since the zeroth-coordinate is $x^0 = c t$, a derivative with respect to $x^0$ is ${\cal{O}}(c^{-1})$ smaller than a spatial derivative.

Now that the equations have been derived, let us solve them.
Equations (\ref{EQ:phi_0}) and (\ref{EQ:A}) have simple source functions, and the Green's function associated with the Klein-Gordon differential operator is well-known. We can thus write the solutions immediately, namely 
\begin{align}
    \phi_{(0)} =&~  q_1 \frac{e^{-r_1/\lambda}}{r_1} + (1 \leftrightarrow 2)\,,\nonumber\\
    A^i_{(1)} =&~  q_1 \frac{e^{-r_1/\lambda}}{r_1}v_1^i + (1 \leftrightarrow 2)\,.
\end{align} 
We see that the field profile is Yukawa-suppressed by the Compton wavelength of the dark photon mass. 

Equation~(\ref{EQ:phi_2}) is significantly more difficult to solve than the other two. To make the problem tractable, we will enforce the assumption that the mass of the Proca field $m_A$ is small, or more concretely, that the length scale of the Proca interaction is much larger than the orbital separation of the binary $r/\lambda\ll1$. This assumption implies $r_1/\lambda\ll1$ and $r_2/\lambda\ll1$ because, when computing the acceleration $\mathbf{a}_1$, we evaluate the fields at the location of body 1, thus taking the limits $r_1\rightarrow 0$ and $r_2\rightarrow r$, and vice versa when computing $\mathbf{a}_2$. 

One may be concerned that the assumption $r/\lambda\ll1$ may be inconsistent with the PN assumption $Gm/(rc^2)\ll1$. Indeed, implicit in the PN approximation is the assumption of a sufficiently large binary separation so that all gravitational fields are weak. Fortunately, when considering gravitational-wave data analysis, the separation of a binary cannot be arbitrarily large because then the GW frequency would be arbitrarily low and, thus, outside the sensitivity band of ground-based detectors. Our simultaneous expansions (large separation and large interaction length scale) imply a hierarchy of scales, $\lambda\gg r \gg Gm/c^2$. We explore the regime of validity of these approximations in Sec. \ref{SEC:DM_params}, and we will apply them throughout the remainder of this paper.

Using the above approximation, we find a resummed solution for $\phi_{(2)}$ that satisfies Eq.~(\ref{EQ:phi_2})
through $\order{\lambda}{-2}$.
\begin{widetext}
\begin{align}
\label{EQ:ProcaSol}
    \phi =&~  q_1 \frac{e^{-r_1/\lambda}}{r_1}\nonumber\\
+&~\frac{q_1}{c^2}\frac{e^{-r_1/\lambda}}{r_1}\bigg\{ G \left(\frac{m_1}{ r_1}+\frac{m_2}{r_2}\right)-\frac{r_1}{2}\left(\frac{\dot r_1^2}{\lambda}-\ddot r_1\right) +\left(2 \frac{Gm_2}{r} +\frac{1}{2}v_1^2\right) +\frac{G m_2}{r}\frac{r_1}{r_2}
e^{-(r_2+r-r_1)/\lambda}\nonumber\\&~
+\frac{G m_2}{\lambda} \bigg[e^{\frac{2 r_1}{\lambda }} \text{Ei}\left(-\frac{2 }{\lambda }\left(r_1+r_2+r\right)\right)-\log \left(\frac{1}{2r}(r_1+r_2+r)\right) - 2e^{-(r_1+r_2+r)/\lambda} \nonumber\\
&~- \text{Ei}\left(-\frac{4 r}{\lambda }\right)+ 2e^{-2r/\lambda}\bigg] +\frac{G m_1}{\lambda} \bigg[e^{2 r_1/\lambda } \text{Ei}\left(-\frac{2 r_1}{\lambda }\right)-\log \left(2\frac{r_1}{\lambda}\right)+\gamma_E\bigg]\bigg\}+(1\leftrightarrow2)+\mathcal{O}(c^{-4},\lambda^{-3})\nonumber\\
 A^i =&~  \frac{q_1}{c} \frac{e^{-r_1/\lambda}}{r_1}v^i_1 +(1\leftrightarrow2)+\mathcal{O}(c^{-3})\,,
\end{align}
\end{widetext}
where  $\gamma_E$ is Euler's constant. We outline the steps taken to arrive at this solution in Appendix \ref{SEC:APP_phi}.   
\subsection{Modified binary equation of motion}
With the Proca field in hand, we are now ready to compute the acceleration of body 1 due to that Proca field and the curvature of the space-time. The four-velocity of body 1 in the presence of the Proca field obeys 
\begin{align}\label{EQ:geodesic}
u_{1}^\beta\nabla_\beta u_{1}^\alpha = - \hbar q_1 \, u_{1}^\beta F^{\alpha}{}_\beta
\end{align}
where $u_{1}^\beta$ is the body's four-velocity and 
\begin{align}
    F^{\mu\nu} = \nabla^\mu A^\nu - \nabla^\nu A^\mu\,  
\end{align} 
is the field strength tensor. 

We recall the definition of the four-velocity $u_{1}^\beta$ in terms of the spatial velocity $\mathbf{v}_1$, and from Eq.~\eqref{EQ:geodesic}, we infer that the components of the spatial acceleration $\mathbf{a}_1 = d\mathbf{v}_1/dt$ are
\begin{widetext}
\begin{align}
    a_1^{i} =& \partial_i U + \frac{1}{c^2}\bigg[-\left(3\partial_tU+3v_1^j\partial_jU+v_1^ja_1^j\right)v_1^i-(2U+v_1^2)a_1^{i}-2(U-v_1^2)\partial_i U-4v_1^j(\partial_iU^j-\partial_jU^i)+4\partial_t U^i +\partial _i \Psi\bigg]\eqbreak
    +q_1 \hbar c\left\{ -\partial_i\phi+\frac{1}{c}\left[v_1^j(\partial_iA^j-\partial_jA^i)-\partial_t A^i\right]+\frac{1}{c^2}\left[\left(3U-\frac{1}{2}v_1^2\right)\partial_i\phi+2\phi\partial_i U\right]\right\}+\order{c}{-3}
\end{align} 
\end{widetext}
In the 1PN terms, which are proportional to $c^{-2}$, we can replace $a_1^i$ with the Newtonian expression $a^i_1 = \partial_i U -q_1 \hbar c\partial_i\phi+\order{c}{-1}$,
to obtain
\begin{widetext}
\begin{align}\label{EQ:a1_gen}
    a_1^{i} =& \partial_i U + \frac{1}{c^2}\bigg[-\left(3\partial_tU+4v_1^j\partial_jU\right)v_1^i+(v_1^2-4U)\partial_i U-4v_1^j(\partial_iU^j-\partial_jU^i)+4\partial_t U^i +\partial _i \Psi\bigg]\eqbreak
    +q_2 \hbar c\left\{ -\partial_i\phi+\frac{1}{c}\left[v_1^j(\partial_iA^j-\partial_jA^i)-\partial_t A^i\right]+\frac{1}{c^2}\left[\left(5U+\frac{1}{2}v_1^2\right)\partial_i\phi+2\phi\partial_i U+v_1^j\partial_jv_1^i\right]\right\}+\order{c}{-3}
\end{align} 
\end{widetext}
We recognize the first line of the above equation as the standard PN expansion for the acceleration of a body in the presence of a matter distribution \cite{Poisson2014-cj}. The Newtonian and .5PN terms in the second line are the usual Lorentz force experienced by a particle moving through an electromagnetic field in flat space. We find there is a 1PN correction to the Lorentz force due to the curvature of the spacetime. We substitute in the gravitational and Proca field potentials for the distribution of point particles at hand and find 
\begin{widetext}
\begin{align}
    \mathbf{a}_1=&~-\frac{Gm_2}{r^2}\mathbf{n}   +\alpha\frac{Gm_2}{r^2}\left(1-\frac{1}{2}\frac{r^2}{\lambda^2}\right)\mathbf{n}\eqbreak+\frac{Gm_2}{c^2r^2}\left\{\left[5 \frac{G m_1 }{r}+4 \frac{G m_2 }{r}+\frac{3}{2} (\mathbf{n}\cdot\mathbf{v}_2)^2-\mathbf{v}_1^2+4 (\mathbf{v}_1\cdot\mathbf{v}_2)-2 \mathbf{v}_2^2\right]\mathbf{n}+\left[4 (\mathbf{n}\cdot\mathbf{v}_1)-3 (\mathbf{n}\cdot\mathbf{v}_2)\right]\mathbf{v} \right\}\eqbreak
    +\alpha\frac{Gm_2}{c^2r^2}\Bigg\{\Bigg[\left(-3 \frac{G m_1 }{r}-5 \frac{G m_2 }{r}-\frac{3}{2} (\mathbf{n}\cdot\mathbf{v}_2)^2-\frac{1}{2}\mathbf{v}_1^2-(\mathbf{v}_1\cdot\mathbf{v}_2)+\mathbf{v}_2^2\right)+\left(\frac{3}{2}\frac{G m_1}{r}+(2 \gamma_E +1)\frac{G m_2}{r}  \right)\frac{r}{\lambda}\eqbreak ~~~~~~~~~~~~~~~+\left( -5 \frac{G m_1 }{r}-2 \frac{G m_2 }{r}+\frac{1}{4}(\mathbf{n}\cdot\mathbf{v}_2)^2+\frac{1}{4}\mathbf{v}_1^2+\frac{1}{2} (\mathbf{v}_1\cdot\mathbf{v}_2)-\frac{1}{2} \mathbf{v}_2^2\right)\frac{r^2}{\lambda^2} \Bigg]\mathbf{n}-\left(1-\frac{1}{2}\frac{r^2}{\lambda^2}\right)(\mathbf{n}\cdot\mathbf{v}_2)\mathbf{v}\Bigg\}\eqbreak+\mathcal{O}(c^{-3},\lambda^{-3},\alpha^2)\,,
\end{align}
\end{widetext}
where we recall the definition of $\alpha$ from Eq.~(\ref{eq:alpha-gamma-mu-def}),
\begin{align}
\alpha = \frac{{\hbar c}}{G}\frac{q_1}{m_1}\frac{q_2}{m_2}.
\end{align}

In computing $\mathbf{a}_1$, we are required to evaluate the potentials and their derivatives in Eq. \eqref{EQ:a1_gen} at $\mathbf{x} = \mathbf{y}_1$. In doing so, $r_2 \rightarrow r$ and  $r_1 \rightarrow 0$.  We find that there are several terms containing inverse powers of $r_1$ that diverge when taking this limit. However, this divergence is not physical but rather a result of our choice to work in the point-particle approximation. We, therefore, follow PN theory conventions and set these terms to zero via Hadamard regularization \cite{Blanchet:2000nu}. 

Now that we have calculated $\mathbf{a}_{1}$, it is straightforward to compute the binary equation of motion. First, we infer $\mathbf{a}_{2}$ from $\mathbf{a}_{1}$ by exchanging all particle labels $1 \leftrightarrow 2$, recalling that $\mathbf{n}$ and $\mathbf{v}$ pick up minus signs under this exchange.  
The relative acceleration is then 
\begin{widetext}
\begin{align}\label{EQ:total_a}
 \mathbf{a}=&~\mathbf{a}_{1}-\mathbf{a}_{2} \neweq
 =&~
 -\frac{Gm}{r^2}\mathbf{n}
    -\frac{G m}{c^2r^2}\bigg\{\left[(1+3\eta)v^2-\frac{3}{2}\eta\dot r^2-2(2+\eta)\frac{G m}{r}\right]\mathbf{n}-2(2-\eta)\dot r \mathbf{v}\bigg\}\eqbreak 
    +\alpha \frac{Gm}{r^2}\left(1-\frac{1}{2}\frac{r^2}{\lambda^2}\right)\mathbf{n} +\alpha\frac{G m}{c^2r^2}\bigg\{(2\eta-1)\left(1-\frac{1}{2}\frac{r^2}{\lambda^2}\right)\dot r \mathbf{v}\eqbreak+\left[\frac{1}{2} (7 \eta -1) v^2-\frac{3}{2} \eta\dot r^2 +(4 \eta -5)\frac{ G m}{r} +\eta_\gamma \frac{G m}{r}\frac{r}{\lambda}+ \left(\frac{1}{4} (1-7 \eta ) v^2+\frac{1}{4}\eta  \dot r^2-\frac{1}{2}(3 \eta +1) \frac{ G m}{r}\right)\frac{r^2}{\lambda^2}\right]\mathbf{n}\bigg\}
    \eqbreak+\mathcal{O}(c^{-3},\lambda^{-3},\alpha^2)\,
\end{align}
\end{widetext}
We define the quantity $\eta_\gamma =  1+\eta + (2-4 \eta )\gamma_\mathrm{E}$ for convenience because it shows up throughout the calculation. We see that the contribution to the acceleration by the DM interaction is proportional to the dimensionless parameter $\alpha$, which,  therefore, quantifies the relative strength of the correction to the equations of motion.

As we continue through the calculation of the waveform, we will work in the small deformation limit, i.e.~assuming $\alpha\ll1$ and discarding all but linear-in-$\alpha$ terms. We make this choice for two reasons. Because we are taking $q_1$ and $q_2$ to have the same sign, we are considering a repulsive correction to the force, i.e., $\alpha>0$. We then require that the repulsive acceleration due to the DM be smaller than the attractive acceleration due to gravity --  otherwise, the binary would not inspiral at all. Furthermore, working in the small-deformation limit is consistent with our choice to discard terms quadratic in the Proca field in Einstein's equation.
We have now introduced three approximations into our calculation: (1) The PN assumption of slow speeds and weak fields; (2) The assumption that the Proca interaction length scale is small compared to the orbital separation; (3) The assumption that the deviation from GR is small. As we proceed through the calculation of the waveform, we will keep terms through $\mathcal{O}(c^{-2},\lambda^{-2},\alpha)$.


\section{Modified Binary Inspiral}\label{SEC:Inspiral}
In this section, we compute a PN approximation to the frequency-domain GW signal produced as the binary system at hand inspirals in a quasicircular orbit. 
The total orbital energy of the binary $E$ and the rate at which that energy is carried away by radiation $\mathcal{P}$ are important ingredients for computing the waveform.
Each quantity consists of a purely gravitational part that is a standard PN result and a correction due to the DM interaction that is also expressed in terms of a PN expansion. 
The leading PN order dark-matter corrections have previously been computed for this system \cite{Alexander:2018qzg}.
We here compute the next-to-leading-order corrections. %

The leading-order DM correction to $E$ occurs at the same order as the Newtonian gravitational term, so the correction we compute here is 1PN.
The presence of the dark vector field interaction introduces new vector radiation modes to $\mathcal{P}$.
The dominant vector radiation term is an electric dipole mode that enters at -1PN order relative to the Newtonian gravitational term.
We, therefore, compute the electric quadrupole mode that modifies the waveform at Newtonian order. 
The gravitational radiation modes are also modified by the DM interaction; we compute these corrections to 1PN. 
We do not compute 1PN corrections to vector radiation modes.
With this omission, our waveform will not be a complete 1PN waveform, but it will still remain the most accurate model constructed to date in this framework. 
Note that even when the DM correction is dominant to the gravitational term, we use ``Newtonian'' to refer to the order of the dominant gravitational term and refer to DM corrections with negative PN orders where relevant.  

\subsection{Orbital energy}\label{SEC:energy}
Here, we compute the orbital energy of the binary.
We begin by computing the conserved orbital energy for a general orbit, and then we apply the quasicircular limit.
To apply the quasicircular limit, we must first show that the orbit is confined to a plane. 
To this end, we also compute the conserved angular momentum. 
As discussed in the introduction to this section, the orbital energy can be broken up into a purely gravitational part $E_\mathrm{grav}$ that is a well-known PN result \cite{Blanchet:2013haa} and a correction due to the DM interaction that we call $E_\mathrm{dark}$. The Newtonian term of $E_\mathrm{dark}$ has been previously computed \cite{Alexander:2018qzg}, and we rederive that result here for completeness. We compute the 1PN correction to  $E_\mathrm{dark}$ here for the first time. 
\subsubsection{Conserved orbital energy and angular momentum}

The PN two-body problem, in the absence of DM, admits the following conserved energy and angular momentum 
\begin{align}
    E_\mathrm{grav} =&~ \frac{1}{2}m\eta v^2 - \frac{Gm^2\eta}
    {r}+\frac{m\eta}{c^2}\bigg[\frac{3}{8}(1-3\eta)v^4\eqbreak+\frac{1}{2}(3+\eta)v^2\frac{G m}{r}+\frac{\eta}{2}\dot r^2\frac{G m}{r}+\frac{1}{2}\left(\frac{G m}{r}\right)^2\bigg]\eqbreak+\order{c}{-3}\, ,\neweq
    \mathbf{L}_\mathrm{grav} =&~ m \eta\bigg\{1+\frac{1}{c^2}\bigg[\frac{1}{2}(1-3\eta)v^2+(3+\eta)\frac{G m}{r}\bigg]\eqbreak~~~~~~+\order{c}{-3}\bigg\} (\mathbf{r}\times\mathbf{v})\, .
\end{align}
We wish to compute the corrections to these quantities due to the Proca interaction. 
Following the calculation in Chapter 10 of \cite{Poisson2014-cj}, we construct Ans\"atze for the corrections and fix constant coefficients by enforcing $dE/dt = 0$ and $d\mathbf{L}/dt =0$. Just like $E_\mathrm{grav}$, at Newtonian order, $E_\mathrm{dark}$ can contain terms proportional to $v^2$, $Gm/r$, and $\dot{r}^2 = (\mathbf{v}\cdot\mathbf{n})^2$, and, at 1PN order, it can contain terms proportional to $v^4$, $v^2Gm/r$, $v^2\dot{r}^2$, $\dot{r}^2Gm/r$, $\dot{r}^4$, and $(Gm/r)^2$. 
In addition to a PN expansion, we also expand to linear order in $\alpha$ and quadratic order in $r/\lambda$. 
With this in mind, we construct the following Ansatz: 
\allowdisplaybreaks[4]
\begin{widetext}
\begin{align}
    E_\mathrm{dark} = &~
    \alpha \left(d_{1}+ d_{2}\frac{r}{\lambda} +  d_{3}\frac{r^2}{\lambda^2}\right)v^2
    +\alpha \left(d_{4}+ d_{5}\frac{r}{\lambda} +  d_{6}\frac{r^2}{\lambda^2}\right)\dot r^2 
    +\alpha \left(d_{7}+ d_{8}\frac{r}{\lambda} +  d_{9}\frac{r^2}{\lambda^2}\right)\frac{Gm}{r}\eqbreak
    +\frac{\alpha}{c^2}   \bigg[ \left(d_{10}+ d_{11}\frac{r}{\lambda} +  d_{12}\frac{r^2}{\lambda^2}\right)+\left(d_{13}+ d_{14}\frac{r}{\lambda} +  d_{15}\frac{r^2}{\lambda^2}\right)\log\left(\frac{r}{\lambda}\right)\bigg ]v^4\eqbreak
     +\frac{\alpha}{c^2}   \bigg[ \left(d_{16}+ d_{17}\frac{r}{\lambda} +  d_{18}\frac{r^2}{\lambda^2}\right)+\left(d_{19}+ d_{20}\frac{r}{\lambda} +  d_{21}\frac{r^2}{\lambda^2}\right)\log\left(\frac{r}{\lambda}\right)\bigg ]v^2\dot r^2\eqbreak
      +\frac{\alpha}{c^2}   \bigg[ \left(d_{22}+ d_{23}\frac{r}{\lambda} +  d_{24}\frac{r^2}{\lambda^2}\right)+\left(d_{25}+ d_{26}\frac{r}{\lambda} +  d_{27}\frac{r^2}{\lambda^2}\right)\log\left(\frac{r}{\lambda}\right)\bigg ]v^2\frac{Gm}{r}\eqbreak
       +\frac{\alpha}{c^2}   \bigg[ \left(d_{28}+ d_{29}\frac{r}{\lambda} +  d_{30}\frac{r^2}{\lambda^2}\right)+\left(d_{31}+ d_{32}\frac{r}{\lambda} +  d_{33}\frac{r^2}{\lambda^2}\right)\log\left(\frac{r}{\lambda}\right)\bigg ]\dot r^4\eqbreak
        +\frac{\alpha}{c^2}   \bigg[ \left(d_{34}+ d_{35}\frac{r}{\lambda} +  d_{36}\frac{r^2}{\lambda^2}\right)+\left(d_{37}+ d_{38}\frac{r}{\lambda} +  d_{39}\frac{r^2}{\lambda^2}\right)\log\left(\frac{r}{\lambda}\right)\bigg ]\dot r^2\frac{Gm}{r}\eqbreak
         +\frac{\alpha}{c^2}   \bigg[ \left(d_{40}+ d_{41}\frac{r}{\lambda} +  d_{42}\frac{r^2}{\lambda^2}\right)+\left(d_{43}+ d_{44}\frac{r}{\lambda} +  d_{45}\frac{r^2}{\lambda^2}\right)\log\left(\frac{r}{\lambda}\right)\bigg ]\left(\frac{Gm}{r}\right)^2\eqbreak+ \mathcal{O}(c^{-3},\lambda^{-3},\alpha^2)\,,
\end{align}
\end{widetext}
where the 45 constant coefficients $d_i$ must be determined by enforcing $dE/dt=0$, where $E=E_\mathrm{grav}+E_\mathrm{dark}$. When taking this derivative, we recall that $d\mathbf{r}/dt = \mathbf{v}$, $d\mathbf{v}/dt = \mathbf{a}$, and $\mathbf{a}$ is given by Eq. (\ref{EQ:total_a}). 
After enforcing energy conservation, some of the coefficients remain unfixed. As is standard in PN theory, 
we fix them by requiring that, in the limit ${G m}/{r}\rightarrow0$,  the energy matches the PN expansion of the special relativistic energy for two bodies $E = c^2 \left[\gamma (v_1) m_1+\gamma(v_2) m_2\right]$, with $\gamma(v_A) = \left(1-v_A^2/c^2\right)$.  This assumption is still valid in the case at hand because, when the bodies become infinitely separated, the dark-matter bound inside each body does not interact.  
We must therefore ensure that $E_\mathrm{dark}$ vanishes as $\frac{G m}{r}\rightarrow0$. By doing so, we then obtain 
\begin{widetext}
\begin{align}\label{EQ:E_gen}
    E = &~ \frac{1}{2}m\eta v^2- \frac{G m^2\eta}{r}+\alpha \frac{G m^2\eta}{r}\left(1-\frac{r}{\lambda}+\frac{1}{2}\frac{r^2}{\lambda^2}\right)\eqbreak
    +\frac{m\eta}{c^2}\left[\frac{3}{8}(1-3\eta)v^4+\frac{1}{2}(3+\eta)v^2\frac{G m}{r}+\frac{1}{2}\eta\dot r^2\frac{G m}{r}+\frac{1}{2}\left(\frac{G m}{r}\right)^2\right]\eqbreak
    +\alpha\frac{m\eta}{c^2}\bigg\{
    -\frac{1}{2} \eta\left(1-\frac{1}{2}\frac{r^2}{\lambda^2}\right)\dot r^2\frac{G m}{r}
    +\left[-1 +\frac{7 }{2}\eta +\eta_\gamma  \frac{r}{\lambda}+(3 \eta +2)\log \left(\frac{r}{\lambda}\right)\frac{r^2}{\lambda^2} \right]\left(\frac{G m}{r}\right)^2
    \bigg\}\eqbreak+\mathcal{O}(c^{-3},\lambda^{-3},\alpha^2)\,.
\end{align}  
\end{widetext}
As expected, we find that at Newtonian order, the expression above matches the energy given in \cite{Alexander:2018qzg} when expanded for $r/\lambda\ll1$. 

We construct a similar Ansatz for $\mathbf{L}_\mathrm{dark}$ expressed in terms of the vector basis [$\mathbf{n}$, $\mathbf{v}$, $\mathbf{r}\times\mathbf{v}$]
\begin{widetext}
\begin{align}
    \mathbf{L}_\mathrm{dark} = &~\alpha\left(d_{1}+ d_{2}\frac{r}{\lambda} +  d_{3}\frac{r^2}{\lambda^2}\right)\mathbf{n}+\alpha\left(d_{4}+ d_{5}\frac{r}{\lambda} +  d_{6}\frac{r^2}{\lambda^2}\right)\mathbf{v}+\alpha\left(d_{7}+ d_{8}\frac{r}{\lambda} +  d_{9}\frac{r^2}{\lambda^2}\right)(\mathbf{r}\times\mathbf{v})\eqbreak
    +\frac{\alpha}{c^2}  \left[ \left(d_{10}+ d_{11}\frac{r}{\lambda} +  d_{12}\frac{r^2}{\lambda^2}\right)\mathbf{n}+\alpha\left(d_{13}+ d_{14}\frac{r}{\lambda} +  d_{15}\frac{r^2}{\lambda^2}\right)\mathbf{v}+\alpha\left(d_{16}+ d_{17}\frac{r}{\lambda} +  d_{18}\frac{r^2}{\lambda^2}\right)(\mathbf{r}\times\mathbf{v})\right]v^2\eqbreak
    +\frac{\alpha}{c^2}  \left[ \left(d_{19}+ d_{20}\frac{r}{\lambda} +  d_{3}\frac{r^2}{\lambda^2}\right)\mathbf{n}+\alpha\left(d_{21}+ d_{22}\frac{r}{\lambda} +  d_{23}\frac{r^2}{\lambda^2}\right)\mathbf{v}+\alpha\left(d_{24}+ d_{25}\frac{r}{\lambda} +  d_{26}\frac{r^2}{\lambda^2}\right)(\mathbf{r}\times\mathbf{v})\right]v^2\log\left(\frac{r}{\lambda}\right)\eqbreak
     +\frac{\alpha}{c^2}  \left[ \left(d_{27}+ d_{28}\frac{r}{\lambda} +  d_{29}\frac{r^2}{\lambda^2}\right)\mathbf{n}+\alpha\left(d_{30}+ d_{31}\frac{r}{\lambda} +  d_{32}\frac{r^2}{\lambda^2}\right)\mathbf{v}+\alpha\left(d_{33}+ d_{34}\frac{r}{\lambda} +  d_{35}\frac{r^2}{\lambda^2}\right)(\mathbf{r}\times\mathbf{v})\right]\dot r^2\eqbreak
    +\frac{\alpha}{c^2}  \left[ \left(d_{36}+ d_{37}\frac{r}{\lambda} +  d_{38}\frac{r^2}{\lambda^2}\right)\mathbf{n}+\alpha\left(d_{39}+ d_{40}\frac{r}{\lambda} +  d_{41}\frac{r^2}{\lambda^2}\right)\mathbf{v}+\alpha\left(d_{42}+ d_{43}\frac{r}{\lambda} +  d_{44}\frac{r^2}{\lambda^2}\right)(\mathbf{r}\times\mathbf{v})\right]\dot r^2\log\left(\frac{r}{\lambda}\right)\eqbreak
    +\frac{\alpha}{c^2}  \left[ \left(d_{45}+ d_{46}\frac{r}{\lambda} +  d_{47}\frac{r^2}{\lambda^2}\right)\mathbf{n}+\alpha\left(d_{48}+ d_{49}\frac{r}{\lambda} +  d_{50}\frac{r^2}{\lambda^2}\right)\mathbf{v}+\alpha\left(d_{51}+ d_{52}\frac{r}{\lambda} +  d_{53}\frac{r^2}{\lambda^2}\right)(\mathbf{r}\times\mathbf{v})\right]\frac{Gm}{r}\eqbreak
    +\frac{\alpha}{c^2}  \left[ \left(d_{54}+ d_{55}\frac{r}{\lambda} +  d_{56}\frac{r^2}{\lambda^2}\right)\mathbf{n}+\alpha\left(d_{57}+ d_{58}\frac{r}{\lambda} +  d_{59}\frac{r^2}{\lambda^2}\right)\mathbf{v}+\alpha\left(d_{60}+ d_{61}\frac{r}{\lambda} +  d_{62}\frac{r^2}{\lambda^2}\right)(\mathbf{r}\times\mathbf{v})\right]\frac{Gm}{r}\log\left(\frac{r}{\lambda}\right)\eqbreak
    +\mathcal{O}(c^{-3},\lambda^{-3},\alpha^2).
\end{align}
\end{widetext}
Again, we determine the constant coefficients by enforcing  $d\mathbf{L}/dt =0$ where $\mathbf{L} = \mathbf{L}_\mathrm{grav}+\mathbf{L}_\mathrm{dark}$. We find 
\begin{align}
    \mathbf{L}
     =&~ m \eta\bigg\{
     1+\frac{1}{c^2}\bigg[\frac{1}{2}(1-3\eta)v^2+(3+\eta)\frac{G m}{r}\bigg]\eqbreak
    -\frac{\alpha}{c^2}\eta\bigg(1+\frac{r^2}{\lambda^2}\bigg)\frac{G m}{r}
    +\mathcal{O}(c^{-3},\lambda^{-3},\alpha^2) \bigg\} (\mathbf{r}\times\mathbf{v})\,.
\end{align}
As with the energy, we fix constants by enforcing that this correction vanishes as ${G m}/{r}\rightarrow0$. We see that, as is the case in the absence of DM, $\mathbf{L}$ is parallel to $\mathbf{r}\times\mathbf{v}$, indicating that the orbit is restricted to a fixed plane.

\subsubsection{The quasicircular limit}\label{SEC:circular}
As noted in the previous section, the orbital motion occurs in a fixed plane. This allows us to describe the motion of the binary in terms of its orbital separation $r$ and orbital angle $\varphi$. Continuing to follow Chapter 10 in \cite{Poisson2014-cj}, we introduce the basis $\mathbf{n} = [-\sin\varphi,\cos\varphi,0]$, $\boldsymbol{\varphi} = [\cos\varphi,\sin\varphi,0]$, and $\mathbf{e}_z = [0,0,1]$. In this new basis, the separation vector is $\mathbf{r} =r \mathbf{n}$, the velocity is $\mathbf{v} = \dot{r}\mathbf{n} + r \dot\varphi \boldsymbol{\varphi}$, and the acceleration is 
\begin{align}\label{EQ:EOM_gen}
    \mathbf{a} = (\ddot r -r\dot\varphi^2)\mathbf{n}+\frac{1}{r}\frac{d}{dt}(r^2\dot\varphi)\boldsymbol{\varphi}\,.
\end{align}
We can put Eq. (\ref{EQ:total_a}) in terms of this new basis and read off the following equations of motion
\begin{widetext}
\begin{align}
   \ddot r -r\dot\varphi^2=&~-\frac{G m}{r^2}\bigg\{1-\alpha\left(1+\frac{1}{2}\frac{r^2}{\lambda^2}\right)+\frac{1}{c^2}\left((1+3 \eta) v^2-\frac{1}{2} (8-\eta) \dot r^2-2 (2+\eta ) \frac{G m}{r}\right)\eqbreak+\frac{\alpha
   }{c^2}\bigg[\frac{1}{2} (1-7 \eta) \left(1-\frac{1}{2} \frac{r^2}{\lambda^2}\right)v^2 
 + \left(1-\frac{1 }{2}\eta-\frac{1}{4} (2-3 \eta) \frac{r^2}{\lambda^2}\right){\dot r}^2\eqbreak
 ~~~~~~~~+ \left(5-4 \eta -\eta_\gamma  \frac{r}{\lambda}+\frac{1}{2} (1+3 \eta) \frac{r^2}{\lambda^2}\right)\frac{G m}{r}\bigg]\bigg\} +\mathcal{O}(c^{-4},\lambda^{-3},\alpha^2)\,,\label{EQ:radial}\\
   \frac{d}{dt}(r^2\dot\varphi)=&~  \frac{G m}{rc^2}\left[(4-2 \eta ) -\alpha  (1-2 \eta)  \left(1-\frac{1}{2}\frac{r^2}{\lambda^2}\right) \right]\dot r \dot \varphi
    +\mathcal{O}(c^{-4},\lambda^{-3},\alpha^2)\,.\label{EQ:angular}
\end{align}
\end{widetext}

From here onward, we apply the quasicircular limit, enforcing $\dot r =0$. 
We see from Eq. (\ref{EQ:angular}) that, in this limit, the orbital frequency $\dot\varphi$ is a constant that we will call $\omega$.
We apply this and $v^2 = r^2\omega^2$ to Eq. (\ref{EQ:radial}) and solve for the corrected version of Kepler's law 
\begin{align}\label{EQ:kepler}
    \omega^2 =&~ \frac{G m}{r^3} \bigg\{1-\alpha\left(1-\frac{1}{2}\frac{r^2}{\lambda^2}\right)
    -\frac{G m}{rc^2}\bigg[(3-\eta)\eqbreak
    +\alpha\bigg(\frac{3}{2} (3-7 \eta )- \eta_\gamma  \frac{r}{\lambda}+\frac{1}{4} (3+19 \eta) \frac{r^2}{\lambda^2}\bigg)\bigg]\bigg\}\eqbreak+\mathcal{O}(c^{-4},\lambda^{-3},\alpha^2)\,.
\end{align}   
We invert Kepler's law perturbatively, first in $\alpha$, and then in  $c^{-1}$,  discarding terms of $\order{\alpha}{2}$ and $\order{c}{-3}$, to obtain 
\begin{align}\label{EQ:kepler_inv}
     r=&~\frac{G m}{c^2 x}\bigg\{1-\frac{\alpha}{3}    \left(1-\frac{1}{2}\frac{\mu^2}{x^2}\right)-x\left(1-\frac{1}{3}\eta\right)\nonumber\\&~
     +\alpha x\bigg[\frac{1}{6} (3-19 \eta )-\frac{1}{3}\eta_\gamma  \frac{\mu}{x}+\frac{5}{36} (11 \eta +3) \frac{\mu^2}{x^2}\bigg]\eqbreak+\mathcal{O}(c^{-4},\mu^{3},\alpha^2)\bigg\}\,,
\end{align}
where $x = ({G m \omega }/c^3)^{2/3} \ll 1$ is a dimensionless PN parameter, and  we recall the definition of $\mu$ from Eq.~(\ref{eq:alpha-gamma-mu-def}), $\mu = G m m_A/(\hbar c)$.
From Eq.~(\ref{EQ:kepler_inv}) we see that \begin{align}
    \frac{r}{\lambda} = \frac{\mu}{x} +\mathcal{O}(c^{-2},\alpha)\,,
\end{align} 
and as we proceed, we keep terms through $\mathcal{O}(\mu^2/x^2)$ in our small-photon-mass approximation.

We are finally ready to compute $E$ in the quasicircular limit. Setting $\dot r= 0$, Eq. (\ref{EQ:E_gen}) becomes
\begin{align}
   E =&~-\frac{1}{2}  \eta m c^2 x\bigg\{ 1 
    - \alpha \left( \frac{2}{3}-2 \frac{\mu}{x}+  \frac{5}{3}\frac{\mu^2}{x^2}\right)\eqbreak
    - x\left(\frac{3}{4}+\frac{1}{12}\eta\right) 
    +\alpha x \bigg[\frac{31}{9}\eta-\frac{2 }{3}\eta_\gamma \frac{\mu}{x}\eqbreak-\left(\frac{5}{3}(1+3 \eta) +(4+6 \eta) \log (\mu/x)\right)\frac{\mu^2}{x^2}\bigg]\eqbreak+\mathcal{O}(c^{-4},\mu^{3},\alpha^2)\bigg\}\,,
\end{align}
where we have expressed this final result in terms of the PN parameter $x$, using $v^2 = r^2 \omega^2 = r^2 x^{3/2}/m$ and $r$ given by Eq. (\ref{EQ:kepler_inv}).

\subsection{Gravitational and Proca field radiation}\label{SEC:radiation} 
As the binary inspirals, energy is radiated as waves in both the gravitational and Proca fields. 
The power radiated through GWs in terms of mass and mass current multipole moments is \cite{Blanchet:2001aw}
\begin{align}
    \mathcal{P}_\mathrm{GW} =&~ \frac{G}{c^5}\Bigg[\frac{1}{5}\dddot I^{<ij>} \dddot I^{<ij>}+\frac{1}{c^2}\bigg(\frac{1}{189}\ddddot I^{<ijk>} \ddddot I^{<ijk>}\eqbreak+\frac{16}{45}\dddot J^{<ij>} \dddot J^{<ij>}\bigg)+\order{c}{-4}\Bigg]  \,,
\end{align}
where $<>$ indicates the symmetric trace-free part of the given tensor.

The moments depend on the positions of the bodies. Velocity can be related to orbital separation through Eq. (\ref{EQ:kepler}) and the definition of velocity $v = r \omega$, namely
\begin{align}\label{EQ:vsquared}
    v^2 = \frac{G m}{r}\left[1-\alpha\left(1-\frac{1}{2}\frac{r^2}{\lambda^2}\right)\right]+\mathcal{O}(c^{-2},\lambda^{-3},\alpha^2)\,.
\end{align}
Inserting this expression in the center-of-mass positions, given in Eq. \eqref{EQ:positions}, one finds

\begin{align}\label{EQ:1PN_positions}
    \mathbf{y}_1 =& \frac{m_2}{m}\mathbf{r} -\alpha \frac{\eta \Delta}{2 c^2}\frac{G m}{r}\left(1-\frac{1}{2}\frac{r^2}{\lambda^2}\right)\mathbf{r}+\mathcal{O}(c^{-4},\lambda^{-3},\alpha^2)\,,\neweq
    \mathbf{y}_2 =& -\frac{m_1}{m}\mathbf{r} -\alpha \frac{\eta \Delta}{2 c^2}\frac{G m}{r}\left(1-\frac{1}{2}\frac{r^2}{\lambda^2}\right)\mathbf{r}+\mathcal{O}(c^{-4},\lambda^{-3},\alpha^2)\,.
\end{align}
In the absence of DM, $\alpha = 0$, and the 1PN corrections to the positions vanish for circular orbits. However, for the system at hand, these linear-in-$\alpha$, 1PN corrections to the center-of-mass positions must be included when computing $I_{ij}$. 

Taking into account Eqs. \eqref{EQ:vsquared} and \eqref{EQ:1PN_positions}, we find 
\begin{align}
        I^{<ij>} = &~m \eta r^{<i}r^{j>} + \frac{m\eta}{c^2}\bigg\{-\frac{G m}{r }\frac{1}{42}\bigg[ (1+39\eta)\eqbreak-\alpha41  (1-3\eta) \left(1-\frac{r^2}{\lambda^2}\right)\bigg] r^{<i}r^{j>}\eqbreak+r^2\frac{11}{21}(1-3\eta)v^{<i}v^{j>} \bigg\}\eqbreak+\mathcal{O}(c^{-4},\mu^{3},\alpha^2)\,.
\end{align}
The details of this calculation are presented in Appendix \ref{SEC:APP_I}.
The mass octupole and current quadrupole, which enter the power at 1PN order, are needed only to Newtonian order
\begin{align}
    I^{<ijk>} = &~m \eta \Delta r^{<i}r^{j}r^{k>} +\order{c}{-2}\,,\\
    J^{<ij>} = &~- m \eta \Delta \sqrt{G m r} e_z^{<i} r^{j>}+\order{c}{-2}\,,
\end{align}
where we recall that $e_z^{i} = [0,0,1]$. 

We compute the necessary derivatives of the moments, recalling that $d\mathbf{r}/dt = \mathbf{v}$, $d\mathbf{v}/dt = \mathbf{a}$, and $\mathbf{a}$ is given by Eq. (\ref{EQ:total_a}). 
We find that the gravitationally radiated power in terms of the PN parameter $x$ is
\begin{align}
   \mathcal{P}_\mathrm{GW} =&~\frac{c^5}{G}\frac{32}{5} \eta^2 x^5\bigg\{ 1 
    - \alpha \left( \frac{4}{3}-\frac{2}{3} \frac{\mu^2}{x^2}\right)\eqbreak
    - x\left(\frac{1247}{336}-\frac{35}{12}\eta\right)
    +\alpha x \Bigg[\frac{1739}{504}-\frac{29 } {18}\eta-\frac{4}{3}\eta_\gamma\frac{\mu}{x}\eqbreak+\left(\frac{1933}{1008}-\frac{197 }{84} \eta\right) \frac{\mu^2}{x^2}\Bigg]+\mathcal{O}(c^{-3},\mu^{3},\alpha^2)\bigg\}\,.
\end{align}  
We see that the well-known result for the radiated power in the absence of DM \cite{Blanchet:2001aw} is modified due to the Proca interaction. This arises due to the corrections to the motion of the bodies that impact the evolution of the mass moments.

In addition to the modified gravitational radiation, the Proca interaction will activate vector radiation modes. Following the notation of \cite{Krause:1994ar}, this can also be expressed as the following multipole expansion, including electric dipole, magnetic dipole, and electric quadrupole terms:
\begin{align}
    \mathcal{P}_\mathrm{vector} =  \mathcal{P}_\mathrm{E1} +  \mathcal{P}_\mathrm{M1} + \mathcal{P}_\mathrm{E2} + \order{c}{-4}. 
\end{align}
The magnetic dipole moment for circular motion is constant through \order{c}{-2}, so $\mathcal{P}_\mathrm{M1} =0$. The calculation of $\mathcal{P}_\mathrm{E1}$ and $\mathcal{P}_\mathrm{E2}$ is included in Appendix \ref{SEC:APP_P_dark}, and we reproduce the final results below:
\begin{align}
    \mathcal{P}_\mathrm{E1} =   \frac{c^5}{G} x^4 \eta^2 &\bigg\{ \frac{2}{3}\gamma  + x \bigg[- 2 \alpha  \Delta -\frac{1}{9} \gamma  (3+9 \Delta +2 \eta ) \bigg] \eqbreak+\mathcal{O}(c^{-3},\mu^{3},\alpha^2,\gamma^2,\alpha\gamma)\bigg\}  \Theta( x^{3/2}/\mu - 1)\, ,\eqbreak\,     \neweq
    \mathcal{P}_\mathrm{E2} 
    = \frac{c^5}{G}x^5\eta^2  &\bigg[  -\frac{8}{5} \alpha  (\Delta -1)+\frac{2}{5} \gamma  (1-\Delta)^2 \eqbreak+\mathcal{O}(c^{-3},\mu^{3},\alpha^2,\gamma^2,\alpha\gamma)\bigg]  \Theta(x^{3/2}/\mu - \frac{1}{2}) \,,
\end{align}
where $\Theta$ is the Heaviside function and we recall the definition of $\gamma$ from Eq.~(\ref{eq:alpha-gamma-mu-def}), 
\begin{align}
 \gamma = \frac{\hbar c}{G}\left(\frac{q_1}{m_1}-\frac{q_2}{m_2}\right)\,.   
\end{align} 
We recognize the first term in $\mathcal{P}_\mathrm{E1}$ as the leading-order correction to the power presented in \cite{Alexander:2018qzg}. Just as we work in the small $\alpha$ limit, we also assume $\gamma\ll1$ and discard terms proportional to $\gamma^2$ and $\gamma\alpha$.
While the electric quadrupole radiation is subdominant to the dipole radiation, its associated Heaviside is activated earlier in the inspiral at a lower orbital frequency. 

A higher-order PN expansion of the radiated power would also include electric octupole and magnetic quadrupole moments, which would contribute to the power at 1PN order relative to the leading-order gravitational term. 
However, we have chosen to include only the leading-order and next-to-leading-order terms in the power radiated by the Proca field, which enter $\mathcal{P}$ at -1PN and Newtonian order, respectively.  
Therefore, the 1PN modes are beyond the scope of this paper. 
 
Combining both the gravitational and vector mode  contributions, the total power radiated as the binary inspirals is 
\begin{widetext}
\begin{align}
   \mathcal{P} =   ~\frac{c^5}{G}\frac{32}{5} \eta^2 x^5&\bigg\{ 1 
    - \alpha \frac{4}{3}\left(1-\frac{1}{2} \frac{\mu^2}{x^2}\right)
    - x\left(\frac{1247}{336}-\frac{35}{12}\eta\right) \nonumber\\&~
    +\alpha x \Bigg[\frac{1739}{504}-\frac{29 } {18}\eta-\frac{4}{3}\eta_\gamma\frac{\mu}{x}+\left(\frac{1933}{1008}-\frac{197 }{84} \eta\right) \frac{\mu^2}{x^2}\Bigg]\eqbreak
   + \frac{5}{32}\left[  \frac{2}{3}\gamma\frac{1}{x} - 2 \alpha  \Delta -\frac{1}{9} \gamma  (3+9 \Delta +2 \eta )\right]  \Theta( x^{3/2}/\mu - 1) \nonumber\\&~ +\frac{5}{32}\left[  -\frac{8}{5} \alpha  (\Delta -1)+\frac{2}{5} \gamma  (1-\Delta)^2\right]  \Theta( x^{3/2}\mu - \frac{1}{2}) \eqbreak+\mathcal{O}(c^{-3},\mu^{3},\alpha^2,\gamma^2,\alpha\gamma)\bigg\}\,.
\end{align}
\end{widetext}
With modified expressions for $E$ and $\mathcal{P}$, we are now ready to compute the frequency domain waveform in the next section. 
\subsection{Gravitational waveform}\label{SEC:Waveform} 

Far from the binary, the metric can be expanded about flat space such that $g_{\mu\nu} = \eta_{\mu\nu}+h_{\mu\nu}$, where $\eta_{\mu\nu}$ is the Minkowski metric and $h_{\mu\nu}$ is a small perturbation. Gravitational waves are oscillatory solutions of the linearized Einstein equations for  $h_{\mu\nu}$. The spatial components of the metric perturbation can be written as the familiar post-Minkowskian expansion
\begin{align}
    h^{ij} = \frac{2 G}{c^4d_L} \left[\ddot{I}^{<ij>} + {\cal{O}}(c^{-1})\right]\,,
\end{align}
in terms of the multipole moments of the source. 
The angle brackets indicate the symmetric and trace-free part of the quadrupole moment tensor $I^{ij}$, while $d_L$ is the luminosity distance to the source.
In principle, the GW strain contains an infinite sum of radiation modes, but we keep only the quadrupole term.
We do this in parallel with the construction of the \text{IMRPhenomD} family of models~\cite{Husa:2015iqa,Khan:2015jqa}, which we use for parameter estimation in the next section.

Upon taking derivatives of the quadrupole moment computed in Appendix \ref{SEC:APP_I} and casting the expression in terms of the orbital basis defined in Sec. \ref{SEC:circular}, we find
\begin{align}
    h^{ij} = h_{+}(t)  \left(
\begin{array}{ccc }
 1 & 0 & 0  \\
 0 & -1 &0\\
 0 & 0 &0\\
\end{array}
\right)+h_{\times}(t)\left(
\begin{array}{ccc}
 0 & 1 &0\\
 1 & 0& 0\\
 0 & 0 &0\\
\end{array}
\right),
\end{align}
where $h_{+}(t)=A(t) \cos[2\varphi(t)]$, $h_{\times}(t)=A(t) \sin[2\varphi(t)]$, $\varphi(t)=\omega t$ is the orbital angle,
and 
\begin{align}
    A(t) = &~ -\frac{4 G \eta m }{c^2 d_L}x\bigg\{1-\alpha  \left(\frac{2}{3}-\frac{1}{3}\frac{\mu^2}{\lambda^2}\right)-x\left(\frac{107}{42}-\frac{55  }{42}\eta\right)
    \nonumber \\
    &~
    +\alpha x  \left[\frac{88}{63}-\frac{341}{63} \eta -\frac{2}{3}\eta_\gamma\frac{\mu}{x}+\left(\frac{283}{252}+\frac{95}{84} \eta \right) \frac{\mu^2}{x^2}\right]\eqbreak+~\mathcal{O}(c^{-3},\mu^{3},\alpha^2)\bigg\}
\end{align}
is the amplitude. It is customary to collect $h_{+}$ and $h_{\times}$ into a single scalar
\begin{align}
    h(t) = h_{+}(t) - i h_{\times}(t) = A(t) e^{-2 i \varphi(t)}\,
\end{align}
 known as the GW strain\footnote{The strain observed at a detector, known as the interferometric response, is $h = F_+ h_+ + F_\times h_\times$, where $F_+$ and $F_\times$ are detector response functions.  If the coordinate systems of the source and the observer are misaligned by an angle $\iota$, then $h_{+}(t)=(1/2)(1+\cos^2\iota )A(t) \cos[2\varphi(t)]$ and $h_{\times}(t)=\cos\iota A(t)  \sin[2\varphi(t)]$, in the reference frame of the detector.  }.

Bayesian parameter estimation of ground-based GW data requires the Fourier transform of the strain 
\begin{align}
\label{eq:htilde}
\tilde{h}(f) =  \int A(t) e^{2\pi i ft} e^{-2i\varphi(t)}  dt \,.  
\end{align}
We evaluate this Fourier transform using the stationary-phase approximation (SPA) (see e.g.~\cite{Yunes:2009yz}). For the  integral of an oscillating function,
\begin{align}
I =     \int A(t) e^{i \psi(t)}dt\,,
\end{align}
a stationary point is some $t_0$ where $\dot\psi(t_0) =0$. The integrand oscillates more slowly at the stationary point, and therefore, the integral is dominated by the behavior of the integrand about this point. Two facts must be true in order to apply this approximation. First, the amplitude must vary much more slowly than the phase. Second,
the orbital frequency and the phase must be positive and
monotonically increasing. We see that the integrand in  Eq.~\eqref{eq:htilde} has a stationary point $t_0$ when $\dot{\varphi}(t_0) = \pi f \equiv \omega(t_0)$.

In the stationary phase approximation, the Fourier transform of the GW strain can be written as
\begin{align}
    \tilde{h}(f) 
    &=  \tilde A(f) e^{-i\tilde \varphi(f)}\, ,
\end{align}
where 
\begin{align}
    \tilde A(f) =&~  A(t_0) \left(\frac{\pi}{\dot{\omega}(t_0)} \right)^{1/2}\,,\neweq
    \tilde\varphi(f) = &~2 t_0\omega (t_0) - 2\varphi(t_0) -  \pi/4\,.
\end{align}
Note that $\tilde A(f)$ denotes the amplitude of the Fourier transform and not the Fourier transform of the time-domain amplitude $A(t)$, and similarly for $\tilde\varphi(f)$. In many applications, such as the analysis performed in this paper, the separate Fourier transforms of the polarizations $\tilde h_{+}(f)$ and $\tilde h_{\times}(f)$ are required, rather than the Fourier transform of the strain $\tilde h(f)$. We therefore note that $\tilde h_{+}(f) = \frac{1}{2}\tilde h(f)$ and  $\tilde h_{\times}(f)  = \frac{i}{2}\tilde h(f)$, such that 
    $\tilde h(f)=\tilde h_{+}(f)-i\tilde h_{+}(f)$.
     To understand this, we expand the time domain polarizations, $h_{+}(t)  = A(t) (e^{-2i \varphi(t)}+e^{2i \varphi(t)})/2$ and $h_{\times}(t)  = i A(t)(e^{-2i \varphi(t)}-e^{2i \varphi(t)})/2$. Observe that, in both cases, the term with the positive exponential does not contain a stationary point and, therefore, will not contribute to the Fourier transform.

Let us now compute  $t_0$,  $\varphi(t_0)$, and  $\dot{\omega}(t_0)$ for the system at hand. The balance equation,
\begin{align}
    \mathcal{P} = -\frac{dE}{dt}\,,
\end{align}
tells us that orbital energy lost in the binary system is dissipated as radiation.
The definition of $\omega$, 
\begin{align}
\omega = \frac{d\varphi}{dt} = \frac{c^3}{G m}x^{3/2}\,,
\end{align}
together with the chain rule, allows us to write
\begin{align}
t_0 =&~ t_c -\int dx \frac{E'(x)}{\mathcal{P}(x)}\Big|_{x = x(t_0)}\, ,\\
\varphi(t_0) =&~ \varphi_c  - \frac{c^3}{G m}\int dx x^{3/2} \frac{E'(x)}{\mathcal{P}(x)} \Big|_{x =x(t_0)}\, ,\\
\dot\omega=&~-\frac{3}{2}\frac{c^3}{m G}x^{1/2}\frac{\mathcal{P}(x)}{E'(x)} \Big|_{x =x(t_0)}\, ,
\end{align}
where $x(t_0) =(G\pi f m )^{2/3}/c^2$ and the prime denotes a derivative with respect to the argument of the function.
Combining all of this, we obtain
\begin{align}\label{EQ:waveform_1}
    \tilde A(f) =&~ -\frac{G^2}{c^5} \sqrt{\frac{5\pi}{6}}\frac{m^2\sqrt{\eta}}{d_L}\frac{1}{x^{7/4}}\sum_{n=-2}^{1}\tilde A_{(n)} x^n\, ,\neweq
   \tilde \varphi(f) =&~  2\pi f t_c - 2\varphi_c - \frac{\pi}{4}+\frac{3}{128\eta}\frac{1}{x^{5/2}}\sum_{n=-2}^{1}\tilde \varphi_{(n)} x^n\, ,
\end{align}
where the Fourier amplitude and phase coefficients are 
\allowdisplaybreaks[4]
\begin{widetext}
\begin{align}\label{EQ:waveform_2}
   \tilde A_{(-2)}=&~ \alpha\mu^2\frac{5}{6}\,,  \neweq
   \tilde A_{(-1)}=&~ \alpha\mu^2\left(\frac{1051}{576}+\frac{301}{48} \eta \right)
    -\gamma\frac{5}{96}  \Theta \left(\frac{x^{3/2}}{\mu}-1\right) \,, \neweq
   \tilde A_{(0)}=&~ 1 - \alpha \left[\frac{1}{3} \left(1+\eta_\gamma \mu\right)-\frac{5}{32}\Delta\Theta \left(\frac{x^{3/2}}{\mu}-1\right) +\frac{1}{8} (1-\Delta) \Theta \left(\frac{x^{3/2}}{\mu}-\frac{1}{2}\right)\right]\eqbreak
   -\gamma\left[\left(\frac{5945}{64512}+\frac{1475}{5376}\eta-\frac{5 }{64}\Delta\right)\Theta \left(\frac{x^{3/2}}{\mu}-1\right) +\frac{1}{32} (1-\Delta)^2 \Theta \left(\frac{x^{3/2}}{\mu}-\frac{1}{2}\right)\right]\,,\neweq
   \tilde A_{(1)}=&~ -\left(\frac{323}{224}-\frac{451}{168}\eta\right)+\alpha  \left(\frac{4895}{2016}-\frac{485}{504}\eta \right)  \,, \neweq
   \tilde \varphi_{(-2)}=&~\alpha\mu^2\frac{10}{27}\,,\neweq
   \tilde \varphi_{(-1)}=&~    
    \alpha\mu^2\left(\frac{311}{126}+\frac{264}{49} \eta\right)
    -\gamma\frac{5}{84}\Theta \left(\frac{x^{3/2}}{\mu}-1\right)\,,\neweq
   \tilde \varphi_{(0)}=&~    
    1+\alpha\left[\frac{2}{3}\left(1+\eta_\gamma \mu\right)+\frac{5 }{16}\Delta  \Theta \left(\frac{x^{3/2}}{\mu}-1\right)-\frac{1}{4} (1-\Delta) \Theta \left(\frac{x^{3/2}}{\mu}-\frac{1}{2}\right)\right]\eqbreak
    -\gamma\left[\left(\frac{4555}{8064}+\frac{5 }{9} \eta-\frac{5}{32}\Delta\right)\Theta \left(\frac{x^{3/2}}{\mu}-1\right)+\frac{1}{16} (1-\Delta)^2 \Theta \left(\frac{x^{3/2}}{\mu}-\frac{1}{2}\right)\right]\,,\neweq
   \tilde \varphi_{(1)}=&~    
    \left(\frac{3715}{756}+\frac{55 }{9}\eta\right)+\alpha\left(\frac{355}{81}+\frac{2540}{81}\eta \right)\,.
\end{align}  
\end{widetext}
Up until now, we have expressed quantities in a way that makes explicit which terms originate from the PN expansion, for which $x\ll1$, and which terms originate from the small photon mass expansion, for which $x/\mu\ll1$. However, both expansion parameters contain the PN parameter $x$. Ultimately, it is useful to express our results in the form of a single PN expansion, collecting powers of $x$, as we have done above. Recalling that $x \propto ~c^{-2}$, we see that our DM modifications manifest as -2PN, -1PN, Newtonian, and 1PN corrections to both the amplitude and the phase of the waveform. As expected, Eqs.  \eqref{EQ:waveform_1} and \eqref{EQ:waveform_2} give us the Fourier transform of the gravitational-wave strain to 1PN order. We note this waveform agrees with the one computed by Alexander \textit{et al}. \cite{Alexander:2018qzg} at leading PN order, when expanded in the small photon mass limit. 

We recognize that when $\alpha=\gamma=0$ in Eq. \eqref{EQ:waveform_2}, we recover the Newtonian and 1PN terms of the amplitude and phase of the IMRPhenomD waveform family \cite{Husa:2015iqa,Khan:2015jqa}.  
When performing the Bayesian parameter estimation analysis in Sec ~\ref{SEC:Bayes}, we incorporate the linear-in-$\alpha$ and linear-in-$\gamma$ terms of Eq. \eqref{EQ:waveform_2} into the inspiral amplitude and phase of the state-of-the-art NS binary frequency domain model \texttt{IMRPhenomD\_NRTidalv2} \cite{Dietrich:2019kaq}. 
The \texttt{IMRPhenomD\_NRTidalv2} model is constructed with a Plank taper that removes the merger and ringdown part of the \texttt{IMRPhenomD} model, taking the waveform smoothly to zero at the end of the inspiral.
Our enhanced waveform model then inherits this taper, allowing us to disregard any corrections to the merger and ringdown of the signal.

\section{Preliminary exploration of the DM parameter space}\label{SEC:DM_params}
In addition to the component masses, three quantities characterize the corrections to the GW signal: the dark charge on each body $q_1$ and $q_2$, and the mass of the Proca field $m_A$. Following \cite{Alexander:2018qzg}, we have thus far chosen a convenient parametrization in terms of three dimensionless parameters: $\alpha$, $\gamma$, and $\mu$, defined in Eq.~(\ref{eq:alpha-gamma-mu-def}). One might wonder if GW observations can be used to place constraints directly on the Lagrangian parameters $\gD$ and $m_\chi$. If each body $I$ has dark-matter-mass-to-total-mass-ratio $f_I \equiv N_I m_\chi/m_I$, then 
\begin{align}
    \label{EQ:params}
    \alpha = &~\frac{{\hbar c}}{G}\frac{g_\mathrm{D}^2}{m_\chi^2}f_1 f_2\,,\qquad
    \gamma =~ \frac{{\hbar c}}{G}\frac{g_\mathrm{D}^2}{m_\chi^2}(f_1 -f_2)^2\,.
\end{align}
We see that $\alpha$ and $\gamma$  are degenerate with respect to $g_\mathrm{D}$, $m_\chi$, and $f_I$. For a given measurement of $\alpha$ and $\gamma$, it would be impossible to determine $g_\mathrm{D}$ and $m_\chi$ without an independent measurement of $f_1$ and $f_2$. Nonetheless, placing constraints on $\alpha$ and $\gamma$ can inform us about the extent to which DM of this sort may be impacting binary inspirals. Independent studies of the DM capture process, which are beyond the scope of this paper, may be used in the future in conjunction with our GW analysis to place constraints on specific DM properties. 

That being said, in this section, we explore the regime of parameter space in which the approximations made while constructing the waveform remain valid. This exploration is important because it informs the boundary of the priors that will be used in our gravitational-wave parameter estimation studies in the next section. We then perform a principal component analysis to determine whether parametrizing the waveform in terms of $\alpha$, $\gamma$, and $\mu$ is the optimal choice for gravitational-wave data analysis. This is important because it is not \textit{a priori} obvious what the optimal directions of parameter space are for a Bayesian analysis. As we shall see in this section, however, $\alpha$, $\gamma$ and $\mu$ are indeed the optimal directions.

\subsection{Restrictions on parameter space}

In constructing the waveform presented earlier in this paper, we have made several assumptions about the magnitude of the aDM parameters $\alpha$, $\gamma$, and $\mu$. First, we assumed that the length scale of the Proca interaction is much greater than the orbital separation of the binary. We established that this assumption is equivalent to assuming $\mu/x \ll 1$, or 
\begin{align}
    \label{eq:mu-cond}
    \mu \ll 5 \times 10^{-3}\left(\frac{f}{\mathrm{10  Hz}}\right)^{2/3}\left(\frac{m}{2 \mathrm{M}_\odot}\right)^{2/3}\,.
\end{align}
When carrying out Bayesian parameter estimation in the next section, we will have to enforce this condition for the entire duration of the signal in the detector band. Going forward, we will assume that the sensitivities of the ground-based detectors we consider are bounded on the lower end by $f_\mathrm{min } =$ 10Hz, and that the NS component masses are bounded by $1 \mathrm{M}_\odot<m_{1,2}<3 \mathrm{M}_\odot$, such that $2 \mathrm{M}_\odot<m<6 \mathrm{M}_\odot$.  We will, therefore, enforce 
\begin{align}\label{EQ:mu_prior}
    \mu < 10^{-4}\,, \qquad {\rm{or}} \qquad
    m_\gamma < 6.7 \sci{-15} \ {\rm{eV}}\, ,
\end{align}
as a small dark photon mass prior, which is guaranteed to satisfy Eq.~\eqref{eq:mu-cond} for all GW frequencies and masses explored. 

The conditions for activating electric dipole and electric quadrupole radiation modes are  $\mu/x^{3/2}<1$ and $\mu/x^{3/2}<2$, respectively. Given our newly-established upper bound on $\mu$, and again assuming a minimum total binary mass of $m =2 \mathrm{M}_\odot$, these conditions are violated below approximately 3.2Hz and 1.6Hz. Therefore, for the signals considered in this paper, both electric radiation modes are activated for the whole duration of a signal in the detector band. Having established this, the unmodified waveform is only recovered when $\alpha = \gamma =0$. The dipole radiation parameter $\gamma$ is positive definite and, due to the capture mechanisms discussed in Sec.~\ref{SEC:Intro}, we assume $\alpha>0$ as well. 

Upper bounds on $\alpha$ and $\gamma$ are enforced by our assumption that the corrections to the waveform constitute small deformations from the unmodified signal. To this end, we discard terms quadratic in $\alpha$ and $\gamma$ when computing the waveform. We enforce the small deformation limit here by requiring that the linear-in-$\alpha$ and linear-in-$\gamma$ corrections to the amplitude and phase of the waveform at Newtonian and positive PN orders are each smaller than the unmodified term at the same PN order. For the negative PN corrections, which do not have unmodified counterparts, we enforce that the corrections be smaller than the Newtonian term for the signal's entire duration in the detector band. The most restrictive bounds  on $\alpha$ and $\gamma$ come from the 1PN and -1PN corrections to the phase, respectively, and are 
\begin{align}\label{EQ:alpha_gamma_prior}
    \alpha<&~\frac{ 2229+2772 \eta}{1988+14224 \eta}\,,\neweq
    \gamma<&~ 7.7 \times 10^{-2}\,.
\end{align}
With this in hand, the maximum value $\alpha$ can take for an NS/NS binary is approximately  $5.9 \times 10^{-1}.$ 
Equations~(\ref{EQ:mu_prior}) and (\ref{EQ:alpha_gamma_prior}) give the regime of validity of our model due to the approximations made while calculating it.

\subsection{Principal component analysis}

We now perform a principal component analysis (PCA) to determine whether the parametrization outlined in Eq.~(\ref{EQ:params}) is, in fact, optimal for parameter estimation.  Following the notation of \cite{Ohme:2013nsa}, we give an overview of the mechanics of a PCA. As an illustration, we rederive the well-known result that the chirp mass is the best constrainable mass parameter of a GW signal. We then return to our newly modified signal to perform a PCA of the DM parameters.

\subsubsection{Review of PCA}

Consider a waveform model $ h =  h(\theta^i)$ that depends on a set of model parameters $\theta^i$. The Fisher information matrix (FIM) of this waveform is  
\begin{align}
    \Gamma_{ij} = \left(\frac{\partial h }{\partial \theta^i}\bigg|\frac{\partial h }{\partial \theta^j}\right)\,,
\end{align}
where the inner product is 
\begin{align}
    (a|b) = 4 \mathrm{Re}\int \frac{\tilde a(f) \tilde b^{*}(f)}{S_n(f)}df\,,
\end{align}
$S_n(f)$ is the spectral noise density of the chosen detector, the tilde indicates the Fourier transform of the function, the asterisks indicate complex conjugation, and $\mathrm{Re}$ is the real part operator.

The FIM is equivalent to the second derivative of the likelihood function evaluated at its maximum, 
\begin{align}
    \Gamma_{ij} = -\frac{\partial^2\mathcal{L}}{\partial \theta^i\partial \theta^j}\bigg|_{h(\theta) = s}\,.
\end{align}
The likelihood  
\begin{align}
    \mathcal{L}(\theta^i) \propto \exp \left[-\frac{1}{2}\left(s - h(\theta^i)|s - h(\theta^i)\right)\right]
\end{align}
gives the probability that the detector response $s$ is observed given the model $h$ and parameters $\theta^i$, $P(s|\theta^i,h)$. The likelihood is maximized when $h(\theta^i) = s$, assuming Gaussian and stationary noise.

For a true signal $h$, the inverse of the Fisher information matrix $\Gamma^{-1}$ gives the variance-covariance matrix of the Bayesian posterior probability density function (PDF) for the true source parameters $\theta^i$, assuming the signal has high SNR, the detector noise is Gaussian and stationary, the model is a linear function of the waveform parameters, and the prior (PDF) on the waveform parameters are uniform~\cite{Vallisneri:2007ev,Porter:2015eha}. %
Therefore, when these conditions are met, the diagonal elements of $\Gamma^{-1}$ give the squares of the 1-$\sigma$ credible intervals of the marginalized posterior PDFs of the model parameters. The off-diagonal elements of $\Gamma^{-1}$ provide a measure of the correlations between parameters. Even when these conditions are only approximately met, the $\Gamma^{-1}$ matrix can still be used to estimate the statistical error that can be optimistically expected from Bayesian parameter estimation of GW signals. 

If  $\Lambda_{ij}$ is the $j$th component of the $i$th eigenvector of $\Gamma_{ij}$, the principal components of the given waveform model are  
\begin{align}
    \mu_i =\Lambda_{ij} \theta^{j}\,.
\end{align}
Since the FIM and its inverse have the same eigenvectors, reparametrizing the model in terms of the principal components will produce a waveform with a diagonal variance-covariance matrix, thus eliminating correlations between model parameters (provided the conditions above are met and $\Gamma^{-1}$ is exactly the covariance matrix). Therefore, the principal components are the linear combinations of parameters that can be best constrained through a Bayesian parameter estimation analysis.
After re-parametrization, the eigenvalues of $\Gamma_{ij}$ become the elements of the newly diagonalized FIM. 
Therefore, the inverse square roots of the eigenvalues provide an estimate of the 1$\sigma$ credible intervals that would be obtained from a Bayesian analysis with the model parametrized in terms of the principle components.
If the model is not a linear function of the original parameters, the PCA parametrization will still tend to minimize correlations. Such minimization, however, is only achieved up to the linearity of the model with respect to its parameters. 

\subsubsection{Implementation of PCA}

To provide an illustration of the above, we rederive a familiar result. Consider the unmodified Newtonian waveform expressed in terms of  $\theta^i = [\log m, \log \eta]$: 
\allowdisplaybreaks[4]
\begin{align}\label{EQ:h_newt}
    \tilde h  =& {(\pi f)^{-7/6}}\exp\bigg\{\frac{5}{6}\left(\log m+\frac{3}{5}\log\eta\right)\bigg\}
    \nonumber \\
    &\exp\bigg\{-i\frac{3}{128}\exp\left[{-\frac{5}{3}\left(\log m+\frac{3}{5}\log\eta\right)}\right](\pi f)^{-5/3}\bigg\}\,.
\end{align}
We choose to work with logarithmic parameters here to enforce the linear parameter dependence of the model.  For the remainder of this section, we take the spectral noise density to be an analytic approximation of the sensitivity of the LIGO detectors during the fifth observing run at design sensitivity \cite{OReilly_2022}. We integrate the inner product of the FIM from 10 Hz to the frequency of the innermost stable circular orbit 
\begin{align}
    f_\mathrm{ISCO} = 4.4\sci{3} \mathrm{Hz} \left(\frac{\mathrm{M}_\odot}{m}\right)\, ,
\end{align}
which is often considered the end of the inspiral. 
Upon performing a PCA we find, unsurprisingly, that the 
principal components are
\begin{align}
\label{eq:PCA-Newt}
    \mu_1 =&~ \log m+\frac{3}{5}\log\eta \,, \neweq
    \mu_2 =&~ \log \eta-\frac{3}{5}\log m \, .
\end{align}
 The quantity $e^{\mu_1} = m \, \eta^{3/5}$ is called the chirp mass $\mathcal{M}$ and is well known to be the best constrainable parameter of a GW signal. 
Parametrized in terms of $\mathcal{M}$, the waveform becomes
\begin{align}
    \tilde h  \propto&~ {(\pi f)^{-7/6}}\mathcal{M}^{5/6}\exp\bigg[-i\frac{3}{128}\mathcal{M}^{-5/3}(\pi f)^{-5/3}\bigg]\,.
\end{align}
In doing the PCA, we have eliminated correlations between the $m$ and $\eta$ parameters because we have expressed the waveform in terms of a single parameter.  

If we include the 1PN corrections to the amplitude and phase, however, the waveform is no longer a function of a single linear combination of the parameters. The PCA is now dependent on the region of parameter space one is in. For example, for a binary with $m_1 = m_2 = 1 \mathrm{M}_\odot$, we find 
\begin{align}
\label{eq:mus-case1}
  \mu_1 =&~\log m+ 0.605\log\eta\,,\neweq
  \mu_2 =&~\log\eta-0.605 \log m\,,
\end{align}
whereas when $m_1 = m_2 = 3 \mathrm{M}_\odot$,
\begin{align}
\label{eq:mus-case2}
  \mu_1 =&~\log m+ 0.610\log\eta\,,\neweq
  \mu_2 =&~\log\eta-0.610 \log m\,.
\end{align} 
We see that, in both cases, we get a result that is close to, but not quite, Eq.~\eqref{eq:PCA-Newt}, with the difference depending on the position in the parameter space where we evaluate the FIM. Because the principal components can vary over the parameter space under consideration, in practice, it makes sense to pick a parametrization that is simple to implement. In either of the cases above, $\mathcal{M}$ is still a reasonable choice for the primary mass parameter. 

Furthermore, because the waveform is no longer a function of a single linear combination of the parameters,  reparametrizing the waveform in terms of the principal components need not completely diagonalize the FIM. As a result, correlations between parameters are not completely eliminated, although they are significantly reduced. We can see this in the 1PN example above. When the 1PN waveform is parametrized in terms of $\theta^i = [\log m, \log \eta]$ and $m_1 = m_2 = 3 \mathrm{M}_\odot$,  we find 
\begin{align}
    \Gamma_{ij} \propto \left(
\begin{array}{cc}
 1 & 0.605 \\
 0.605 & 0.366 \\
\end{array}
\right)\,. 
\end{align}
On the other hand, when the waveform is reparametrized in terms of the principal components of Eq.~\eqref{eq:mus-case2}, we find
\begin{align}
   \Gamma_{ij} \propto   \left(
\begin{array}{cc}
 1. & 1.6\sci{-13} \\
  1.6\sci{-13} &  1.5\sci{-6} \\
\end{array}
\right)\,.
\end{align}
We see similar results when $m_1 = m_2 = 1 \mathrm{M}_\odot$. We see that the correlations between parameters, quantified by the off-diagonal elements of the FIM, are significantly reduced when the waveform is reparametrized in terms of the principal components. However, they are not entirely eliminated as they were in the Newtonian case, where the waveform could be expressed in terms of a single linear combination of parameters. 

Another shortcoming of a PCA is that it returns only linear combinations of the original model parameters. Consider, instead, that the Newtonian waveform in Eq.~(\ref{EQ:h_newt}) is parametrized in terms of $\theta^i = [\log m, \log q]$, where $q = m_2/m_1 = (1-\sqrt{1-4 \eta })/(1+\sqrt{1-4 \eta })$ is the mass ratio. A PCA would \textit{not} return $\log \mathcal{M}$ as one of the principal components because there is no linear combination of $\log m$ and $\log q$ that gives $\log \mathcal{M}$. This structure cannot be captured by a standard PCA.

All that being said, we now return to our modified waveform given in Eqs.~(\ref{EQ:waveform_1}) and (\ref{EQ:waveform_2}). In our PCA, we take $\theta^{i} = [{\log{\alpha}}, {\log{\gamma}},  {\log{\mu}}$], while the other parameters are set at fixed values and not differentiated when computing the FIM. For this analysis we choose $m_1= 1.5\mathrm{M}_\odot$ and $m_2= 1.25\mathrm{M}_\odot$. As we have seen, the results of a PCA are parameter-dependent, and thus, we perform the analysis multiple times along the boundary of the parameter space $[{\log{\alpha}}, {\log{\gamma}},  {\log{\mu}}]$, outlined in Eqs. (\ref{EQ:mu_prior}) and (\ref{EQ:alpha_gamma_prior}). With three parameters, this leaves us with eight analyses to perform.  When carrying out this analysis, we must be careful because each parameter is bounded on the lower end by zero, the log of which approaches negative infinity. We find that, as we lower the minimum value approaching zero, the principal components and their order remain stable.

We collect the results of the PCA in Table \ref{tab:PCA}. We find that for every choice of parameters, the optimal parametrization is either exactly the parametrization we started with or approximately so. We report the principal components in descending order, with the best constrainable parameter coming first. We find in almost all cases, that $\gamma$ is the best constrainable parameter. Based on the results of the PCA, we choose to proceed with the parametrization outlined in Eq.~(\ref{EQ:params}). 
\begin{table*}[th]
    \centering
    \begin{tabular}{c c c |c c c}
    \hline\hline
     \multicolumn{3}{c|}{Parameter values}& \multicolumn{3}{c}{Principal components}\\
    $\alpha$ & $\gamma$ & $\mu$ & 1st& 2nd& 3rd \\\hline
    0 & 0 & 0 &
        $\gamma\alpha^{-6.1\sci{-2}}$&
        $\alpha\gamma^{6.1\sci{-2}}$&
        $\mu$\\
    0 & 0 & $10^{-4}$ & 
        $\gamma \alpha^{-6.1\sci{-2}}\mu ^{9.4\sci{-5}}$ & 
        $\alpha \gamma^{6.1\sci{-2}}\mu ^{-2.6\sci{-3}}$ & 
       $  \mu \alpha^{2.6\sci{-3}} \gamma^{6.4\sci{-5}}$\\
    $5.9 \times 10^{-1}$ & 0 & 0 &$\alpha $ & $\mu$  & $\gamma$ \\
    $5.9 \times 10^{-1}$  & 0 & $10^{-4}$&
        $\alpha\mu ^{-1.6\sci{-3}}$&
        $\mu\alpha ^{1.6\sci{-3}}$&
        $\gamma$ \\
    0 & $7.7 \times 10^{-2}$ & 0& 
        $\gamma$&
        $\alpha$&
        $\mu$\\
    0 & $7.7 \times 10^{-2}$ & $10^{-4}$ &
        $\gamma$&
        $\alpha$&
        $\mu$\\
    $5.9 \times 10^{-1}$  & $7.7 \times 10^{-2}$ & 0 &
        $\gamma \alpha^{-5.7\sci{-1}}$&
        $\alpha  \gamma^{5.7\sci{-1}}$&
        $\mu$\\    
    $5.9 \times 10^{-1}$  & $7.7 \times 10^{-2}$ & $10^{-4}$ &
        $\gamma\alpha ^{-5.7\sci{-1}}\mu^{1.0\sci{-3}}$&
        $\alpha \gamma ^{5.7\sci{-1}}\mu ^{-2.9\sci{-3}}$&
        $\mu\alpha ^{2.6\sci{-3}} \gamma ^{5.1\sci{-4}} $ \\
    \hline\hline
    \end{tabular}
    \caption{\label{tab:PCA} A summary of the results of our PCA conducted on the waveform given in Eqs. (\ref{EQ:waveform_1}) and (\ref{EQ:waveform_2}).  While the PCA was done in terms of the parametrization $\theta^{i} = [{\log{\alpha}}, {\log{\gamma}},  {\log{\mu}}$], we present the exponentiated parameter values and principal components. Observe that for every choice of parameter values, the optimal parametrization is either exactly the parametrization given in Eq. (\ref{EQ:params}) or approximately so. We report the principal components in descending order, with the best constrainable parameter coming first. In almost all cases, $\gamma$ appears first.}
\end{table*}
   
\section{Bayesian Analysis of Gravitational-Wave Data}\label{SEC:Bayes}
In this section, we outline the methodology and results of our Bayesian parameter estimation study. We investigate the constraints that can be placed on the DM parameters $\alpha$, $\gamma$, and $\mu$ from observed LVK NS/NS and NS/BH events. We then look to the future and perform an injection and recovery campaign to understand how these constraints may be improved during the LVK Collaboration's fifth observing run (O5). 

\subsection{Methodology}\label{SEC:Methods}

As detailed in Sec. \ref{SEC:Waveform}, we have incorporated the linear-in-$\alpha$ and linear-in-$\gamma$ terms of Eq. \eqref{EQ:waveform_2} into \texttt{IMRPhenomD\_NRTidalv2}. The \texttt{IMRPhenomD\_NRTidalv2} model is a nonspin-precessing frequency domain inspiral model, which depends on thirteen astrophysical parameters: chirp mass $\mathcal{M}= {(m_1m_2)^{3/5}(m_1+m_2)^{1/5}}$, mass ratio $q=m_2/m_1$, dimensionless spin magnitudes $\chi_{1}$ and $\chi_{2}$, tidal deformabilities $\Lambda_{1}$ and $\Lambda_{2}$,  luminosity distance $d_L$, inclination angle $\iota$, polarization angle $\psi$, right ascension angle $\mathrm{ra}$, declination angle $\mathrm{dec}$, coalescence phase $\varphi_c$, and coalescence time $t_c$.
To this set of parameters, we add our three DM parameters, $\alpha$, $\gamma$, and $\mu$. 
We employ a likelihood function where $\varphi_c$, and $d_L$ have been analytically marginalized over \cite{Farr_2022,Singer:2015ema,Singer:2016eax}, and vary over the remaining 14 (astrophysical and DM) parameters of the model, except when noted.

To perform our analyses, we use the Bayesian inference library \texttt{Bilby} \cite{Ashton:2018jfp, Romero-Shaw:2020owr} and employ nested sampling with \texttt{Dynesty} \cite{Speagle_2020}.
For the NS/NS events,  we choose the \texttt{Bilby} sampler settings \texttt{nlive}= 1000, \texttt{naccept}= 60,
\texttt{dlogz}= 0.1, \texttt{sample}=‘acceptance-walk’, and \texttt{bound}=‘live’ to ensure that our exploration of the likelihood surface has converged. 
For the NS/BH events, we keep the same setting except with \texttt{naccept}= 100.
To verify that the sampler converges on the true posteriors, we perform a set of runs with the settings
\texttt{nlive}= 1500, \texttt{nact}= 10,
\texttt{dlogz}= 0.01, \texttt{sample}=‘rwalk’, and \texttt{bound}=‘live’ and confirm that the recovered distributions are consistent.

Bayesian parameter estimation requires that we assign prior PDFs to our model parameters.
We assign uniform prior PDFs to our DM parameters $\alpha$, $\gamma$, and $\mu$, enforcing the boundaries given in Eqs. (\ref{EQ:mu_prior}) and (\ref{EQ:alpha_gamma_prior}). 
We also must select priors PDFs for the eleven parameters of \texttt{IMRPhenomD\_NRTidalv2} we have not marginalized away.  
When analyzing NS/NS events, we use the same prior PDFs that were used in the original analysis of GW170817 \cite{LIGOScientific:2017vwq}, selecting the low-spin prior corresponding to $ 0 \leq (\chi_1,\chi_2) \leq 0.05$, unless otherwise noted. 
As was done in that analysis, we select prior PDFs for $\mathcal{M}$ and $q$ that correspond to uniform distributions on the component masses $m_1$ and $m_2$, with the restriction $m_1 > m_2$. 
However, when analyzing NS/NS events, we use the narrower bounds $ 1~\mathrm{M}_\odot \leq (m_1,m_2) \leq 3~\mathrm{M}_\odot$, to reflect the expected mass range of typical NSs. 

For NS/BH events, we change the prior range of the primary mass to reflect that of stellar-mass BHs, i.e. $3~\mathrm{M}_\odot \leq m_1\leq 50~\mathrm{M}_\odot$ and we set the prior on $\Lambda_1$ to be a Dirac delta function peaked at zero. Additionally, the no-hair theorem dictates that the BH will not carry massive dark charge. Therefore, for an NS/BH event $\alpha=0$. With $\alpha$ removed from the waveform, $\mu$ only enters through the Heaviside functions that turn on the vector radiation modes. However, we have established that for the range of parameters considered here, the modes will be activated for the entirety of the signal in the detector band, removing any dependence on $\mu$ from our model as well. 
We, therefore, give $\alpha$ and $\mu$ Dirac delta priors peaked at zero for the NS/BH events.

\subsection{Constraints from observed LVK signals}
To date, the LVK Collaboration has observed two confirmed NS/NS binary signals, designated GW170817 \cite{LIGOScientific:2017vwq,LIGOScientific:2018hze} and GW190425 \cite{LIGOScientific:2020aai} and two NS/BH binary signals designated GW200105 and GW200115 \cite{LIGOScientific:2021qlt}.
In this section, we apply our enhanced binary model to perform Bayesian parameter estimation on these events and place constraints on the DM parameters.

GW170817 was observed during the second observing run (O2) by both of the advanced LIGO detectors \cite{LIGOScientific:2014pky} and advanced Virgo \cite{TheVirgo:2014hva}, with a signal-to-noise ratio (SNR) of 32.4.
GW190425 was observed during the third observing run (O3) by only the advanced LIGO detector at Livingston and advanced Virgo because the LIGO detector at Hanford was offline at the time. It was observed with an SNR of 12.9.
Both GW200105 and GW200115 were observed during O3. GW200105 was observed by the advanced LIGO detector at Livingston and advanced Virgo because the LIGO detector at Hanford was offline at the time, while GW200115 was observed by both advanced LIGO detectors and advanced Virgo. The events had SNRs of approximately 13.9 and 11.6, respectively.

For each event, we use the publicly available 4kHz strain data, with a duration of 128s or 64s for the NS/NS and NS/BH events, respectively \cite{LIGOScientific:2019lzm,KAGRA:2023pio}.
During GW170817, there was a prominent glitch in the LIGO detector at Livingston, and we, therefore, use the noise-subtracted version of the data, which removes the glitch.
We employ the spectral noise density curves released by the LVK Collaboration corresponding to each event, except for GW200105, for which we construct the spectral noise density curves from the time-series data immediately preceding the event at each detector.
We set the lower bound on the frequency at 20Hz, as is standard for O2 and O3 events. 
Before analyzing these signals with our enhanced model, we first employ \texttt{IMRPhenomD\_NRTidalv2} with no modifications, verifying that we recover posterior PDFs consistent with the analyses performed by the LVK Collaboration \cite{LIGOScientific:2017vwq,LIGOScientific:2018hze,LIGOScientific:2020aai,LIGOScientific:2021qlt}.
With that verification complete, we next analyze these signals with our enhanced \texttt{IMRPhenomD\_NRTidalv2} model.

For the NS/NS events, we vary over the three DM parameters $\alpha$, $\gamma$, and $\mu$, in addition to the parameters of the \texttt{IMRPhenomD\_NRTidalv2} model.
For the NS/BH events, we follow a similar procedure but do not sample on $\Lambda_1$, $\alpha$, and $\mu$, as noted in the previous section. 
For each of the events, we find that the posterior PDFs on  $\gamma$ and $\alpha$ (in the case of NS/NS binaries) are peaked near zero, allowing us to place constraints on $\gamma$ and $\alpha$. However, we find that no constraint can be placed on the dark photon mass parameter $\mu$.  When $\alpha = \gamma = 0$, no dark charge is present in the stars and the parameter $\mu$ does not appear in the model of Eq.~\eqref{EQ:waveform_2}.  When this is the case,  no information can be gained about the mass of the dark photon. We, therefore, do not include the posterior PDFs on $\mu$ in our discussion to follow.

\begin{figure*}
    \centering
    \includegraphics[width=0.90\textwidth]{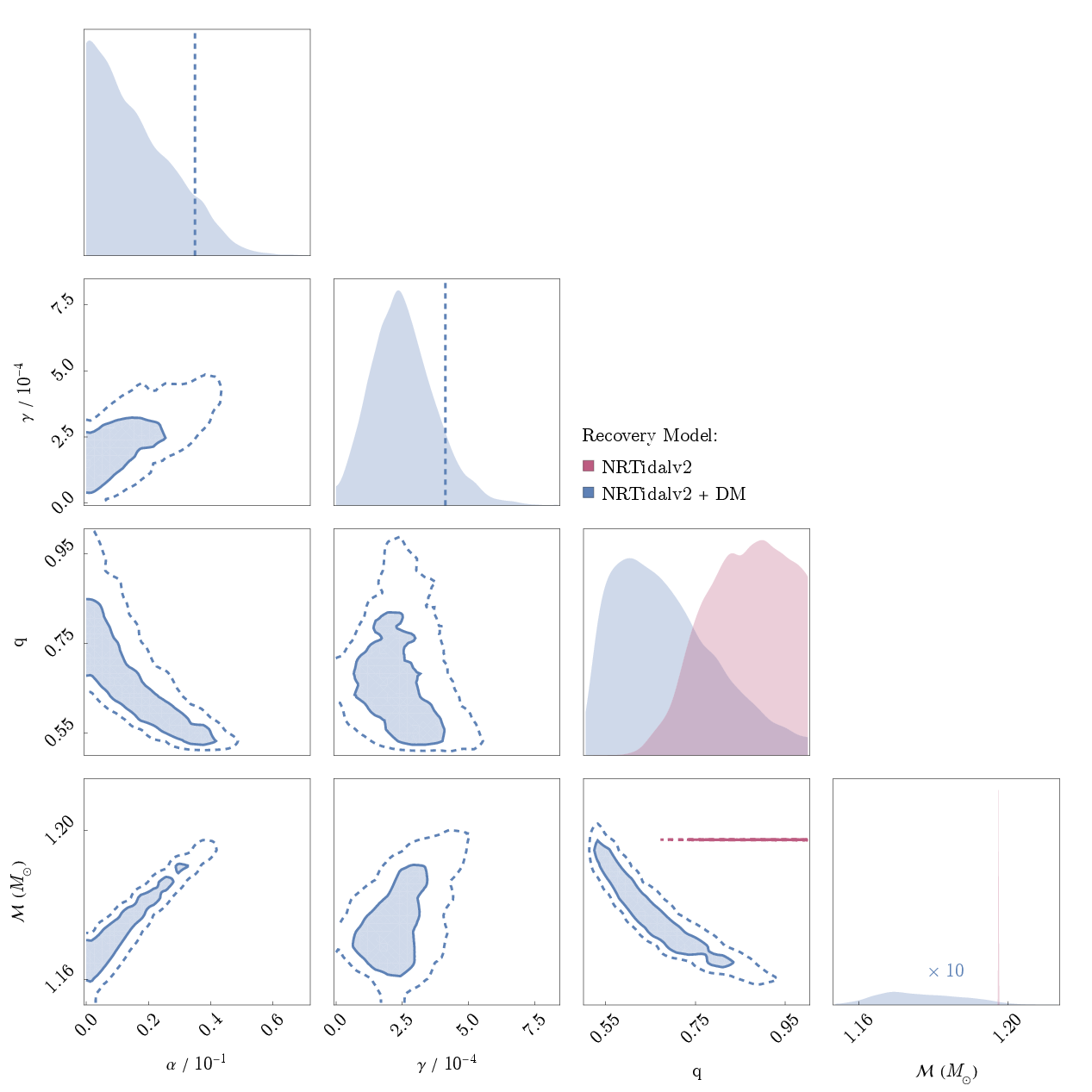}
    \caption{A partial corner plot produced after a parameter estimation analysis of the NS/NS signal GW170817 using unmodified \texttt{IMRPhenomD\_NRTidalv2} (pink) and  \texttt{IMRPhenomD\_NRTidalv2} enhanced with DM corrections (blue). The corner plot includes the marginalized one- and two-dimensional posterior PDFs on the DM and mass parameters. In both the one- and two-dimensional plots, dashed lines represent 90\% credible regions, and in the two-dimensional plot, solid lines enclose the 50\% credible regions. The normalized one-dimensional marginalized posterior PDF on $\mathcal{M}$ obtained from the analysis with the enhanced waveform has been multiplied by a factor of 10 so that it is visible in the plot.  We find that $\alpha<3.5\sci{-2}$ and $\gamma<4.1\sci{-4}$ with $90\%$ credibility.}
    \label{fig:GW170817}
\end{figure*}

Let us first consider GW170817. In Fig.~\ref{fig:GW170817}, we present a partial corner plot that contains one-dimensional marginalized posterior PDFs and the 50\% and 90\% credible region contours of the two-dimensional marginalized posterior PDFs on the DM and mass parameters.
We include posterior PDFs obtained by analyzing the GW170817 signal with both the unmodified \texttt{IMRPhenomD\_NRTidalv2} (plotted in pink) and our enhanced version of \texttt{IMRPhenomD\_NRTidalv2} with the DM parameters included (plotted in blue). 
Observe that the two-dimensional posterior PDF in the $\alpha$--$\gamma$ plane obtained using the enhanced waveform is shifted away slightly from the no-DM point ($\alpha = 0 = \gamma$).
Additionally, the one-dimensional marginalized posterior PDF for $\gamma$ does not peak at $\gamma=0$.
At a glance, one might assume that this indicates a positive detection of DM effects. 
However, there are several reasons that this behavior could occur despite no DM being present, so let us investigate them one by one. 

The behavior of an approximate posterior PDF obtained through Bayesian analysis is governed by many factors including the chosen priors and noise model, the information contained in the likelihood, and the ability to discretely sample the PDFs accurately.
We have verified that our nested sampling exploration of the posterior is indeed robust, as described in Sec.~\ref{SEC:Methods}.
Furthermore, our choice of priors does not impact the posterior PDFs we have obtained, as they are flat in the DM parameters within the range of the corner plot of Fig.~\ref{fig:GW170817}.
With those explanations ruled out, two other possible culprits could cause the behavior in Fig. \ref{fig:GW170817} in the absence of DM.
First, correlations between model parameters, which enter the analysis through the likelihood, can cause the maxima of the marginalized posteriors to be biased from the true values, as we show explicitly with a toy problem in Appendix~\ref{SEC:degeneracies}.
Alternatively, our Gaussian and stationary approximation to the noise model may be insufficient.

To investigate whether these effects could be contributing to the structure we see in Fig.~\ref{fig:GW170817}, we perform two injection and recovery analyses.  
In both analyses, we perform parameter estimation on the same synthetic injection of a GW170817-like event.
We choose injection values consistent with posteriors obtained from our analysis of GW170817 with the unmodified waveform, i.e.~we set $m_1 = 1.5 \mathrm{M}_\odot$, $m_2 = 1.25 \mathrm{M}_\odot$, $\chi_{1}=\chi_{2} =0 $, $\Lambda_{1} = 160$, $\Lambda_{2} = 500$, $d_L =40$Mpc, $\iota = 2.64$ rad, $\psi = 1.8$,  $\mathrm{ra}$ = 3.4, $\mathrm{dec} = -0.401 $,  $\varphi_c = 0$, and  $t_c =1187008882.4$ s.
The injected tidal deformability parameters are chosen to be consistent with both the GW170817 posterior PDFs and the APR equation of state, given the chosen values for the injected masses \cite{Akmal_1998,Schneider:2019vdm}.
We enforce that the signal is produced by a system \textit{without} DM by setting $\alpha = \gamma = \mu = 0$ for the injection.
We assume the noise is Gaussian and stationary, and we employ the spectral noise density estimated for the O2 detectors around the time of the GW170817 event. 
However, we do not insert a particular noise realization or any glitches into the data.
In both parameter estimation analyses, we use the same sampler settings and the same enhanced model that we used when analyzing the true GW170817 event.
However, in each analysis, we implement different priors. 
In the first analysis, we use the same priors that we had used when analyzing GW170817.
In the second analysis, we set the prior PDFs on the mass parameters $q$ and $\mathcal{M}$ to Dirac deltas peaked at the injected values, effectively removing these parameters from the model. 

\begin{figure*}
    \centering
    \includegraphics[width=.579\linewidth]{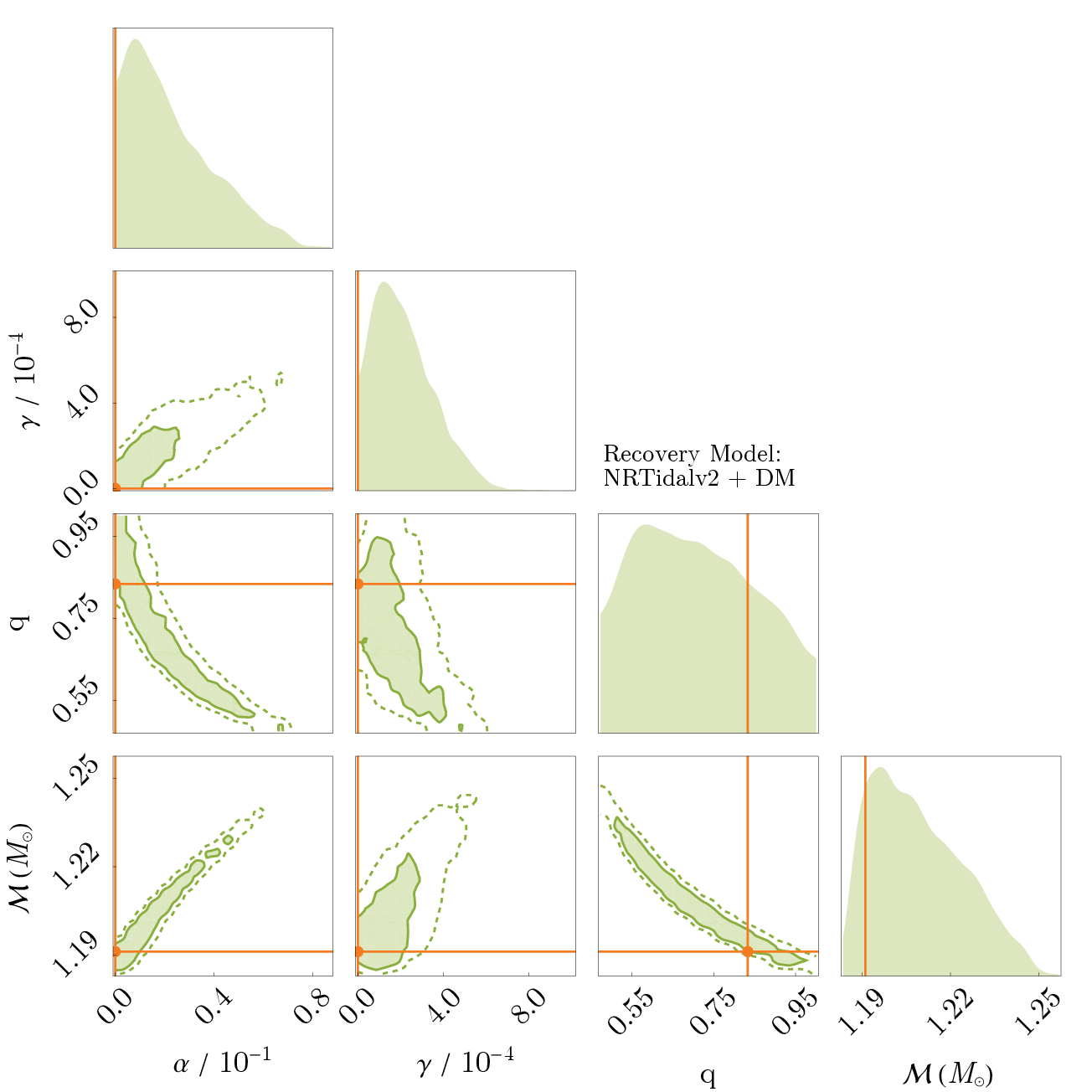}
    \vspace{.01\textwidth}
    \includegraphics[width=.321\linewidth]{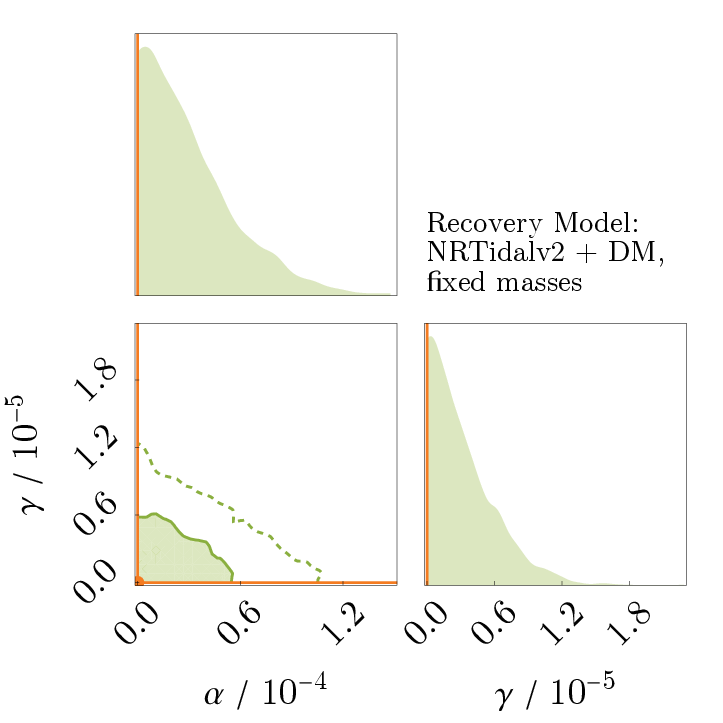}
    \caption{Partial corner plots produced after a two parameter estimation analyses of same synthetic GW170817-like data. Both analyses employed \texttt{IMRPhenomD\_NRTidalv2} enhanced with DM corrections. However, in the analysis used to produce the plot on the right, the prior PDFs on the mass parameters were set to Dirac delta functions peaked at their injected values. In both the one- and two-dimensional plots, dashed lines represent 90\% credible regions, and in the two-dimensional plot, solid lines enclose the 50\% credible regions. Orange lines indicate the injected values. We note that, in the left-hand plot, the peaks in the one-dimensional posterior PDFs do not line up with the injected values. 
    On the other hand, in the right-hand plot, where the mass parameters had been removed from the recovery model, the one-dimensional posterior PDFs do indeed peak at their injected values.}
    \label{fig:O2-injection}
\end{figure*}

In Fig. \ref{fig:O2-injection}, we present partial corner plots obtained from these analyses.
To understand the impact that noise in the signal has on the recovered posterior PDF, we look to the corner plot produced by the first analyses in the left-hand panel. 
We note that when no noise is inserted into the data, the two-dimensional posterior PDF for the DM parameters is indeed centered at $\alpha=0=\gamma$. 
Turning to our analysis of the true GW170817 signal, we recall that there was a glitch during the GW170817 event that the LVK Collaboration removed before analyzing the data \cite{LIGOScientific:2017vwq}. As stated in Sec.~\ref{SEC:Methods}, we have analyzed the GW170817 signal with the glitch removed. However, with any glitch removal procedure, the subtraction is never perfect, and the glitch-subtracted GW170817 data still contains some leftover nonstationarity. We, therefore, hypothesize that the glitch subtraction implemented on the GW170817 data has left behind some additional power that could be mimicked by the DM terms of our enhanced model, thus leading to the posterior PDFs shown in Fig.~\ref{fig:GW170817}. This is strong evidence that the shift in the $\alpha$--$\gamma$ posterior PDF of Fig.~\ref{fig:GW170817} is an artifact of the noise during that event. 

To understand how correlations between model parameters can impact the marginalized posterior PDFs, we look to the second analysis.
With the mass parameters eliminated, we see that the one-dimensional posterior PDFs on $\alpha$ and $\gamma$ are much narrower and do indeed peak at the injected values.
This indicates that the behavior of the one-dimensional posterior PDFs on $\alpha$ and $\gamma$ is strongly affected by correlations between the DM and mass parameters.
This is expected because the corrections we have included in the waveform depend on $q$ and $\mathcal{M}$, in addition to $\alpha$, $\gamma$, and $\mu$, thus inducing nontrivial covariances.
When parameters are strongly correlated, it is possible that the maxima of the marginalized posteriors do not correspond to the maximum of the full posterior\footnote{Incidentally, these covariances also explain the deterioration in the measurement of the chirp mass when using the enhanced model, as one can see in the 1D posterior PDF for the chirp mass in Fig.~\ref{fig:GW170817}.}.

Given the sensitivity of the posterior PDFs on the DM parameters to the effects of correlation and detector noise, it is natural to wonder whether we would be able to distinguish the effects of DM if they are in fact present in the data. We produce eight additional synthetic signals: four for which $\alpha \neq 0$ and $\gamma = 0 $, and then four others for which $\gamma \neq 0$ and $\alpha =0 $. For each parameter, we choose one injection that roughly corresponds to the peak of the one-dimensional posterior PDF on that parameter when no DM is present in the signal, and one injection that is well outside the 90\% credible interval of that posterior PDF. We then select two additional injections between these values. We keep the remaining model parameters the same as in the GW170817-like injections illustrated in Fig. \ref{fig:O2-injection} and use the same sampler settings. We do not inject the signal with a specific noise realization and we allow the masses to vary in these analyses. 

\begin{figure*}[]
    \centering
    \includegraphics[width=.45\textwidth]{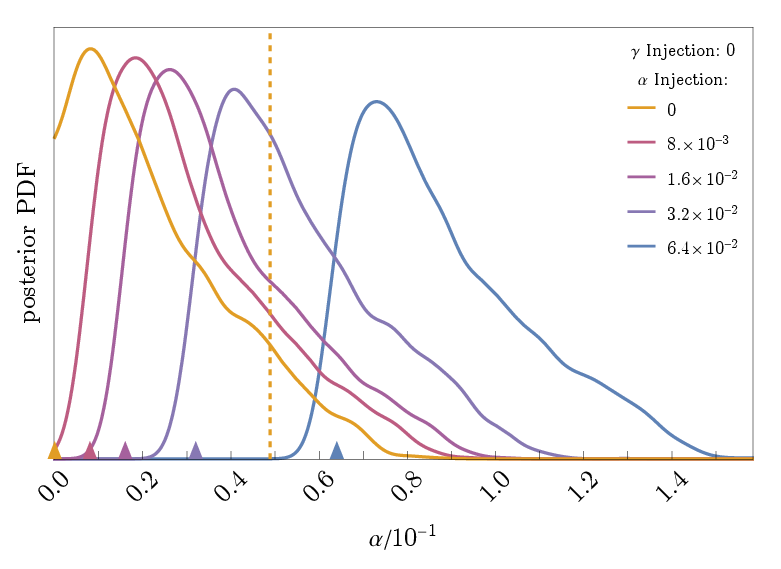}
    \vspace{.01\textwidth}
    \includegraphics[width=.45\textwidth]{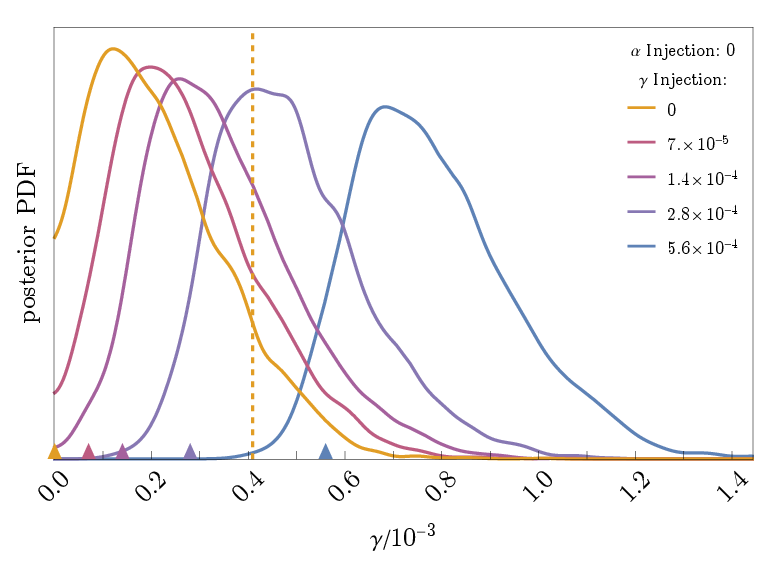}
    \caption{Marginalized posterior PDFs produced after nine parameter estimation analyses of synthetic NS/NS binary data. Each signal was given a different DM injection, indicated in the key and with triangles along the bottom of the plots. The analyses employed \texttt{IMRPhenomD\_NRTidalv2} enhanced with DM corrections. In each plot, the dashed vertical line indicates the 90\% credible interval of the no-DM injection ($\alpha=0=\gamma$).}
    \label{fig:DM_inj}
\end{figure*}

Figure~\ref{fig:DM_inj}, shows the one-dimensional marginalized posterior PDFs on either $\alpha$ or $\gamma$ obtained from these analyses. We also include the marginalized posterior PDFs obtained from the analysis above, where $\alpha=0=\gamma$. We find that the maxima of the marginalized posterior PDFs are shifted from their injected values (indicated with triangles at the bottom of the plot) by roughly constant intervals ($\Delta \alpha \sim 1\sci{-2}$ and $\Delta\gamma\sim 1\sci{-4}$). We also note that while the peaks of the marginalized PDFs from the no-DM injection are shifted from zero, they have significantly more support at zero than those of any of the injections with DM effects present. However, as we have seen in Fig. \ref{fig:GW170817}, when nonstationary noise effects are present in the signal, it is possible for the peaks of the (both the one- and two-dimensional) marginalized posterior PDFs to be shifted from the no-DM point, mimicking the presence of DM in the signal.

While these observations seem discouraging, we note that the marginalized posterior PDFs on $\alpha$ and $\gamma$ are significantly more informative than the wide uniform priors we began with.  With all of these factors taken into account, we proceed to use the one-dimensional marginalized-posterior PDFs in Fig.~\ref{fig:GW170817} to place constraints on $\gamma$ and $\alpha$. We find that $\gamma<4.1\sci{-4}$ and $\alpha<3.5\sci{-2}$ with 90\% credibility, which we indicate in Fig. \ref{fig:GW170817} with dashed vertical lines.
 
\begin{figure}[h]
    \centering
    \includegraphics[width=0.45\textwidth]{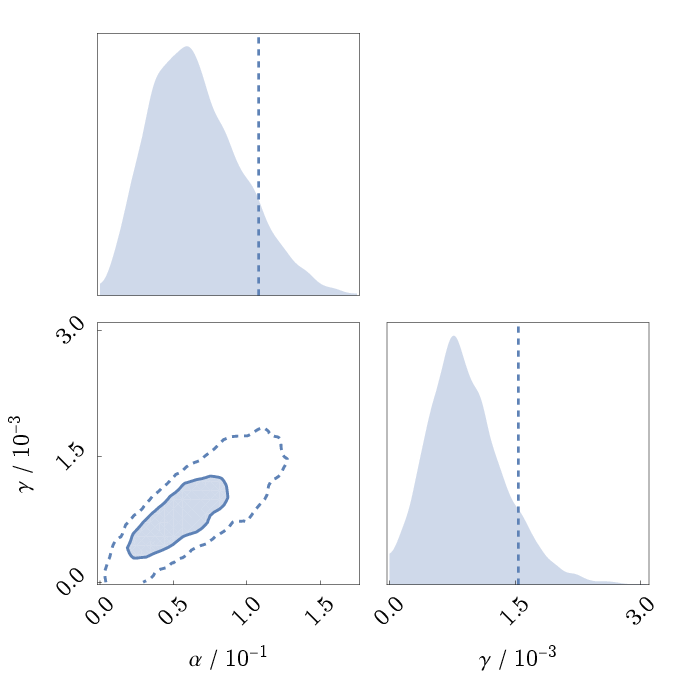}
    \caption{A partial corner plot produced after a parameter estimation analysis of the NS/NS signal GW190425 with the enhanced \texttt{IMRPhenomD\_NRTidalv2}. The corner plot includes the marginalized one- and two-dimensional posterior PDFs on the DM parameters. In both the one- and two-dimensional plots, dashed lines represent 90\% credible regions, and in the two-dimensional plot, solid lines enclose the 50\% credible regions. We find that $\alpha<1.5\sci{-3}$ and $\gamma<1.1\sci{-1}$ with $90\%$ credibility.}
    \label{fig:GW190425}
\end{figure}

\begin{figure}
    \centering
    \includegraphics[width = .45\textwidth]{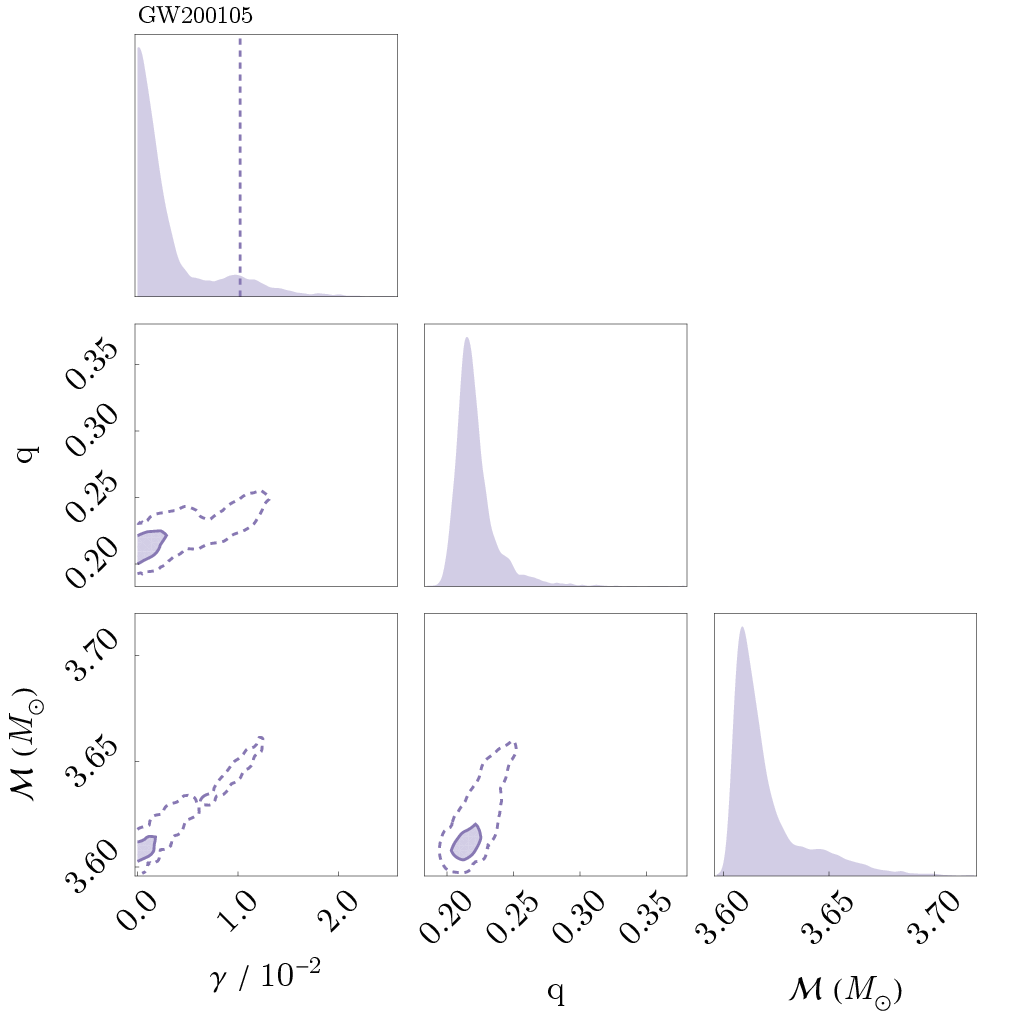}
    \includegraphics[width = .45\textwidth]{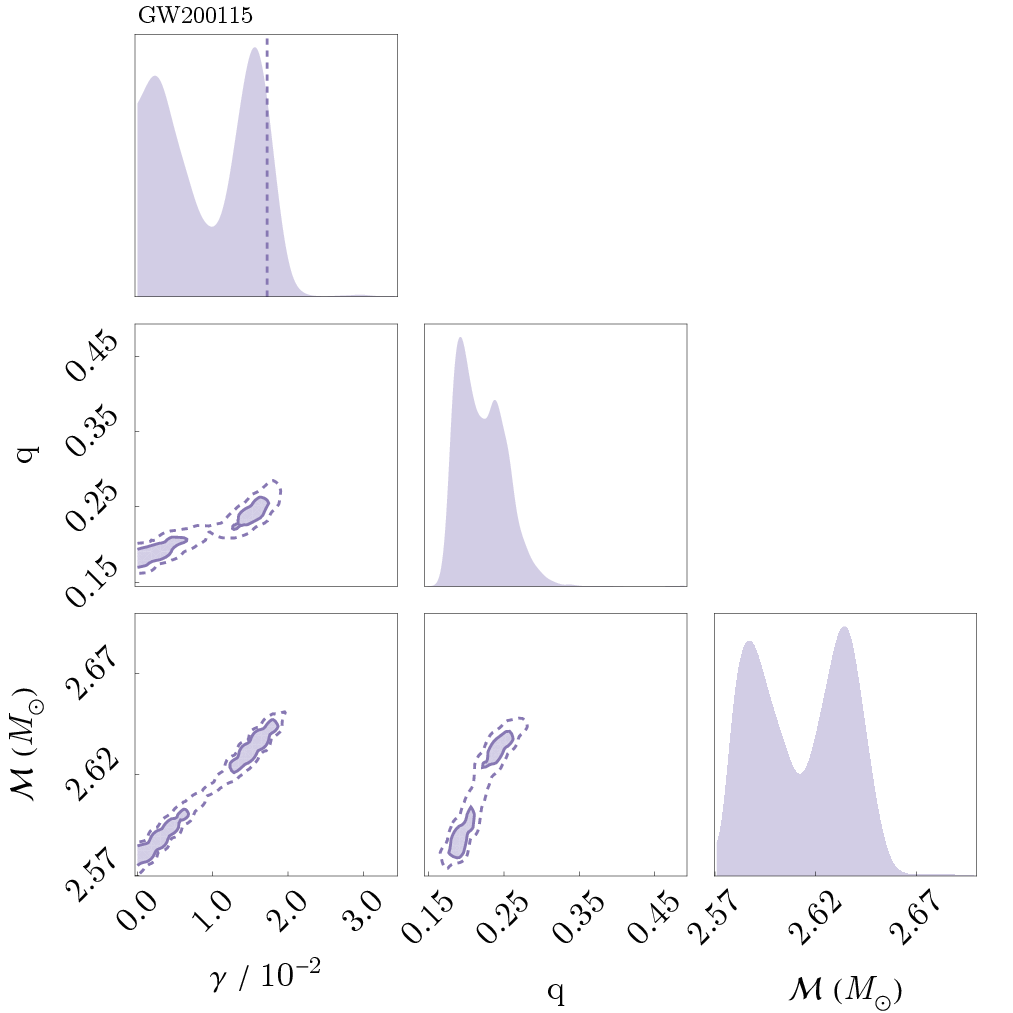}
        \caption{Partial corner plots obtained from parameter estimation analysis of the NS/BH signals GW200105 (top) and GW200115 (bottom). The parameter $\gamma$ is the only DM parameter that enters the waveform for NS/BH binaries. The dashed vertical lines represent 90\% credible bounds. We find that for GW200105 (GW200115),  $\gamma<1.0\sci{-2}$  ($\gamma<1.7\sci{-2}$) with $90\%$ credibility.} 
    \label{fig:NS/BH}
\end{figure}

We now turn to the remaining events. In Fig. \ref{fig:GW190425}, we present a partial corner plot obtained from our analysis of the NS/NS binary GW190425. We see here very similar behavior to what was found when analyzing the GW170817 event. This time, however, the 2D posterior PDF is indeed aligned with the no-DM point ($\alpha = 0 = \gamma$) because the properties of the noise during the GW190425 event were better behaved. In Fig. \ref{fig:NS/BH}, we present partial corner plots obtained from our analysis of the NS/BH binaries GW200105 and GW200115, including both the DM parameter $\gamma$ and the mass parameters. The bimodal behavior of the posterior PDFs on $\gamma$ (which is particularly pronounced for the GW200115 event) arises from correlations with the mass parameters. 

\clearpage

These correlations are even more significant with $\alpha$ removed from the waveform.
In both cases, the maximum of the full posterior corresponds to the peak in the marginalized PDF that is closer to $\gamma =0$. We collect the 90\% credible bounds on $\gamma$ and $\alpha$ for all events in Table \ref{tab:constraints}.  We find that, as predicted by our PCA in Sec. \ref{SEC:DM_params}, the most stringent constraints are placed on $\gamma$.
\FloatBarrier
\begin{table}[]
    \centering
    \begin{tabular}{c c c}
    \hline\hline
    Event name &$\gamma$&$\alpha$\\
    \hline
    GW170817&4.1\sci{-4}&3.5\sci{-2}\\
    GW190425&1.5\sci{-3}&1.1\sci{-1}\\
    GW200105&1.0\sci{-2}&\textbf{-}\\
    GW200115&1.7\sci{-2}&\textbf{-}\\
    \hline\hline
    \end{tabular}
\caption{\label{tab:constraints}The 90\% credible bounds on $\gamma$ and $\alpha$ obtained after parameter estimation analysis of LVK NS/NS and NS/BH events. These bounds are indicated with dashed vertical lines in Figs. \ref{fig:GW170817}, \ref{fig:GW190425} and \ref{fig:NS/BH}.}
\end{table}
\subsection{Projected constraints from future observations}
Now that we have seen what can be learned from observed signals, we look to understand what may be learned in the future. The fifth observing run (O5) for the LVK Collaboration is set to begin in 2027, and it is predicted to have a significantly improved range over O4, potentially leading to an influx of signals observed with high sensitivity \cite{KAGRA:2013rdx}.
With the enhanced sensitivity this observing run will bring, we can hope to place even more stringent constraints on dark-sector effects. 

Given this, we perform a second analysis of the synthetic GW170817-like signal detailed in the previous section. 
However, instead of using the O2 detector sensitivities,
we use analytic approximations of the detector design sensitivities during the O5 run \cite{OReilly_2022}. 
We assume that the signal is observed by the LIGO, Virgo, and KAGRA detectors. 

We perform a preliminary recovery with a 32s signal beginning at 40 Hz and with an SNR of 139. We verify that, as with the analysis of the LVK data during O2 and O3, we are unable to place constraints on $\mu$. Given this, we remove it as a model parameter and repeat our analysis, this time with a 512s-long signal, beginning at 15 Hz and with an SNR of 153. Despite the computationally expensive nature of such an analysis, the negative PN corrections in our model mean that the lower frequencies of the signal are important for constraining the DM parameters. To speed up the calculation further, we use a likelihood where $t_c$ has been analytically marginalized over in addition to $\phi_c$ and $d_L$ \cite{Farr_2022} and we do not vary over the spin parameters. 

\begin{figure}
    \centering
    \includegraphics[width=0.45\textwidth]{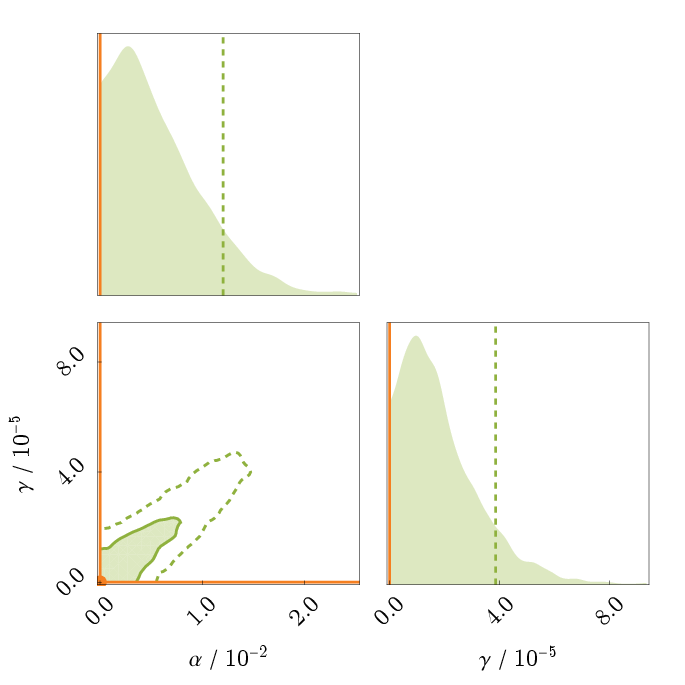}
    \caption{A partial corner plot produced after a parameter estimation analysis of synthetic binary NS data as observed by the O5 detector network at design sensitivity. The corner plot includes the marginalized one- and two-dimensional posterior PDFs on the DM parameters. In both the one- and two-dimensional plots, dashed lines represent 90\% credible regions, and in the two-dimensional plot, solid lines enclose the 50\% credible regions. The orange lines indicate the injected values. We find that $\alpha< 1.2\sci{-2}$ and $\gamma<3.9\sci{-5}$ with $90\%$ credibility. }
    \label{fig:O5_injection}
\end{figure}

In Fig. \ref{fig:O5_injection}, we present a partial corner plot that includes the posterior PDFs on $\alpha$ and $\gamma$. We find that the constraints have improved compared to the analysis of GW170817. In particular, we find that  $\alpha< 1.2\sci{-2}$ and $\gamma<3.9\sci{-5}$ with $90\%$ credibility. This is an improvement of about a factor of three and an order of magnitude, respectively.

\FloatBarrier
\section{Discussion}\label{SEC:discussion}
In this paper, we considered the effect of aDM with long-range self-interactions on NS/NS and NS/BH binaries. Assuming that the NS(s) have accumulated some amount of net dark charge, we have extended the leading PN order analysis of the waveform computed by Alexander, \textit{et al.}~\cite{Alexander:2018qzg}.
Our waveform includes corrections to the orbital energy and power radiated through GWs to 1PN order. 
We also included corrections to the power due to electric dipole and quadrupole radiation modes of the massive dark photon, which enter the waveform at -1PN and Newtonian orders, respectively.
We included these modifications in the {\tt {IMRPhenomD\_NRTidadlv2}} model, thus developing an enhanced {\tt {IMRPhenomD\_NRTidadlv2}} model that includes dark-matter effects.

With this model at hand, we then carried out a Bayesian parameter estimation analysis of the observed LVK NS/NS binary events GW170817 and GW190425 and NS/BH binary events GW200105 and GW200115. 
For all events, we were able to place constraints on the model parameter $\gamma$. 
For the NS/NS events, we were also able to place constraints on the model parameter $\alpha$. 
In particular, for GW170817, we find that $\alpha < \mathcal{O}(10^{-1})$ and $\gamma < \mathcal{O}(10^{-4})$, and for GW190425, we find $\alpha < \mathcal{O}(10^{-2})$ and $\gamma<\mathcal{O}(10^{-3})$ with 90\% credibility.
For both NS/BH events, where $\alpha = 0$, we find that $\gamma < \mathcal{O}(10^{-2})$ with 90\% credibility. 
However, because the posterior PDFs of these parameters are constrained so closely to zero, we were unable to place any constraints on the dark photon mass parameter $\mu$.

We also performed an injection-recovery analysis on synthetic data with injected parameters consistent with the GW170817 event, but observed with a detector network simulating the LVK Collaboration's fifth observing run. 
We found that the constraints on $\alpha$ and $\gamma$ improve by a factor of three and an order of magnitude, respectively. 
This result is encouraging for the prospects of future, more sensitive observing runs to place stricter constraints on the effects of this DM framework. 

As noted in Sec.~\ref{SEC:DM_params}, an improved understanding of capture mechanisms is a requirement for placing constraints on the parameters of the DM model. In particular, it will be important to delineate the parameter space for which such a large amount of DM may be captured in the lifetime of the binary system and ensure that such a model is consistent with stringent cosmological and astrophysical constraints on long-range forces in the dark sector, such as those studied in Ref.~\cite{Kopp:2018jom}. For sufficiently large gauge coupling $\gD$, the effects of DM self-capture may be important, as are the in-medium effects of the dense matter in the NS if the dark photon has a kinetic mixing with the Standard Model photon~\cite{DeRocco:2022rze}. Furthermore, the presence of the oppositely-charged fermion $\chi'$ in the atomic DM model may lead to screening effects that may modify the capture process, the dominant dark force between the binary companions, or both. We leave all of these interesting directions for future work and look forward to the continued exploration of DM through the upcoming data-rich era of GW astronomy.

\section*{Acknowledgments}
We thank Rohit Chandramouli, David Curtin, Ian Harris, Justin Ripley, Abhishek Hegade K. R., and Yiqi Xie for helpful conversations. C.B.O.~ and N.~Y.~ acknowledge support from the Simons Foundation through Award No. 896696, the NSF through Grant No.~PHY-2207650, and NASA through Grant No.~80NSSC22K0806. C.B.O is also supported by ERC Starting Grant No.~945155--GWmining, 
Cariplo Foundation Grant No.~2021-0555, 
MUR PRIN Grant No.~2022-Z9X4XS, 
MUR Grant ``Progetto Dipartimenti di Eccellenza 2023-2027'' (BiCoQ),
and the ICSC National Research Centre funded by NextGenerationEU.  Y.~K.~is supported in part by DOE grant DE-SC0015655. A.~T.~is supported by NSF Grant No. 2212983. We also acknowledge the Illinois Campus Cluster Program, the Center for Astrophysical Surveys (CAPS), and the National Center for
Supercomputing Applications (NCSA) for the computational resources that were used to produce the results presented in this paper.

\bibliographystyle{apsrev4-1}
\bibliography{citations.bib}

\appendix
\section{The Lorenz Gauge Condition for a Massive Vector Field in Curved Space}\label{SEC:app_lorenz}
In this appendix, we show that while the Lorenz gauge condition cannot, in general, be applied to a massive vector field, we recover  $\nabla_\alpha  A^\alpha=0$ to the PN order considered in this paper as a constraint enforced by DM charge conservation.  
Consider the Lagrangian in Eq. (\ref{EQ:lagrangian}) in the limit that $m_A\rightarrow0$ 
\begin{align}
    \mathcal{L}_\mathrm{DS} =&~ -\frac{1}{4}F_{\mu\nu}F^{\mu\nu}+\bar\chi\left(i\gamma^\mu D_\mu-\frac{c}{\hbar} m_\chi\right)\chi \,\,,
\end{align}
where recall that $D_\mu = \nabla_\mu + i g_\mathrm{D} A_\mu$. This Lagrangian is invariant under the simultaneous transformation
\begin{align}\label{EQ:transform}
    A_\mu \rightarrow &~ A_\mu +\frac{1}{g_\mathrm{D}} \nabla_\mu \beta(x^\nu)\,,\neweq
    \chi \rightarrow&~ e^{-i\beta(x^\nu)}\chi\,.
\end{align}
One way we can fix this gauge invariance is by imposing the Lorenz gauge $\nabla_\mu A^\mu =0$, which can be accomplished by choosing $\beta(x^\nu)$ such that $\nabla_\mu\nabla^\mu\beta(x^\nu) = -\nabla_\mu A^\mu$. 

However, if we introduce a mass term proportional to $m_A A^{\mu}A_{\mu}$, the Lagrangian is no longer invariant under the transformation in Eq.~(\ref{EQ:transform}), and we are no longer free to apply the Lorenz gauge to remove the first term in the Proca equation of Eq.~\eqref{eq:proca-eq-massaged}.  

Instead, we turn to DM charge conservation $\nabla_\alpha J^\alpha$ = 0, which is guaranteed by the Bianchi identities. Acting $\nabla_\alpha $ on both sides of the Proca equation gives us 
\begin{align}
\label{eq:newgauge}
\nabla_\alpha  A^\alpha  = -\lambda^2 R_{\alpha\beta} \nabla^\beta A^\alpha\,,
\end{align}
because the commutator of the d'Alembertian with the covariant derivative acting on a vector vanishes by the symmetry of the Ricci tensor.
We see that in flat space, when $R_{\alpha\beta} \rightarrow0$, we recover the Lorenz ``gauge condition'' as a constraint due to DM charge conservation. 

The Ricci tensor, however, does not vanish on the background of a binary system. Applying the constraint in Eq.~\eqref{eq:newgauge} to  Eq. \eqref{eq:proca-eq-massaged} gives us 
\begin{align}
-\lambda^2\nabla^\alpha\left(R_{\beta\gamma} \nabla^\beta A^\gamma \right)+R^\alpha_{~\beta} A^\beta &\nonumber\\-\nabla_\beta\nabla^\beta A^\alpha +\frac{1}{\lambda^2}A^\alpha=&~\frac{4\pi}{c}J^\alpha\, .
\end{align}
Fortunately, the first term in the above equation is $\mathcal{O}(c^{-3})$ in the time component and $\mathcal{O}(c^{-2})$ in the spatial components relative to the terms in the second line. 
Such a term will not impact our calculation of the equation of motion or energy at 1PN order, allowing us to consider $\nabla_\alpha  A^\alpha=0$ approximately in this paper.

\section{Calculation of $\phi_{(2)}$}\label{SEC:APP_phi}
Here, we solve for the 1PN correction to the scalar potential $\phi_{(2)}$, which was found in Sec. \ref{SEC:Proca} to obey Eq.~\eqref{EQ:phi_2}. 
We massage the source term $S$ defined in Eq.~\eqref{EQsource} into a more convenient form 
\begin{align}
    S= &~\left[\nabla^2-\frac{1}{\lambda^2}\right]\left[\phi_{(0)}U-\frac{\lambda}{2}\partial_t^2(r_1\phi_{(0)})\right]\eqbreak+U\left[\nabla^2-\frac{1}{\lambda^2}\right]\phi_{(0)}+\phi_{(0)}\nabla^2U -4\pi \bar J^0_{(2)} \eqbreak+\frac{2}{\lambda^2}\phi_{(0)}U\\
    =&~\left[\nabla^2-\frac{1}{\lambda^2}\right]\left[\phi_{(0)}U-\frac{\lambda}{2}\partial_t^2(r_1\phi_{(0)})\right]\eqbreak-4\pi\left(U \bar J^0_{(0)}+\bar J^0_{(2)}+\phi_{(0)}\rho\right)  \eqbreak+\frac{2}{\lambda^2}\phi_{(0)}U\,,
\end{align}
where we applied the equations of motion for $\phi_{0}$ and $U$ in Eqs.~\eqref{EQ:Ueq} and~\eqref{EQ:phi_0} to get from the first to the second line. 
We  break both $\phi_{(2)}$ and $S$ into three terms 
such that Eq.~\eqref{EQ:phi_2} becomes
\begin{align}
     &\sum_{i=1}^3\left[\nabla^2-\frac{1}{\lambda^2}\right]f_i = \sum_{i=1}^3S_i\,, 
\end{align}
where 
\begin{align}
    \label{eq:S1}
  S_1 =&~\left[\nabla^2-\frac{1}{\lambda^2}\right]\left[\phi_{(0)}U-\frac{\lambda}{2}\partial_t^2(r_1\phi_{(0)})\right]\, ,\\
     S_2 =&~-4\pi\left(U \bar J^0_{(0)}+\bar J^0_{(2)}+\phi_{(0)}\rho^{*}\right)\, ,\\
     S_3 =&~\frac{2}{\lambda^2}\phi_{(0)}U\, ,
\end{align}
and $\phi_{(2)} = f_1 + f_2 +f_3$. We solve each equation in the sum separately.

We can simplify this problem further by noting that the Proca equation is linear in $A^\alpha$, allowing us to solve for the field produced by each body separately and then add the results to obtain the total field. For the remainder of this appendix, we will consider the Proca field produced by the only first body, i.e. 
\begin{align}
\bar J^\alpha = \frac{q_1}{c}u_1^\alpha \delta^3(\mathbf{r}_1)  \quad \mathrm{and}\quad \phi_{(0)} =  q_1 \frac{e^{-r_1/\lambda}}{r_1}\,.
\end{align}
However, we cannot disregard the impact of the second body on the gravitational field background so we use 
\begin{align}
    \rho = m_1\delta^3(\mathbf{r}_1)+m_2\delta^3(\mathbf{r}_2)\quad \mathrm{and}\quad U = \frac{Gm_1}{r_1}+\frac{Gm_2}{r_2}\,.
\end{align}

With this simplification in place, the first equation is solved trivially, 
\begin{align}
 f_1 = &~\phi_{(0)}U-\frac{\lambda}{2}\partial_t^2(r_1\phi_{(0)})\nonumber\\
    =&~ G q_1\left(\frac{m_1}{ r_1}+\frac{m_2}{r_2}\right)\frac{e^{-r_1/\lambda}}{r_1}-\frac{q_1}{2}\left(\frac{\dot r_1^2}{\lambda}-\ddot r_1\right)e^{-r_1/\lambda}\,,
\end{align}
because we have written $S_1$ in terms of the differential operator in Eq.~\eqref{eq:S1}, which is a well-known strategy in PN theory (see e.g.~\cite{Alsing:2011er}). 

The second source term can be expressed explicitly in terms of the mass and DM charge densities as
\begin{align}
    S_2
    =&-4\pi q_1\Bigg[2G \left(\frac{m_1}{ r_1}+\frac{m_2}{r_2}\right) +\frac{1}{2}v_1^2+\frac{m_1}{r_1}e^{-r_1/\lambda}\Bigg]\delta^3(\mathbf{r}_1)\eqbreak-4\pi q_1\frac{m_2}{r_1}e^{-r_1/\lambda}\delta^3(\mathbf{r}r_2)\,.
\end{align}
The presence of Dirac delta functions means the equation for $f_2$ can be easily integrated via the Green's function, following e.g.~Chapter 6 of \cite{Poisson2014-cj}. Because we will ultimately evaluate the solution at the location of the second body, we seek a solution for a near-zone field point.
The solution can be broken into a near-zone part and a wave-zone part
\begin{align}
f_2=f_2^{\mathcal{N}}+ f_2^{\mathcal{W}}
\end{align}
where
\begin{align}\label{EQ:PsiNear}
    f_2^{\mathcal{N}} = \sum_l \frac{(-1)^l}{l!c^l}\left(\frac{\partial}{\partial t}\right)^l\int S_2(t,\mathbf{x'})|\mathbf{x}-\mathbf{x'}|^{l}G(\mathbf{x}-\mathbf{x'})d^3x'.
\end{align}
and $f_2^{\mathcal{W}} =0$, because $S_2$ has no support in the wave-zone. We keep only the leading-order term of $f_2^{\mathcal{N}}$, because $\phi_{(2)}$ is proportional to $c^{-2}$, giving us 
\begin{align}\label{EQ:f2}
    f_2 = \int S_2(t,\mathbf{x'}) G(\mathbf{x}-\mathbf{x'})d^3x' \, .
\end{align}
The Green's function for the Klein Gordon operator is 
\begin{align}
    G(\mathbf{x}-\mathbf{x'}) = -\frac{1}{4\pi}\frac{e^{-|\mathbf{x}-\mathbf{x'}|/\lambda}}{|\mathbf{x}-\mathbf{x'}|}\,.
\end{align}
When evaluating the Dirac delta function $\delta^3({\mathbf{r}_1})$, we apply $|\mathbf{x}-\mathbf{x'}| \rightarrow r_1$, $r_2\rightarrow r$, and $1/r_1\rightarrow 0$ via Hadamard regularization \cite{Blanchet:2000nu}. We apply analogous rules when evaluating $\delta^3({\mathbf{r}_2})$. Once the Dirac delta functions have been integrated, we obtain the solution
\begin{align}\label{EQ:f2-again}
    f_2
    =&~  
    q_1 
   \left(2G \frac{m_2}{r} +\frac{1}{2}v_1^2\right)\frac{e^{-r_1/\lambda}}{r_1} +G q_1 m_2
    \frac{e^{-r/\lambda}}{{r}}\frac{e^{-r_2/\lambda}}{r_2}\,.
\end{align}

We are left with solving the last equation for $f_3$. The third source term can be written explicitly as
\begin{align}
    S_3 = &\frac{2Gq_1}{\lambda^2} \left(\frac{m_1}{ r_1}+\frac{m_2}{r_2}\right) \frac{e^{-r_1/\lambda}}{r_1} \,.
\end{align}
We break this source term up further such that 
$f_3 = g_1 + g_2$, where $g_{1,2}$ satisfy  
\begin{align}
    \left[\nabla^2-\frac{1}{\lambda^2}\right]g_1=&~ \frac{2Gm_1q_1}{\lambda^2} \frac{e^{-r_1/\lambda}}{r_1^2}\,,\label{EQ:Green1}\\
     \left[\nabla^2-\frac{1}{\lambda^2}\right]g_2= &~\frac{2Gm_2q_1}{\lambda^2} \frac{e^{-r_1/\lambda}}{r_1r_2}\,.\label{EQ:Green2}
\end{align}
The source in Eq.~(\ref{EQ:Green1}) 
depends on the coordinates only through $r_1$, so we assume the same is true for the solution, i.e. $g_1=g_1(r_1)$. 
With this assumption, Eq.~(\ref{EQ:Green1}) becomes an ordinary differential equation 
\begin{align}
    g_1''(r_1)+\frac{2}{r_1}g_1'(r_1) - \frac{1}{\lambda^2}g_1(r_1)=  \frac{2Gm_1q_1}{\lambda^2} \frac{e^{-r_1/\lambda}}{r_1^2}\,,
\end{align}
where the primes represent derivatives with respect to argument $r_1$. We require $g_1\rightarrow0$ as $r_1\rightarrow\infty$ and obtain the family of solutions
\begin{align}
\label{eq:g1solution}
    g_1 =&~ \frac{Gm_1q_1}{\lambda} \left[e^{2 r_1/\lambda } \text{Ei}\left(-\frac{2 r_1}{\lambda }\right)-\log \left(\frac{r_1}{r_{0,1}}\right)\right]\frac{e^{-r_1/\lambda}}{r_1}\, ,
\end{align}
where $r_{0,1}$ is an integration constant with units of length that we must fix later.

Equation~(\ref{EQ:Green2}), on the other hand, cannot be recast as an ordinary differential equation in the same way because its source term depends both on $r_1$ and $r_2$. Instead, we assume that $g_2$ is a function of the positions of both bodies  $g_2=f_2(r_1,r_2)$.
We solve Eq.~(\ref{EQ:Green2})  perturbatively in the limit that the length scale of the Proca interaction is much larger than the separation of the binary $r/\lambda\ll1$. This is equivalent to the assumption that the photon mass is small relative to the scales of the system. We therefore assume that $g_2(r_1,r_2)$ can be expanded in powers of $r/\lambda$
\begin{align}
    g_2(r_1,r_2) = &~g_2^{(0)}(r_1,r_2)+\frac{r}{\lambda}g_2^{(1)}(r_1,r_2)
    \nonumber \\
    &+\frac{r^2}{\lambda^2}g_2^{(2)}(r_1,r_2)+ {\cal{O}}\left(\frac{r^3}{\lambda^3}\right)\,,
\end{align}
where $g_2^{(n)}(r_1,r_2)$ is the coefficient of the $\mathcal{O}(\lambda^{-n})$ term in $g_2(r_1,r_2)$.
We substitute this Ansatz into Eq~(\ref{EQ:Green2}), expand the exponential on the right-hand side, and collect powers of $r/\lambda$. We obtain the following set of coupled differential equations\footnote{For a function $f=f(r_1,r_2)$, the Laplacian operator becomes 
\begin{align}\label{EQ:g2}
\nabla^2f = &~f^{(2,0)}+f^{(0,2)}+\frac{{r_1}^2+{r_2}^2-{r}^2}{ {r_1} {r_2}} f^{(1,1)}\eqbreak+\frac{2}{r_1}f^{(1,0)}+\frac{2}{r_2}f^{(0,1)}
\end{align}
where  $f^{(n,m)} = (\partial_{r_1})^n (\partial_{r_2})^m f(r_1,r_2)$, and where we have applied the identity $\mathbf{r}_1\cdot\mathbf{r}_2 = (r_1^2 + r_2^2 -r^2)/2$.}
\begin{align}\label{EQ:g2_more}
    \nabla^2 g_2^{(n)} -\frac{1}{r^2} g_2^{(n-2)} = &~2Gm_2q_1 \frac{(-1)^n}{(n-2)!}\frac{1}{r_1^3 r_2 }\left(\frac{r_1}{r}\right)^n\,,~  n\ge2\,,\neweq
    \nabla^2 g_2^{(n)}  =&~ 0\,,~~~ n=0,1\,.
\end{align}

Equation~\eqref{EQ:g2_more} is significantly simpler than what we started with, so we search for a solution of the form
\begin{align}
    g_2^{(n)} = \sum_{a,b}\left[c_{n,ab} + d_{n,ab}\log (r_1+r_2+r)\right]r_1^ar_2^b\,,
    \label{eq:g2ans}
\end{align}
based on analogy with the solution for $g_1$ in Eq.~\eqref{eq:g1solution}. 
We find that when $g_1$ is expanded for $r/\lambda\ll1$ in a similar way, the coefficients take the form
\begin{align}
    g_1^{(n)} = \sum_{a}\left[c_{n,a} + d_{n,a}\log (r_1)\right]r_1^a\,.
\end{align}
Since $\nabla^2 \log (r_1) = {1}/{r_1^2}$ and $\nabla^2 \log (r_1+r_2+r) = {1}/({r_1 r_2})$, we recognize the denominators of Eq. (\ref{EQ:Green1}) and Eq. (\ref{EQ:Green2}), respectively, and make the educated guess that, if the expanded solution to  Eq. (\ref{EQ:Green1}) contains $\log (r_1)$, the expanded solution of (\ref{EQ:Green2}) will contain $\log (r_1+r_2+r)$.

We determine the values of the coefficients $c_{n,ab}$ and $d_{n,ab}$ by substituting Eq.~\eqref{eq:g2ans} into Eq.~(\ref{EQ:g2_more}), collecting powers of $r_1$, $r_2$, and $\log(r_1+r_2+r)$, and requiring that each term vanishes independently. We obtain a family of solutions in terms of certain integration constants $c_{n,ab}$.
Again, based on analogy with the solution $g_1$, we choose to fix those $c_{n,ab}$ so that we are able to construct the following (nonunique) resummed and approximate solution
\begin{align}
    g_2(r_1,r_2) = \frac{Gm_2q_1}{\lambda} \bigg\{&~e^{\frac{2 r_1}{\lambda }} \text{Ei}\left[-\frac{2 }{\lambda }\left(r_1+r_2+r\right)\right]\eqbreak-\log \left[\frac{1}{r_{0,2}}(r_1+r_2+r)\right]  \eqbreak-2e^{-(r_1+r_2+r)/\lambda}\bigg\}\frac{e^{-r_1/\lambda}}{r_1} \,,
\end{align}
where $r_{0,2}$ is an integration constant.
The above solution contains an infinite number of terms when expanded in $r/\lambda \ll 1$, but it is compatible with the family of perturbative solutions we found only through $\order{\lambda}{-2}$, so that when $g_2$ is acted upon by the operator, we obtain 
the same as in Eq.~\eqref{EQ:Green2} to ${\cal{O}}(\lambda^{-2})$.
As we shall see, such an accuracy is sufficient for the GW data analysis purposes of this paper.

Collecting results, we have the small photon mass solution
\begin{align}
    f_3 =&~ \frac{G q_1}{\lambda} \bigg\{m_1e^{2 r_1/\lambda } \text{Ei}\left(-\frac{2 r_1}{\lambda }\right)-m_1\log \left(\frac{r_1}{r_{0,1}}\right)\eqbreak +  m_2e^{\frac{2 r_1}{\lambda }} \text{Ei}\Bigg[-\frac{2 }{\lambda }\left(r_1+r_2+r\right)\Bigg]-\eqbreak m_2\log \left[\frac{1}{r_{0,2}}(r_1+r_2+r)\right] \eqbreak- 2m_2e^{-(r_1+r_2+r)/\lambda}\bigg\}\frac{e^{-r_1/\lambda}}{r_1} \,,
\end{align}
but we must be careful at $r_A = 0$. In deriving $f_3$ above, we have implicitly been assuming that $\nabla^2 r_A^{-1} = 0$, when in reality $\nabla^2 r_A^{-1} = -4\pi \delta^3(\mathbf{r}_A)$.  
Therefore, when the  operator acts on the solution above, we actually obtain
\begin{widetext}
\begin{align}
\left[\nabla^2-\frac{1}{\lambda^2}\right] 
    f_3
    &~=\frac{2Gq_1}{\lambda^2} \left(\frac{m_1}{ r_1}+\frac{m_2}{r_2}\right) \frac{e^{-r_1/\lambda}}{r_1} -4\pi\delta^3(r_1)\frac{G q_1}{\lambda} \bigg\{m_1e^{2 r_1/\lambda } \text{Ei}\left(-\frac{2 r_1}{\lambda }\right)-m_1\log \left(\frac{r_1}{r_{0,1}}\right)\nonumber \\&~+  m_2e^{\frac{2 r_1}{\lambda }} \text{Ei}\left[-\frac{2 }{\lambda }\left(r_1+r_2+r\right)\right]-m_2\log \left[\frac{1}{r_{0,2}}(r_1+r_2+r)\right] - 2m_2e^{-(r_1+r_2+r)/\lambda}\bigg\}e^{-r_1/\lambda}\,.
\end{align}
\end{widetext}
We must then subtract the term proportional to the Dirac delta function from the source and integrate it via the Green's function to correct the $f_3$ solution. The part proportional to $m_1$ is set to zero via Hadamard regularization, but the part proportional to  $m_2$ remains. Our final expression is then 
\begin{widetext}
\begin{align}
    f_3 =&~ \frac{3}{2}\frac{G q_1}{\lambda} \bigg\{m_1e^{2 r_1/\lambda } \text{Ei}\left(-\frac{2 r_1}{\lambda }\right)-m_1\log \left(\frac{r_1}{r_{0,1}}\right)
    +  m_2e^{\frac{2 r_1}{\lambda }} \text{Ei}\left[-\frac{2 }{\lambda }\left(r_1+r_2+r\right)\right]
    \nonumber \\&~
    +  m_2e^{\frac{2 r_1}{\lambda }} \text{Ei}\left[-\frac{2 }{\lambda }\left(r_1+r_2+r\right)\right]-m_2\log \left[\frac{1}{2 r}(r_1+r_2+r)\right] - 2e^{-(r_1+r_2+r)/\lambda} \nonumber\\
    &~-m_2 \text{Ei}\left(-\frac{4 r}{\lambda }\right)+ 2m_2e^{-2r/\lambda}\bigg\}\frac{e^{-r_1/\lambda}}{r_1}\,.
\end{align}
\end{widetext}
Note that any dependence on $r_{0,2}$ has dropped out of the solution, but $r_{0,1}$ remains. To fix this length scale, we turn to the finite size limit, where the bodies are no longer approximated as point particles. This calculation can be found in Appendix \ref{SEC:app_finitesize}, where we find that $r_{0,1}= \frac{\lambda}{2}e^{\gamma_E}$. We collect the solution for $\phi_{(2)} = f_1+f_2+f_3$ in Sec.~\ref{SEC:Proca}. We infer the field produced by body 2 by exchanging all body labels $1\leftrightarrow2$ and add this to our solution to obtain the field produced by both bodies. 
\section{Calculation of $I^{<ij>}$}\label{SEC:APP_I}
We compute the mass quadrupole moment following the notation of \cite{Blanchet:2001aw}. To first PN order,  
\begin{align}
I^{ij} = \int d^3x \Bigg\{&~x^{<i}x^{j>}\sigma+\frac{1}{c^2}\Bigg[\frac{1}{14}\partial_t^2\left( x^{<i}x^{j>}\sigma x^2\right)\eqbreak-\frac{20}{21}x^{<i}x^{j}x^{k>}\partial_t\left(\sigma_k\right)\Bigg]\Bigg\}+\mathcal{O}(c^{-4})
\end{align}
where the symmetric trace-free part of a tensor $ x^{ij}$ is represented by $ x^{<ij>}$. The relevant matter source densities are
\begin{align}
    \sigma(\mathbf{x},t) = &~\tilde\mu_1 \delta^3(\mathbf{r}_1)+(1\leftrightarrow2)\,,\\
    \sigma^k(\mathbf{x},t) = &~\mu_1 v_1^k \delta^3(\mathbf{r}_1)+(1\leftrightarrow2)\,,
\end{align}
in terms of the following effective masses
\begin{align}
\mu_1 =&~ m_1\left[1+\frac{1}{c^2}\left(\frac{1}{2}\left(\frac{m_2}{m}\right)^2v^2-\frac{G m_2}{r}\right)\right]\,,\\
\tilde\mu_1 =&~ \mu_1\left(1+\frac{v_1^2}{c^2}\right) =m_1\left[1+\frac{1}{c^2}\left(\frac{3}{2}\left(\frac{m_2}{m}\right)^2v^2-\frac{G m_2}{r}\right)\right]\,.
\end{align}
where we have applied $v_1 =-m_2v/m  +\mathcal{O}(c^{-2})$. The effective masses $\mu_2$ and $\tilde\mu_2$ can be obtained via the substitution $(1\leftrightarrow2)$. We use the relationship between orbital separation and velocity given in Eq.~\eqref{EQ:vsquared} and the positions in Eq. (\ref{EQ:1PN_positions}) to obtain
\begin{align}
        I^{<ij>} = &~m \eta r^{<i}r^{j>}+ \frac{m\eta}{c^2}\bigg\{-\frac{G m}{r }\frac{1}{42}\bigg[ (1+39\eta)\eqbreak-\alpha41  (1-3\eta) \left(1-\frac{r^2}{\lambda^2}\right)\bigg] r^{<i}r^{j>}\eqbreak+r^2\frac{11}{21}(1-3\eta)v^{<i}v^{j>} \bigg\}\eqbreak+\mathcal{O}(c^{-4},\mu^{3},\alpha^2)\,.
\end{align}
\section{Calculation of $\mathcal{P}_\mathrm{dark}$}\label{SEC:APP_P_dark}
Here, we compute the radiation modes sourced by the Proca field. The leading-order contribution to the power is the electric dipole mode, which enters at -1PN. We compute the next-to-leading-order corrections to the dipole radiation mode and the electric quadrupole mode here, which enter at Newtonian order. We show that there is no magnetic dipole contribution. 

Following the notation of \cite{Krause:1994ar}, 
the modes are given in terms of the Fourier coefficients of the multipole moments, 
\begin{widetext}
\begin{align}
    \mathcal{P}_\mathrm{E1} =&~ \frac{\hbar}{c^2}\frac{4}{3}\sum_{n>n_0}\left\{n^4\omega^4|\mathbf{d}_n|^2\left[1-\left(\frac{n_0}{n}\right)^2\right]^{1/2}\left[1+\frac{1}{2}\left(\frac{n_0}{n}\right)^2\right]\right\}\,,\label{EQ:PE1_gen}\\
    \mathcal{P}_\mathrm{M1} =&~ \frac{\hbar}{c^2}\frac{4}{3}\sum_{n>n_0}n^4\omega^4|\mathbf{m}_n|^2\left[1-\left(\frac{n_0}{n}\right)^2\right]^{3/2}\,,\\
    \mathcal{P}_\mathrm{E2} =&~ \frac{\hbar}{c^4}\frac{1}{90}\sum_{n>n_0}n^6\omega^6\left[1-\left(\frac{n_0}{n}\right)^2\right]^{3/2}\left\{|(Q_{ij})_n|^2+\left(\frac{n_0}{n}\right)\left[\frac{4}{3}|(Q_{ij})_n|^2 +\frac{2}{3}Q^2|\langle R^2\rangle_n|^2\right]\right\}\label{EQ:PE2_gen}\,,
\end{align}
\end{widetext}
where $n_0 = {c}/({\lambda\omega}) = \mu/x^{3/2}$, $Q$ is the total charge of the configuration, and $\mathbf{d}_n$, $\mathbf{m}_n$, $(Q_{ij})_n$, and $\langle R^2\rangle_n$ are the Fourier coefficients of the electric and magnetic dipole moments, the electric quadrupole moment, and the mean-square-charge radius of the configuration, respectively. Vertical bars indicate the magnitude of the complex quantity.  We now compute these quantities for a quasicircular binary of point DM charges. 

\subsection{Electric dipole radiation}
The electric dipole moment is 
\begin{align}
    \mathbf{d} =&~\frac{1}{c}\int \mathbf{x}J^{0} d^3x
\end{align}
where 
\begin{align}
    J^0 =&~   q_1 c[1 + \frac{1}{2 c^2}\left(v_1^2+2U\right)+\order{c}{-4}]\delta^3(\mathbf{r}_1)\eqbreak+ (1\leftrightarrow2)
\end{align}
is the zeroeth component of the four-current density. In addition to the 1PN order four-current density, we must also use the 1PN-corrected center of mass positions, given in Eq. (\ref{EQ:1PN_positions}). Doing so, we find 
\begin{align}
        \mathbf{d}&=m\eta \sqrt{\frac{G}{\hbar c}}\sqrt{\gamma_{-}} \mathbf{r}\eqbreak+\frac{m\eta}{2c^2}\sqrt{\frac{G}{\hbar c}}\bigg\{\left(\sqrt{\gamma_{-}}-\sqrt{\gamma_{+}}\Delta\right)\frac{G m}{r} 
        \nonumber \\ \nonumber 
        & 
        +\frac{1}{2} \left[(1-2\eta)\sqrt{\gamma_{-}}-\sqrt{\gamma_{+}}\Delta\right]v^2\\&~+\frac{\alpha}{2}(\sqrt{\gamma_{-}}+\sqrt{\gamma_{+}}\Delta)\Delta\frac{G m}{r}\left(1-\frac{1}{2}\frac{r^2}{\lambda^2}\right)\bigg\}\mathbf{r} \eqbreak+\mathcal{O}(c^{-3}). 
\end{align}    
We have here introduced the following dimensionless parameters: 
\begin{align}
    \gamma_{-} = &~\frac{\hbar c}{G}(\frac{q_1}{m_1}-\frac{q_2}{m_1})^2\nonumber\,,\\
    \gamma_{+} = &~\frac{\hbar c}{G}(\frac{q_1}{m_1}+\frac{q_2}{m_1})^2 = 4  \alpha + \gamma_{-}\,.
\end{align} 

In the quasicircular limit where $\dot r =0$, the time dependence of $\mathbf{d}$ is contained in $\mathbf{r} =r[ \cos(\varphi) , \sin(\varphi),0]$, with $\varphi = \omega t$. Therefore, to obtain the Fourier coefficients $\mathbf{d}_n$, we require the Fourier coefficients   
\begin{align}
    (\mathbf{r})_n = \frac{1}{2\pi}\int_0^{2\pi}e^{i n\varphi}\mathbf{r} d\varphi\,.
\end{align} 
We find that there is only one nonzero coefficient for $n>0$, namely
\begin{align}
    (\mathbf{r})_1 = \frac{r}{2}[1,i,0]\,.
\end{align}
Substituting the above expressions into Eq.~\eqref{EQ:PE1_gen}, we obtain
\begin{align}
    \mathcal{P}_\mathrm{E1} &=\frac{G}{c^3}\frac{2}{3}m^2\eta^2\omega^4 r^2\bigg[\gamma+\gamma\left(1-\Delta\right)\frac{G m}{rc^2}
    \nonumber \\ \nonumber 
    &+\frac{1}{2} \gamma\left(1-2\eta-\Delta\right)\frac{v^2}{c^2}-\frac{\alpha}{c^2} G \Delta (2\frac{G m}{r} - v^2)\eqbreak
    + \mathcal{O}(c^{-3},\alpha^2,\gamma^2,\alpha\gamma)\bigg]
    \nonumber \\  
    &\times
    \Theta(1/n_0 - 1)
\end{align}
where we have applied
\begin{align}
\sqrt{\gamma_{-}}\sqrt{\gamma_{+}} = \gamma_{-} +2  \alpha +\mathcal{O}(\alpha^2) 
\end{align}
and defined $\gamma = \gamma_{-}$.
We have also used the fact that $\left(1-n_0^2\right)^{1/2}\left(1+\frac{1}{2}{n_0}^2\right)$ can be approximated by the Heaviside function $\Theta(1/n_0 - 1)$~\footnote{The choice to approximate this nontrivial frequency dependence as a Heaviside function, here and in the electric quadrupole mode below, will lead to discontinuities in the waveform that must be mitigated by introducing terms at 2.5PN and 4PN (see Appendix A in \cite{Alexander:2018qzg}). While these terms are small relative to the 1PN corrections we consider in this paper, they could prove important when analyzing real data.  However, as we show in Sec. \ref{SEC:DM_params}, for the signals considered in this paper, the Heaviside functions that modify the electric dipole and quadrupole vector radiation modes are activated for the entirety of a signal's duration in the band. When this is the case, these additional 2.5PN and 4PN terms will completely degenerate with the coalescence time $t_c$ and phase $\phi_c$, which enter the waveform at 2.5PN and 4PN, respectively.}. At leading order, this matches the result from \cite{Alexander:2018qzg} and is off by a factor of 2 from what is in  \cite{Croon:2017zcu}. This discrepancy is noted in \cite{Alexander:2018qzg}. 
In terms of the PN parameter x, the electric dipole radiation mode is 
\begin{align}
    \mathcal{P}_\mathrm{E1} =   \frac{c^5}{G} x^5 \eta^2 &\bigg[  \frac{2}{3}\gamma\frac{1}{x} - 2 \alpha  \Delta\eqbreak -\frac{1}{9} \gamma  (3+9 \Delta +2 \eta )\eqbreak+\mathcal{O}(c^{-3},\mu^{3},\alpha^2,\gamma^2,\alpha\gamma)\bigg]  \Theta( x^{3/2}/\mu - 1)\,, 
\end{align}

\subsection{Magnetic dipole radiation}
The magnetic dipole moment is 
\begin{align}
    \mathbf{m} =&~\frac{1}{2c}\int (\mathbf{x}\times\mathbf{J}) d^3x
\end{align}
where 
\begin{align}
        \mathbf{J} =&~  q_1\mathbf{v}_1 \delta^3(\mathbf{r}_1)+ q_2\mathbf{v}_2\delta^3(\mathbf{r}_2)+\order{c}{-2}
\end{align}
is the spatial part of the four-current density for the circular binary of point DM charges at hand. We find that the magnetic dipole moment is 
\begin{align}
    \mathbf{m}
    =&~\frac{r^2\omega}{2c}m^2\eta^2\left(\frac{q_1} {m_1^2} + \frac{q_2} {m_2^2}\right)\hat{z}\,.
\end{align}
This quantity does not vary in time (in the orbital timescale) and, therefore,  $\mathcal{P}_{M1} = 0$. This result differs from the nonzero result presented in \cite{Croon:2017zcu}. 
\subsection{Electric quadrupole radiaton}
The electric quadrupole enters the radiated power at Newtonian order relative to the leading gravitational contribution. However, this is 1PN order relative to the leading-order vector mode contribution. Because we are computing only the leading- and next-to-leading-order corrections to the vector radiation modes, we need the electric quadrupole moment only at Newtonian order
\begin{align}
\label{eq:quad}
    Q_{ij} = \int d^3x J^{0} (3 x_i x_j- r^2\delta_{ij}).
\end{align}
For our configuration, the mean-square-charge radius
\begin{align}
    \langle R^2\rangle =\frac{1}{Q}\int d^3x  J^{0} r^2
\end{align}
has no time dependence and, therefore,  will not contribute to the radiation. 

The quadrupole moment only has one nonzero Fourier coefficient for $n>0$, the square of which is 
\begin{align}
    |(Q_{ij})_2|^2 = \frac{9}{4} \frac{r^4}{m^4}\left(m_2^2 q_1+m_1^2 q_2\right) +\order{c}{-2}\,.
\end{align}
With this, the electric quadrupole power is 
\begin{align}
    \mathcal{P}_\mathrm{E2} 
    =&~ \frac{c^5}{G}x^5\eta^2  \bigg[  -\frac{8}{5} \alpha  (\Delta -1)+\frac{2}{5} \gamma  (1-\Delta)^2\eqbreak
    + \mathcal{O}(c^{-2},\alpha^2,\gamma^2,\alpha\gamma)\bigg]  \Theta(x^{3/2}/\mu - \frac{1}{2})  \,, 
\end{align}
where we have approximated $\left(1 + \frac{1}{3}n_0^2\right)\left(1 - \frac{1}{4}n_0^2\right)^{3/2}$ as $\Theta(x^{3/2}/\mu - 1/2)$. Again, the result presented in this paper corrects that presented in \cite{Croon:2017zcu}. 

\section{Finite-Size Effects}\label{SEC:app_finitesize}
In computing the binary waveform, we have worked in the point particle approximation in which each body is taken to be infinitesimally small. In order to explore the effects of finite size, we now consider a simplified system: a single ball at rest in flat space. We take the ball to have radius $R$, with mass $m_0$ and DM charge $q_0$ evenly distributed throughout. Its mass and current densities are 
\begin{align}
    \rho(r) =& \frac{3}{4\pi}\frac{m_0}{R^3}\Theta(R-r)\\
    J^\alpha(r) = &\frac{3}{4\pi}\frac{q_0  }{ R^3}c\delta^\alpha_t\Theta(R-r)
\end{align}
The gravitational and Proca field potentials obey 
\begin{align}
    \nabla^2 U=& -4\pi G \rho(r)\, ,\\
    \left[\nabla^2 - \frac{1}{\lambda^2}\right] A^\alpha=& -\frac{4\pi}{c}J^\alpha(r)\,.
    \label{eq:FSAeq}
\end{align}
The gravitational potential U inside and outside the ball are 
\begin{align}
    U =& -\frac{Gm_0 r^2}{2 R^3} + \frac{c_1}{r} + c_2\,, \qquad r\leq R\,,\neweq
    U =& \frac{c_3}{r} + c_4\,, \qquad \qquad r > R\,,
\end{align}
where the constants $c_i$ must be set so that the solution satisfies the chosen boundary conditions. It is standard to require that $U\rightarrow0$ as $r\rightarrow\infty$ and that $U$ is regular as $r\rightarrow 0$. This requires $c_1=c_4=0$. The remaining two constants are chosen to enforce continuity and differentiability at the boundary $r=R$, leaving us with 
\begin{align}
    U =&~ \frac{Gm_0}{r}\left(\frac{3}{2}\frac{r}{R}-\frac{1}{2}\frac{r^3}{R^3}\right)\,,  \qquad r\leq R\,,\neweq
     U =&~\frac{Gm_0}{r}\,,  \qquad r> R\,.
\end{align}
Note that this recovers the well-known result for a spherically symmetric mass distribution: outside of the ball, the gravitational potential depends only on the mass.

Now let us consider $\phi$, the first component of the four-potential $A^\alpha = (\phi,\mathbf{A})$. The solution to Eq.~\eqref{eq:FSAeq} for $\phi$ is 
\begin{align}
    \phi =&~ 3q_0\frac{ \lambda ^2 }{R^3}+\frac{c_1 e^{-\frac{r}{\lambda }}}{r}+\frac{c_2 e^{r/\lambda }}{r}\,, \qquad r\leq R\,,\neweq
    \phi =&~ \frac{e^{-\frac{r}{\lambda }}}{r} \left(\frac{3}{2}q_0 \left(\frac{ \lambda ^2}{ R^2}-\frac{ \lambda ^3}{ R^3}\right)e^{R/\lambda }+c_3\right)\eqbreak+\frac{e^{\frac{r}{\lambda }}}{r} \left(\frac{3}{2}q_0 \left(\frac{ \lambda ^2}{ R^2}+\frac{ \lambda ^3}{ R^3}\right)e^{-R/\lambda }+c_4\right)\,,  \; r> R\,.
\end{align}
To enforce the boundary conditions as  $r\rightarrow\infty$ and $r\rightarrow0$, we choose $c_4$ so the term proportional to $e^{r/\lambda }$ is set to zero and set $c_1 = -c_2$. Again, the remaining two constants are chosen to enforce continuity and differentiability at the boundary $r=R$. The final result is 

\begin{align}
    \phi =&~3q_0\left[\frac{ \lambda ^2}{R^3}- \frac{\sinh \left(r/\lambda\right)}{r} \left(\frac{ \lambda ^3}{ R^3}+\frac{ \lambda ^2}{R^2}\right)e^{-\frac{R}{\lambda }}\right]\,,  \; r\leq R\,,\neweq
    \phi =~&  3 q_0\left[\frac{ \lambda ^2 }{R^2}\cosh \left(R/\lambda\right)-\frac{ \lambda ^3 }{R^3}\sinh \left(R/\lambda\right)\right]\frac{e^{-\frac{r}{\lambda }}}{r}\,,  \; r> R\,.
\end{align}
We find that because the field is massive, the field outside the body differs from the point particle solution and depends on the radius of the ball as well as the charge.  However, when expanding for $R/\lambda\ll1$, the point particle result is recovered to leading order outside of the ball,
\begin{align}
\phi =q_0\frac{e^{-\frac{r}{\lambda }}}{r} \left[1+ \mathcal{O}\left(\frac{R^2}{\lambda^2}\right)\right]\,,  \; r> R\,. 
\end{align}
As we compute the waveform, we enforce the assumption that the length scale $\lambda$ is much larger than the orbital separation of the binary, so it follows that, for each body, $R/\lambda\ll1$. The point particle approximation is, therefore, sufficient.

However, the finite-size calculation can inform on the value of the integration constant $r_{0,1}$ that is left undefined in our calculation of the 1PN correction to the Proca field scalar potential in Appendix \ref{SEC:APP_phi}. The constant $r_{0,1}$ appears in the solution to a differential equation of the form 
\begin{align}
    \left[\nabla-\frac{1}{\lambda^2}\right]f = \frac{1}{\lambda^2}\phi \, U \,.
\end{align}
For the toy problem studied here, the external solution to this equation is 
\begin{align}
  f =&~   \frac{3}{2}m_0 q_0 \frac{\lambda^2}{R^3}  \left[\frac{R}{\lambda} \cosh (R/\lambda)-\sinh (R/\lambda)\right]\eqbreak\times\left\{\left[c_1+\text{Ei}\left(-\frac{2 r}{\lambda }\right)\right]\frac{e^{r/\lambda }}{r} -\log(r/c_2) \frac{e^{-r/\lambda }}{r}\right\}\,, \eqbreak r> R\,. 
\end{align}
We set $c_1=0$ to enforce $f\rightarrow0$ as $r\rightarrow\infty$.  
The integration constant $c_2$ corresponds to the constant $r_{0,1}$ we found in Appendix \ref{SEC:APP_phi}. We can determine its value by enforcing continuity and differentiability with the internal solution (which we do not reproduce here because it is lengthy and uninformative) at $r=R$. In the large $\lambda$ limit, any dependence on $R$ drops out of the constant, and to leading order 
\begin{align}
    c_2 = \frac{\lambda}{2}e^{\gamma_E}+\mathcal{O}\left(\frac{R}{\lambda}\right)\,,
\end{align}
and our external solution is 
\begin{align}
    f = &~ \frac{Gm_0q_0}{\lambda} \bigg[\gamma_E+e^{2 r/\lambda } \text{Ei}\left(-\frac{2 r}{\lambda }\right)\eqbreak-\log \left(\frac{2 r}{\lambda }\right)\bigg]\frac{e^{-r/\lambda}}{r}+\mathcal{O}\left(\frac{R}{\lambda}\right)\,, \eqbreak r> R\,.
\end{align}    
In Appendix \ref{SEC:APP_phi}, we set $r_{0,1} = \lambda e^{\gamma_E}/{2}$.
\section{Degeneracies in Bayesian Parameter Estimation}\label{SEC:degeneracies}
In this appendix, we explore the impact that parameter degeneracies in a waveform can have on the posterior PDF recovered during parameter estimation.
We begin with a brief review of Bayesian statistics and then consider an illustrative toy model. 

Through Bayesian parameter estimation, we construct an approximation of the posterior PDF
\begin{align}
    P(\theta^i|s,h) = \frac{ P(\theta^i)}{P(s)}P(s|\theta^i,h)
\end{align}
which is a multidimensional PDF on the model parameters $\theta^i$, given the observed signal $s$ and the chosen model $h=h(\theta^i)$. The posterior PDF depends on the prior PDF $P(\theta^i)$, which describes beliefs about the model parameters before considering the signal. The posterior also depends on the evidence $P(s)$, which is the probability of observing the signal $s$, and serves as a normalization factor. The posterior PDF is also proportional to the likelihood PDF,
\begin{align}
    P(s|\theta^i,h)\propto \exp \left[-\frac{1}{2}\left(s - h(\theta^i)|s - h(\theta^i)\right)\right]\,,
\end{align}
which is the probability that $s$ is observed given $h$ and $\theta^i$. The difference $n =s - h(\theta^i)$ is the detector noise, and when $n=0$, the likelihood is maximized when $s =h(\theta^i)$. 

Consider the Newtonian frequency domain waveform
\begin{align}\label{EQ:hnewt}
    \tilde h  \propto&~ {(\pi f)^{-7/6}}\left(m\eta^{3/5}\right)^{5/6}\eqbreak\times\exp\bigg[-i\frac{3}{128}\left(m\eta^{3/5}\right)^{-5/3}(\pi f)^{-5/3}\bigg]\,,
\end{align}
written in terms of total mass $m = m_1+m_2$ and symmetric mass ratio  $\eta = (m_1 m_2)/(m_1+m_2)^2$. 
In Sec. \ref{SEC:DM_params}, we showed that this is not the optimal choice of parametrization and that it is much better to write this waveform in terms of a single parameter, the chirp mass $\mathcal{M} = m\eta^{3/5}$. However, for this exploration of the impact of degeneracy on parameter estimation, we choose to work with the waveform parametrized in terms of $m$ and $\eta$, as presented in Eq.~\eqref{EQ:hnewt}.  

We use the above model to construct synthetic data for a light stellar-mass BH binary with $m_1=15 M_\odot$ and $m_2 = 5 M_\odot$, corresponding to $m=20 M_\odot$ and $\eta= 3/16$. In addition to $m$ and $\eta$, the observed signal will also depend on several extrinsic parameters, most importantly the inclination angle $\iota$ and the luminosity distance $d_L$.  We set $\iota = 0$ and $d_L=100$Mpc for the synthetic injection and for the recovery model, so we do not sample on the extrinsic parameters in this analysis. We assume zero noise realization in the data, thus modeling the noise only through its spectral density. We consider Virgo and both LIGO detectors and use analytic approximations of the detector design sensitivities during the O5 run \cite{OReilly_2022}. We integrate the signal from 20Hz to $f_\mathrm{ISCO}=220$Hz, yielding an SNR of 70. We employ the same sampler settings that we used for the NS/NS binaries in Sec. \ref{SEC:Bayes}. Initially, we choose simple uniform priors:  $10~\mathrm{M}_\odot \leq m\leq 100~\mathrm{M}_\odot$ and $10/121\leq \eta\leq 1/4$ based on the assumption that for stellar-mass BHs, $5~\mathrm{M}_\odot \leq m_I\leq 50~\mathrm{M}_\odot$. However, we do not yet apply any constraints on the component mass values in our prior.

\begin{figure}[H]
    \centering
\includegraphics[width=0.42\textwidth]{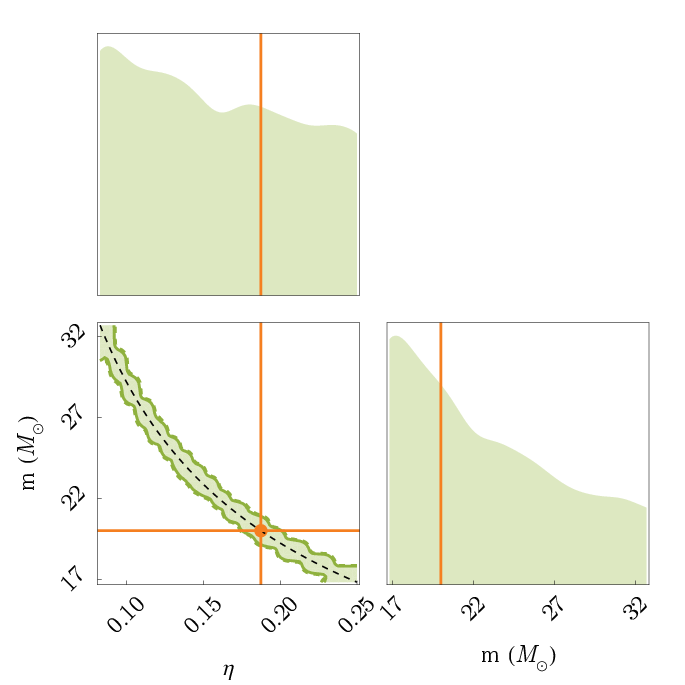}
    \caption{Corner plot obtained from a parameter estimation analysis of synthetic gravitational-wave data using the Newtonian model in Eq~\eqref{EQ:hnewt} and a uniform prior on $m$ and $\eta$. This figure includes the full two-dimensional posterior PDF and the marginalized one-dimensional distributions. In the two-dimensional panel, the solid (dashed) curve encloses the 50\% (90\%) credible region. Orange lines indicate the injected values. The black dashed line indicates the curve $m\eta^{3/5} = 20 (3/16)^{3/5} \mathrm{M}_\odot$.}
    \label{fig:degeneracies_uniform}
\end{figure}

We recover parameters with the same Newtonian model used to produce the synthetic data, eliminating any systematic effects. The full corner plot produced in this analysis is shown in Fig. \ref{fig:degeneracies_uniform}. We begin by considering the full two-dimensional posterior PDF in the bottom left panel. When the prior PDF is uniform, the posterior PDF is equivalent to the likelihood PDF in the region of parameter space supported by the prior. Because we have injected no noise realization into the data, the likelihood PDF, and therefore the posterior PDF, is maximized when the signal matches the model. As we might have expected, while we injected $m= 20\mathrm{M}_\odot\,,~\eta=3/16,$ the posterior PDF is maximized for a family of solutions for which $m\eta^{3/5} = 20 (3/16)^{3/5} \mathrm{M}_\odot$. Any solution along this curve, indicated with a dotted black line in Fig. \ref{fig:degeneracies_uniform}, is equally likely given the exact posterior PDF.  However,  rather than computing the exact posterior PDF, we constructed it approximately through discrete sampling. We therefore find that the two-dimensional posterior PDF does have a ``maximum'' (at $m \sim 17.69 \mathrm{M}_\odot$ and $\eta\sim 0.23$) that is a result of sampling error and has no relation to the injection values. 

We now consider the marginalized one-dimensional posterior PDFs in Fig. \ref{fig:degeneracies_uniform}. We see that while there is a family of equally probable solutions in the full posterior PDF, the marginalized distributions are not uniform. This illustrates the way that correlations between parameters can lead to the accumulation of ``artificial'' peaks in the marginalized distributions. Such peaks can be biased from the injection parameters (indicated in orange) even in the absence of systematic error. Note that the nonuniform structure of the marginalized PDFs is not an artifact of discrete sampling, but rather a consequence of the slope of the degeneracy region in the $m$-$\eta$ two-dimensional posterior.  

\begin{figure}[H]
    \centering
        \includegraphics[width=0.45\textwidth]{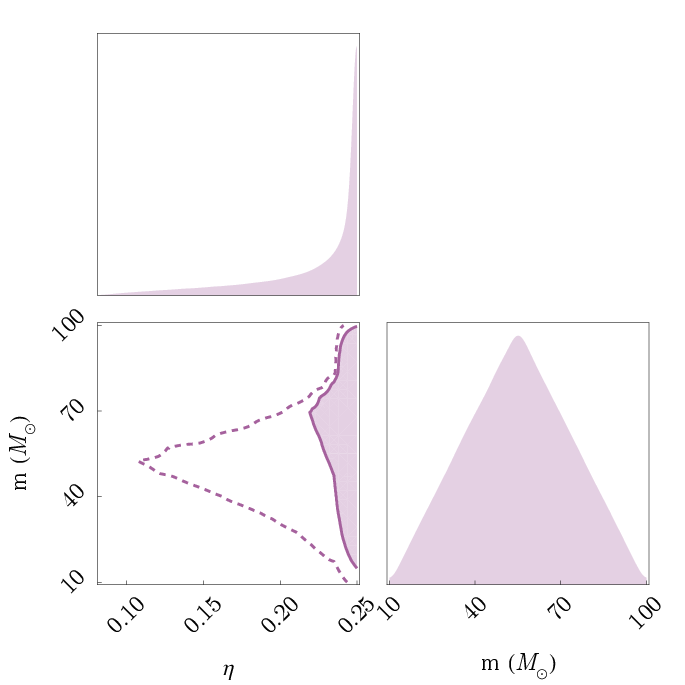}
    \caption{Corner plot depicting the joint priors on the total mass $m$ and the symmetric mass ratio $\eta$, assuming a uniform distribution in the component masses, with the restrictions $5~\mathrm{M}_\odot \leq m_I\leq 50~\mathrm{M}_\odot$ and $m1>m2$.  In the two-dimensional panel, the solid (dashed) curve encloses the 50\% (90\%) credible region.}
    \label{fig:priors}
\end{figure}

\begin{figure}[H]
    \centering
    \includegraphics[width=0.45\textwidth]{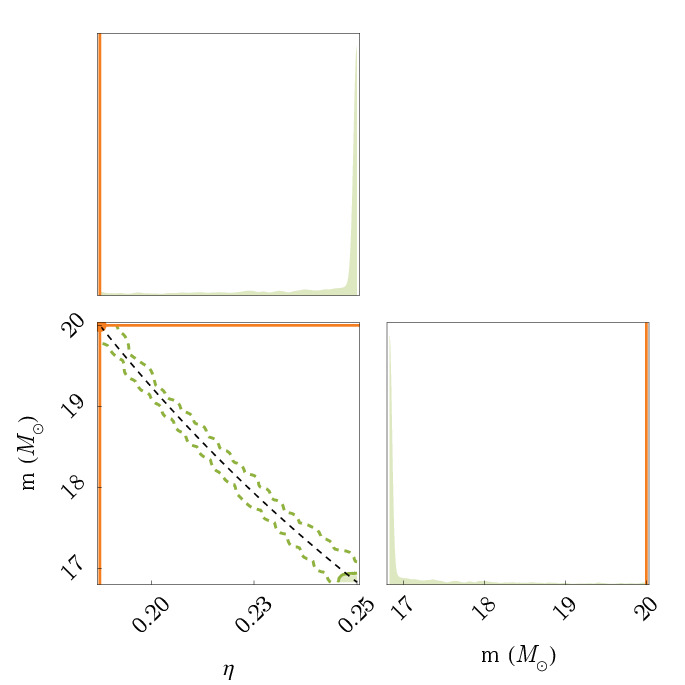}
    \caption{Corner plot produced after a parameter estimation analysis of synthetic gravitational-wave data using the Newtonian model in Eq~\eqref{EQ:hnewt} and the priors of Fig.~\ref{fig:priors}. We show the full two-dimensional posterior PDF and the marginalized one-dimensional distributions. In the two-dimensional panels, the solid (dashed) curve encloses the 50\% (90\%) credible region. Orange lines indicate the injected values. The black dashed line indicates the curve $m\eta^{3/5} = 20 (3/16)^{3/5} \mathrm{M}_\odot$}.
    \label{fig:degeneracies_nonuniform}
\end{figure}

The previous analysis employed a uniform prior PDF on $m$ and $\eta$, but it is better physically motivated to use a prior that corresponds to a uniform distribution in the component masses, with the restrictions $5~\mathrm{M}_\odot \leq m_I\leq 50~\mathrm{M}_\odot$ and $m_1>m_2$. This leads to nonuniform priors in $m$ and $\eta$, as illustrated in Fig.~\ref{fig:priors}. We see that this prior PDF strongly favors equal-mass binaries ($\eta = 1/4$) and total masses in the middle of the allowed range. We repeat the recovery analysis with the new prior PDF. The corner plot produced in this analysis is shown in Fig. \ref{fig:degeneracies_nonuniform}. We see that now the posterior PDF is maximized for an equal mass binary, which corresponds to $m\sim 16.83 \mathrm{M}_\odot$ in the family of solutions along the dotted black curve. We see that because the likelihood PDF is so degenerate, the behavior of the posterior PDF is dominated by the behavior of the prior PDF and not by the likelihood PDF. 
In actual parameter estimation analyses of GW data, we never use waveforms with parametrizations as degenerate as the one used in this toy model. However, the DM corrections we have computed in Sec.~\ref{SEC:Waveform}, depend on both the dark-matter parameters and the mass parameters, and there is some amount of degeneracy between them.
Through this exploration, we have seen three ways in which parameter degeneracies can lead to bias in recovered parameters:  (i) sampling error can lead to a biased maximum in the full posterior, (ii) correlations between parameters can lead to biased peaks in the marginalized posterior PDFs, and (iii) the behavior of the posterior PDF can be dominated by the prior PDF, rather than the likelihood PDF.

\end{document}